\documentclass[11pt,a4paper,twoside]{article}
\newif\ifdraft \global\drafttrue

\pdfoutput=1

\usepackage[expansion=true]{microtype}

\usepackage[T1]{fontenc}
\usepackage[utf8]{inputenc}
\usepackage{times}
\usepackage{graphicx}
\usepackage{mathrsfs}
\usepackage{subfig}

\usepackage{hyperref}
\usepackage{xcolor}
\colorlet{darkblue}{blue!50!black}
\hypersetup{
    colorlinks,%
    citecolor=darkblue,%
    filecolor=red,%
    linkcolor=darkblue,%
    urlcolor=magenta,%
    pdfnewwindow=true,%
    pdfstartview={FitH}
}
\usepackage{fancyhdr}
\usepackage{latexsym}
\usepackage{amsmath}
\usepackage{amsthm}

\usepackage{bbm}
\usepackage{bm}
\usepackage{amssymb}
\usepackage{amsfonts}
\usepackage{enumerate}
\usepackage{cancel}
\usepackage[shortlabels]{enumitem}
\usepackage{xspace}
\setlist{noitemsep,topsep=0pt,parsep=5pt,partopsep=0pt}
\usepackage{stmaryrd}
\usepackage{etoolbox}
\usepackage{needspace}

\ifdraft
\definecolor{refkey}{gray}{.25} 
\definecolor{labelkey}{gray}{.25}
\usepackage[left=3cm, right=3cm, bottom=3cm, top=3cm, marginparwidth=2.6cm, marginparsep=2mm]{geometry}
\else
\usepackage[margin=3cm]{geometry}
\fi

\newcounter{smallarabics}

\newcounter{smallroman}

\newcommand{\ben}{\begin{enumerate}[{\rm (i)}]}
\newcommand{\een}{\end{enumerate}}

\newtheorem{theorem}{Theorem}[section]
\newtheorem{proposition}[theorem]{Proposition}
\newtheorem{lemma}[theorem]{Lemma}

\newtheorem{definition}[theorem]{Definition}

\theoremstyle{definition}
\newtheorem{remark}[theorem]{Remark}

\AtBeginEnvironment{theorem}{\Needspace{5\baselineskip}}
\AtBeginEnvironment{proposition}{\Needspace{5\baselineskip}}

\usepackage{marginnote}

\makeatletter
\def\namedlabel#1#2{\begingroup
   \def\@currentlabel{#2}%
   \label{#1}\endgroup
}
\makeatother

\newenvironment{assumption}[1]{\begin{quote}{\bf Assumption #1} \namedlabel{ass:#1}{#1}}{\end{quote}}
\newcommand{\assref}[1]{{\bf \ref{ass:#1}}}

\def\rr{{\mathbb R}}
\def\zz{{\mathbb Z}}
\def\qq{{\mathbb Q}}
\def\cc{{\mathbb C}}
\def\nn{{\mathbb N}}

\def\textsl{{}}

\def\c0inf{C_0^\infty}
\def\s{{\rm s}}

\def\proof{\noindent {\bf Proof.}\ \ }

\def\cH{{\cal  H}}

\def\cL{{\cal L}}
\def\PP{\mathbb{P}}
\def\EE{\mathbb{E}}
\def\QQ{\mathbb{Q}}
\def\II{\mathbb{I}}

\def\cB{{\cal B}}

\def\i{{\rm i}}

\newcommand{\bigtimes}{\mathop{\hbox{\mathsurround=0mm\LARGE$\times$}}}

\newcommand{\e}{\mathrm{e}}
\renewcommand{\i}{\mathrm{i}}

\renewcommand{\d}{\mathrm{d}}

\newcommand{\ep}{\mathrm{ep}}
\newcommand{\defeq}{\mathrel{\mathop:}=}
\newcommand{\eqdef}{=\mathrel{\mathop:}}

\newcommand{\atanh}{\operatorname{argth}}
\renewcommand{\sinh}{\operatorname{sh}}
\renewcommand{\cosh}{\operatorname{ch}}
\renewcommand{\tanh}{\operatorname{th}}

\newcommand{\sign}{\operatorname{sign}}

\makeatletter
\newcommand{\pushright}[1]{\ifmeasuring@#1\else\omit\hfill$\displaystyle#1$\fi\ignorespaces}
\newcommand{\pushleft}[1]{\ifmeasuring@#1\else\omit$\displaystyle#1$\hfill\fi\ignorespaces}
\makeatother

\def\qed{\hfill$\Box$\medskip}
\def\cS{{\cal S}}
\def\cP{{\cal P}}
\def\cJ{{\cal J}}

\def\cT{{\cal T}}
\def\cF{{\cal F}}
\def\cI{{\cal I}}

\def\cA{{\cal A}}

\def\cK{{\cal K}}

\def\bar{\overline}
\def\ubar{\underline}

\def\12{\frac{1}{2}}

\def\supp{\operatorname{supp}}

\def\e{{\rm e}}

\def\d{{\rm d}}
\def\Ran{{\rm Ran}\,}

\def\one{{\mathbbm 1}}

\def\cH{{\cal H}}

\def\r{{\overline{\rm  v}}}

\def\sp{{\rm sp}}
\def\cS{{\cal S}}

\def\cX{{\cal X}}

\def\fJ{\mathfrak{J}}

\def\s{{\rm s}}
\def\tr{{\operatorname{tr}}}

\newcommand{\ds}{\displaystyle}
\def\ie{{\sl i.e., }}

\def\wrho{{\widehat{\rho}}}
\def\wcJ{\widehat{\cJ}}
\def\wP{{\widehat{\mathbb P}}}

\newcommand{\beq}[1]{\begin{equation}\label{#1}}
\newcommand{\eeq}{\end{equation}}

\newcommand{\lbr}{\llbracket}
\newcommand{\rbr}{\rrbracket}
\numberwithin{equation}{section}

\begin{document}
\author{T. Benoist$^{1}$, N. Cuneo$^{2}$, V. Jak\v{s}i\'c$^{3}$,  C-A. Pillet$^{4}$\\ \\
\\ \\
$^1$Institut de Math\'ematiques de Toulouse, UMR5219, Universit\'e de Toulouse,\\
CNRS, UPS IMT, F-31062 Toulouse Cedex 9, France
\\ \\
$^2$Laboratoire de Probabilit\'es, Statistique et Mod\'elisation (LPSM)\\  Université de Paris - CNRS  - Sorbonne Université, F-75205 Paris Cedex 13, France
\\ \\
$^3$Department of Mathematics and Statistics,
McGill University, \\
805 Sherbrooke Street West,
Montreal,  QC,  H3A 2K6, Canada\\ \\
$^4$Aix Marseille Univ, Université de Toulon, CNRS, CPT, Marseille, France
}

\title{On entropy production of repeated quantum measurements II.\\
Examples}
\date{}
\maketitle

\bigskip
\centerline{\large \bf Dedicated to Joel  Lebowitz on the occasion of his 90th birthday}

\bigskip
\bigskip
\bigskip

\thispagestyle{empty}

{\small\textbf{Abstract.} We illustrate the mathematical theory of entropy production in repeated quantum measurement processes developed in a previous work by studying examples of quantum instruments displaying various interesting phenomena and singularities. We emphasize the role of the thermodynamic formalism, and give many examples of quantum instruments whose resulting probability measures on the space of infinite sequences of outcomes (shift space) do not have the (weak) Gibbs property. We also discuss physically relevant examples where the entropy production rate satisfies a large deviation principle but fails to obey the central limit theorem and the fluctuation--dissipation theorem. Throughout the analysis, we explore the connections with other, a priori unrelated topics like functions of Markov chains, hidden Markov models, matrix products and number theory.


}
\tableofcontents

\setlength{\parskip}{4pt}

\section{Introduction}

This paper is a companion to~\cite{BJPP1} and its goal is to complement the general theory of repeated quantum measurement processes developed therein. In the present paper, we  explore the rich class of invariant probability measures on shift spaces that result from such repeated quantum measurement processes and  show that, depending on the physical models and their parameters (sometimes even on the number-theoretic properties of the parameters), very interesting and singular properties arise. We discuss connections with classes of measures that have been widely studied in the literature (Gibbs and weak Gibbs measures, Markov measures, hidden Markov models, matrix product measures, ...). In many physically motivated examples, we obtain explicit expressions for the mean entropy production and the associated entropic pressure.

We refer the reader to the general introduction of~\cite{BJPP1} for historical  perspectives and physical motivations. We also note that a repeated measurement process
can be viewed as a singular kind of repeated interaction process. Recently, the latter have been intensively studied
as alternative to stochastic modeling of thermodynamic behavior in open quantum systems (see the review~\cite{BJM} and 
references therein). Irrespective  of  physical motivations, 
many examples analyzed here might be of interest to researchers in the fields of dynamical systems, probability and statistics; see Section~\ref{sec-hmcp}.

The paper is organized as follows. In Section~\ref{sec-direct-measurements}, we
briefly discuss the nature of time evolution and its reversibility in quantum mechanics, complementing  Section 1.1 in~\cite{BJPP1}.
The following  Sections~\ref{sec-setup}--\ref{sec-probes} are devoted to a review of the setup and results of~\cite{BJPP1}, combined with refinements 
established in~\cite{CJPS-2017}. 
The role of the thermodynamic formalism in the general theory is discussed in Section~\ref{sec-thermo}. In Section~\ref{sec-remarks} we  finish the introduction with some additional remarks. 
In Section~\ref{sec-examples} we outline the main results of the paper.
The proofs are given in Sections~\ref{sec:flip--keep}, \ref{spin-main}, and~\ref{sec-ex-ro}. We postpone a more detailed description of the structure of Sections~\ref{sec-examples}--\ref{sec-ex-ro} to the beginning of Section~\ref{sec-examples}, after some notation and concepts are introduced.

This paper will be followed by the works~\cite{Dima1,DetBal}, dealing respectively with the statistical mechanics of repeated quantum measurements\footnote{Many additional examples are considered in~\cite{Dima1}.} and  with the quantum detailed balance condition.

It is an honour and pleasure to dedicate this work to Joel Lebowitz on the occasion of his 90th birthday. Throughout the years the two senior authors VJ and CAP  learned  greatly from Joel and benefited from  his wisdom, generosity and friendship. They will always remain grateful for that. 

\medskip\noindent{\bf Acknowledgments.} 
This research was supported by
the \textit{Agence Nationale de la Recherche\/} through the grant NONSTOPS
(ANR-17-CE40-0006-01, ANR-17-CE40-0006-02, ANR-17-CE40-0006-03), and the CNRS
collaboration grant \textit{Fluctuation theorems in stochastic systems\/}. 
Additionally, this work received funding by the CY Initiative of Excellence (grant ``Investissements d'Avenir'' ANR-16-IDEX-0008) and was developed during VJ's stay at the CY Advanced Studies, whose support is gratefully acknowledged.  The research of TB was supported by ANR-11-LABX-0040-CIMI within the program ANR-11-IDEX-0002-02. NC and TB were also supported by the ANR grant QTRAJ (ANR-20-CE40-0024-01). VJ acknowledges the support of NSERC. The work
of CAP has been carried out in the framework of the Labex Archim\`ede
(ANR-11-LABX-0033) and of the A*MIDEX project (ANR-11-IDEX-0001-02), funded by
the ``Investissements d'Avenir'' French Government programme managed by the
French National Research Agency (ANR).  We wish to thank D. Roy and D. Jakobson for useful discussions.

\subsection{Direct measurements and the time irreversibility of the projection postulate}
\label{sec-direct-measurements}

According to most introductory textbooks on quantum mechanics (see, e.g., \cite[Chapter~III]{CTDL1}), and following the initial formulation by von~Neumann~\cite{vonNeumann}, there are two ways the state of a quantum system can evolve. If the system is kept isolated during some period of time, then its state (represented by a density matrix\footnote{A density matrix $\rho$ is a non-negative operator whose trace satisfies $\tr(\rho)=1$.} $\rho$) is updated according to the deterministic rule
\beq{eq:unitary_evol}
\rho\mapsto\rho'=U\rho U^\ast,
\eeq
where $U$ is a unitary operator related to the total energy of the system through the time-dependent Schrödinger equation. The second one is related to measurements: if the system is subject to the direct measurement of a physical observable, then its state after the measurement will depend on the outcome of the measurement, which is a random quantity. A measurement allowing only a finite number of distinct outcomes labeled by a finite alphabet $\cA$ can be modeled by a resolution of the identity indexed by $\cA$, \ie a family $(P_a)_{a\in\cA}$ of mutually orthogonal projections summing up to the identity. By the projection postulate, the measurement updates the system state according to the stochastic rule\footnote{Such an instantaneous direct measurement is, of course, an idealization, most real measurements are indirect and require some finite time to complete (see~\cite{FGH19} for a concrete study).}
\beq{eq:projection_postulate}
\rho\mapsto \widetilde \rho_a=\frac{P_a\rho P_a}{p_a},
\eeq
where $p_a=\tr(P_a\rho)$ is the probability for the outcome $a$ to occur. As a result, for the observer unaware of this outcome,
the state of the system after the measurement is
\beq{eq:projection_postulate2}
\widetilde \rho=\sum_{a\in\cA}p_a\widetilde \rho_a=\sum_{a\in\cA}P_a\rho P_a.
\eeq
As already noticed by von~Neumann, the deterministic evolution~\eqref{eq:unitary_evol} does not increase the statistical uncertainty of the state. More precisely, the von~Neumann entropy, $S(\rho)=-\tr(\rho\log\rho)$, remains constant during this first kind of evolution: $S(\rho')=S(\rho)$. In contrast, the stochastic rule~(\ref{eq:projection_postulate},\ref{eq:projection_postulate2}) increases this statistical uncertainty in the sense that\footnote{See~\cite[Section~V.3]{vonNeumann}. In modern parlance, this follows from the operator-convexity of the function $x\mapsto x\log x$, see, e.g., \cite{carlen-2010}.} $S(\widetilde \rho)\geq S(\rho)$, with strict inequality unless each projection $P_a$ commutes with $\rho$ so that $\widetilde \rho=\rho$.  Inspired by a strong analogy with thermodynamics, the peculiar properties of quantum measurements were sometimes proposed to be at the origin of the thermodynamic arrow of time (see for example~\cite{vonNeumann,Bohm}).
 
Note that when the initial state has maximal entropy, any measurement yields $S(\widetilde \rho)=S(\rho)$, and hence $\widetilde \rho=\rho$. This applies to systems with finite-dimensional Hilbert spaces when $\rho$ is a multiple of the identity. More generally, if $\rho$ maximizes the entropy under some linear constraints $\tr(\rho C_j)=\gamma_j$, it is easy to see that any measurement such that each projection $P_a$ commutes with all $C_j$ gives a $\widetilde \rho$ satisfying the same constraints. Hence, again, $\widetilde \rho=\rho$. This also applies to constraints of the type $\rho=Q_j\rho=\rho Q_j$ where the orthogonal projections $Q_j$ commute with the $P_a$ (thus, in particular, to superselection rules).  This is consistent with the commonly encountered assertion that the low entropy of the initial state of the universe is responsible for the observed irreversibility of its evolution. 
  
In this respect, it is worth noticing that in his well-known discussion of the arrow of time and Boltzmann's entropy~\cite{lebowitz1993macroscopic}, Lebowitz wrote a section on quantum mechanics where he refers to one of his early works with Aharonov and Bergmann~\cite{ABL}. In the latter, building on Wigner's motto {\sl ``the laws of quantum mechanics only furnish probability connections between results of subsequent observations carried out on a system''}~\cite{wigner}, the authors take Born's rule as their starting point and study the reversibility of successive measurements from a statistical ensemble point of view. 
In a spirit similar to~\cite{ABL},  our research program follows the footsteps of Heisenberg and von Neumann, and can be broadly described as the study of the statistical properties of quantum-mechanical probabilistic rules and the resulting irreversibility in the specific setting of repeated quantum measurement processes. Its general description is given in the introduction of~\cite{BJPP1}, and in the sequel we will occasionally further comment on it.

Returning to~\cite{ABL}, in Section \ref{sec-vn-instruments}, we reformulate successive direct measurements in our language, and recover the reversibility of the measurement outcome sequence, that is, the first crucial observation made in \cite{ABL}. From a technical perspective, the other results in~\cite{ABL} are based on the above remarks about states of maximal entropy and on the ability to implement pre- and post-selection mechanisms by fixing the outcome of the first and the last measurement. We do not discuss these observations in the present article. Compared to \cite{ABL}, we are concerned with the repetition of the same direct or indirect measurement with the possibility of added internal and external dynamical contributions. The introduction of these dynamical contributions may result in non-zero entropy production and therefore in irreversibility. In an upcoming publication~\cite{DetBal}, we provide necessary and sufficient conditions on the measurement to generate non-vanishing entropy production and therefore irreversibility.

\subsection{Setup}
\label{sec-setup}

Throughout the paper we will use the following notations and conventions. $\cH$ denotes a Hilbert space of finite dimension $d$.  The inner product of two elements $u,v\in\cH$, denoted by $\langle u\,|\,v\rangle$, is assumed to be linear in its second argument. $\cB(\cH)$ is the $C^\ast$-algebra  of linear operators on $\cH$, and $\one$ denotes its unit. A self-adjoint element $X\in\cB(\cH)$ is positive, written $X\ge0$, if its spectrum $\sp(X)$ is a subset of $[0,\infty{[}$. It is strictly positive, written $X>0$, if $\sp(X)\subset{]}0,\infty{[}$.
A linear map $\Phi:\cB(\cH)\to\cB(\cH)$ preserving positivity is called positive.  We denote by $\Phi^\ast$ the adjoint of $\Phi$ w.r.t.\;the Hilbert--Schmidt inner product $(Y\,|\,X)=\tr(Y^\ast X)$ on $\cB(\cH)$. A positive map $\Phi$  is called  irreducible if the relation $\Phi[P]\le\lambda P$ for some orthogonal projection $P$ and some $\lambda>0$ holds only for  $P\in\{0,\one\}$ (see~\cite{EHK}).
 A positive map $\Phi$ is called positivity improving if $\Phi[X]>0$ for any $X\geq 0$, $X\not=0$.
 $\Phi$ is completely positive whenever $\Phi\otimes\mathrm{Id}$ is positive as a map on $\cB(\cH)\otimes\cB(\cK)\simeq\cB(\cH\otimes\cK)$ for any finite-dimensional Hilbert space $\cK$.

We shall consider measurements whose possible outcomes can be labeled by a finite alphabet $\cA$.  The measurement process itself  is described by a pair $(\cJ,\rho)$, where  $\cJ\defeq(\Phi_a)_{a\in\cA}$ is a quantum instrument  on $\cH$, and   $\rho$ is the initial state of the system.
Here, the $\Phi_a$ are completely positive maps on $\cB(\cH)$ such that
\[
\Phi\defeq\sum_{a\in\cA}\Phi_a
\]
satisfies $\Phi[\one]=\one$,  and  $\rho$ is identified with a density matrix on $\cH$. We write\footnote{For an integer $T$, $\cA^T$ denotes the $T$-fold cartesian product of the set $\cA$ with itself.} $\Omega_T\defeq\cA^T$, and denote its elements by $\omega=(\omega_1,\ldots,\omega_T)$, and sometimes simply $\omega_1\cdots\omega_T$. For $\omega\in \Omega_T$, we define
\[
\Phi_\omega\defeq\Phi_{\omega_1}\circ\cdots\circ\Phi_{\omega_T}.
\]
The probability measure defined on  $\Omega_T$ by the mass function
\beq{eq:defPttr}
	\PP_T(a_1,\ldots,a_T)\defeq\tr\left(\rho\,(\Phi_{a_1}\circ\cdots\circ\Phi_{a_T})[\one]\right)
\eeq
describes  the statistics of the first $T$ outcomes of the repeated measurement process generated by $(\cJ,\rho)$; see~\cite{BJPP1} and Section~\ref{sec-probes} for details. 
By a slight abuse of notation, we shall also use the symbol $\PP_T$
to denote this probability measure.

The set of infinite sequences of measurement outcomes is\footnote{$\nn$ denotes the set of natural integers including $0$, and $\nn^\ast\defeq\nn\setminus\{0\}$. $\cA^{\nn^\ast}$ is the set of sequences $\omega=(\omega_1,\omega_2,\ldots)$.} $\Omega\defeq\cA^{\nn^\ast}$, 
and we  equip it  with  the metric  $d(\omega, \omega^\prime)\defeq\lambda^{k(\omega, \omega^\prime)}$,
where $\lambda \in {]}0,1{[}$ is fixed and $k(\omega, \omega^\prime)\defeq\inf\{t\in\nn^\ast\mid\omega_t\not=\omega_t^\prime\}$. This metric generates the product topology and the pair $(\Omega, d)$ is a compact metric space. $\cF$ denotes the associated Borel $\sigma$-algebra, $\cP(\Omega)$ the set of probability measures on $(\Omega,\cF)$ and $\supp\PP$ the support of $\PP \in \cP(\Omega)$, which consists of all points $\omega \in \Omega$ such that $\PP_T(\omega_1, \ldots, \omega_T) > 0$ for all $T\in \nn^\ast$.  We equip  $\cP(\Omega)$ 
with the topology of weak convergence which turns it into a compact metric space.

The following terminology will also be used in the sequel.  We take the convention that $\Omega_0$ contains only the empty word which has length zero and mass $\PP_0()=1$.  We denote by
\[
\Omega_{\rm fin}\defeq\bigcup_{T\in\nn}\Omega_T
\]
the set of finite words from the alphabet $\cA$. 
For $\omega\in\Omega$ and $S,T\in\nn^\ast$, $S\leq T$, we set\footnote{For any integers $ i\le j$ we set $\lbr i,j\rbr=[i,j]\cap\zz$.}
$$
\omega_{\lbr S, T\rbr}\defeq(\omega_S, \dots,\omega_T)\in\Omega_{T-S+1}.
$$
Finally, for $a\in \cA$, $a^T$ is the word $\omega_1\cdots\omega_T$ where $\omega_i=a$ for all $i\in\lbr1,T\rbr$.

The length of $\eta=(\eta_1, \dots, \eta_T) \in\Omega_{\rm fin}$
is $|\eta|=T$, the cylinder with base $\eta$ is the set\footnote{By convention, the cylinder with empty base is $\Omega$.}
\[
[\eta]\defeq\left\{\omega\in\Omega\mid \omega_{\lbr1,T\rbr}=\eta\right\},
\]
and $\eta\omega$ denotes the concatenation of $\eta$ with $\omega\in\Omega_\mathrm{fin}\cup\Omega$,
\ie
$$
\eta\omega\defeq(\eta_1,\ldots,\eta_T,\omega_1,\omega_2,\ldots).
$$
  
There is a unique $\PP\in\cP(\Omega)$ such that, for all $\eta\in\Omega_{\rm fin}$
\[
\PP([\eta])=\PP_{|\eta|}(\eta).
\]
This measure describes the  complete statistics of the repeated measurement process  generated by $(\cJ,\rho)$.
We shall denote by $\EE$ the expectation w.r.t.\;the measure $\PP$.
With a slight abuse of terminology, we shall sometimes say that the pair $(\cJ,\rho)$ is a quantum instrument and
refer to $(\Omega,\cF,\PP)$ or simply to $\PP$ as its {\em unraveling.} 

Different instruments acting on possibly different Hilbert spaces may have the same unraveling.  Given a probability measure $\PP$ on $(\Omega,\cF)$, we denote by $\fJ_\PP$ the collection of all instruments 
$(\cJ,\rho)$ whose unraveling is $\PP$. $\fJ_\PP$ might be empty, but it is never a singleton.
Although our analysis is focused on unravelings, their instrumental origin is of central importance regarding the formulation, relevance,  and interpretation of our results. 

We shall only consider instruments  $(\cJ,\rho)$  satisfying the following assumption:
\begin{assumption}{(A)} $\Phi^\ast[\rho]=\rho$ and $\rho >0$. 
\end{assumption}

It is worth pointing out that assuming  $\rho >0$ in~\assref{(A)}  is not a restriction. Setting $\cK=\Ran \rho$ and using $\Phi^\ast[\rho]=\rho$, one easily shows that each $\Phi_a$ preserves the subspace $\cB(\cK)$ of $\cB(\cH)$, and that the restriction of the instrument 
$\cJ$ to $\cK$ is also an instrument. By replacing $\cH$ with $\cK$ one then obtains an instrument with the same unraveling  for which~\assref{(A)}  holds.

Assumption~\assref{(A)} has several important consequences. 
\ben
\item The unraveling  $\PP$ of $(\cJ,\rho)$  is invariant under the left-shift
\[
\begin{array}{rccc}
\phi:&\Omega&\to&\Omega,\\
&(\omega_1,\omega_2,\ldots)&\mapsto&(\omega_2,\omega_3,\ldots),
\end{array}
\]
\ie $\PP\circ\phi^{-1}=\PP$. We denote by $\cP_\phi(\Omega)$ the set of $\phi$-invariant elements of $\cP(\Omega)$. 
\item Extending the function $\PP_T$ by setting $\Omega\ni\omega\mapsto\PP_T(\omega)=\PP_T(\omega_1,\ldots,\omega_T)$,
the following upper-decoupling property holds~\cite[Lemma 3.4]{BJPP1}:  for all $T,S\in\nn^\ast$ and $\omega\in\Omega$, 
\beq{dec-i}
\PP_{T+S}(\omega)\leq\lambda_0^{-1}\PP_T(\omega)\PP_S\circ\phi^T(\omega),
\eeq
where $\lambda_0=\min \sp(\rho) \in{]}0,1]$, and $\phi^T$ denotes the $T$-fold composition of $\phi$ with itsef.  
\item The so-called {\em outcome reversal} (abbreviated OR) is well defined.  Its construction involves 
a choice of  involution 
$\theta: \cA\rightarrow\cA$. 
Given $(\cJ,\rho)$ satisfying~\assref{(A)}  with ${\cJ}=(\Phi_a)_{a\in\cA}$, the pair $(\widehat {\cJ}, \widehat \rho)$ with
 $\widehat{\cJ}\defeq(\widehat{\Phi}_a)_{a\in\cA}$ and $\widehat \rho$ defined by 
\beq{eq:canonical_OR_instrument}
\widehat{\Phi}_a[X]\defeq\rho^{-\12}\Phi_{\theta(a)}^\ast\left[\rho^\12X\rho^\12\right]\rho^{-\12},\qquad \widehat{\rho}\defeq\rho,
\eeq
also satisfies~\assref{(A)}, and  the unraveling $\wP$ of $(\widehat {\cJ}, \widehat \rho)$  is determined by 
\beq{eq:PphatTR}
\wP_T(\omega_1,\dots,\omega_T)=\PP_T(\theta(\omega_T), \dots, \theta(\omega_1)).	
\eeq
We shall refer to $(\widehat{\cJ}, \widehat \rho)/\wP/{\fJ}_\wP$ as the OR of $(\cJ, \rho)/\PP/{\fJ}_\PP$. 
For additional discussion of the OR,
see~\cite{Crooks08} and~\cite{BJPP1,DetBal}. We emphasize that the OR depends on the choice 
of the involution $\theta$.
\een

The work~\cite{BJPP1}  concerned a proposal for the study of the emergence of
the quantum arrow of time in repeated quantum measurement processes. This emergence is identified  with a suitable degree of
{\em distinguishability} between $(\cJ,\rho)$ and its OR
$(\widehat{\cJ},\widehat{\rho}\,)$, and is  quantified by the entropic distinguishability of the
respective unravelings  $\PP$ and $\wP$. This entropic distinguishability
is closely linked to notions of  entropy production in non-equilibrium statistical mechanics and hypothesis testing in statistics,  and
was examined in~\cite{BJPP1}  on two levels:
\begin{itemize}
\item[{\bf I}:] Asymptotics of relative entropies and mean entropy production rate,
Stein error exponent.
\item[{\bf II}:] Asymptotics of R\'enyi's relative entropies and fluctuations of entropy
production, large deviation principle (LDP) and fluctuation theorem, Chernoff and Hoeffding
error exponents.
\end{itemize}
The results of~\cite{BJPP1} regarding the LDP and fluctuation theorem were further refined in~\cite{CJPS-2017}.
To state the combined set of  results, we introduce the remaining assumptions of~\cite{BJPP1}. The first of them is a  non-triviality assumption (see Section 2.2
in~\cite{BJPP1}), which we will always assume to hold without further saying:\begin{assumption}{(B)} $\supp\PP_T=\supp \wP_T$ for all $T\in\nn^\ast$. 

\end{assumption}
The last and central  assumption concerns the main technical tool of~\cite{BJPP1}, the non-additive thermodynamic formalism of dynamical
systems\footnote{see~\cite{Bar2011} for a general introduction to this topic.}, and complements the upper-decoupling property~\eqref{dec-i} with a suitable lower decoupling property.  To motivate this  assumption, we recall the following result; see~\cite{BJPP1,feng09}.

\begin{proposition}\label{Irreprop}
 Suppose that the map $\Phi=\sum_{a\in\cA}\Phi_a$ is irreducible. Then there exist $C>0$ and $\tau\ge0$ such that for all
$\omega, \nu\in \Omega_{\rm fin}$ one can find $\xi, {\widehat \xi} \in \Omega_{\rm fin}$ satisfying  $|\xi|\leq \tau$, $|\widehat \xi|\leq \tau$,
 so that
\beq{dec-tou}
\PP([\omega\xi\nu])\geq
C\PP([\omega])\PP([\nu]), \qquad \wP([\omega\widehat \xi\nu])\geq
C\wP([\omega])\wP([\nu]).
\eeq
\end{proposition}
Starting with the lower-decoupling property~\eqref{dec-tou}, one can go quite far in an independent study of  the dynamical systems $(\Omega, \phi, \PP)$ and $(\Omega, \phi,  \wP)$. However, our
simultaneous analysis of the pair $(\PP,\wP)$  requires~\eqref{dec-tou} to hold with $\xi =\widehat \xi$. 
\begin{assumption}{(C)}
There exist $C>0$ and $\tau\ge0$ such that for all
$\omega, \nu\in \Omega_{\rm fin}$ one can find $\xi\in \Omega_{\rm fin}$ satisfying  $0\leq |\xi|\leq \tau$, so that
\beq{dec-int}
\PP([\omega\xi\nu])\geq
C\PP([\omega])\PP([\nu]), \qquad \wP([\omega\xi\nu])\geq
C\wP([\omega])\wP([\nu]).
\eeq
\end{assumption}

Unlike Assumptions \assref{(A)} and \assref{(B)} which are assumed throughout, we shall mention explicitly when Assumption~\assref{(C)} is in force.

\begin{remark}\label{rem:1} Assumption~\assref{(C)} implies that $\PP$ and $\wP$ are $\phi$-ergodic; see~\cite[Lemma~A.2]{CJPS-2017}.
\end{remark}

\begin{remark}  Assumption~\assref{(C)} was phrased differently in~\cite{BJPP1}. However, under Assumption~\assref{(A)} and more precisely condition~\eqref{dec-i}, it is easy to show that the present phrasing and the one of~\cite{BJPP1} are in fact equivalent.
\end{remark}

\begin{remark}  If $\tau=0$, then Assumption~\assref{(C)} reduces to Assumption (D)  of~\cite{BJPP1}. In that case, for all $\omega, \nu\in \Omega_{\rm fin}$, one has
\[
\PP([\omega\nu])\geq
C\PP([\omega])\PP([\nu]), \qquad \wP([\omega\nu])\geq
C\wP([\omega])\wP([\nu]).
\]
\end{remark}

\begin{remark}\label{rem:4} In~\cite[Proposition~2.6]{BJPP1}, it is proven that if the completely positive map ${\Psi:\cB(\cH\otimes\cH)\to\cB(\cH\otimes\cH)}$ defined by
\[
\Psi=\sum_{a\in\cA}\Phi_a\otimes\widehat{\Phi}_a	
\]
is irreducible on $\cB(\cH\otimes\cH)$, then  Assumption~\assref{(C)} holds. 
\end{remark}

\medskip
The central object of study in~\cite{BJPP1} was  the sequence of random variables
\beq{eq:defsigmaT}
\Omega\ni\omega\mapsto\sigma_T(\omega)\defeq\log\frac{\PP_T(\omega_1,\ldots,\omega_T)}{\wP_T(\omega_1,\ldots,\omega_T)},	
\eeq
which  quantifies the irreversibility, or equivalently, the entropy production of the measurement process.\footnote{$\sigma_T$ is often called the  {\em log-likelihood ratio}  in the framework of hypothesis testing, and {\em relative information random variable} in information theory.} We recall that
\[
\EE(\sigma_T)=S(\PP_T|\wP_T)\geq 0
\]
is the relative entropy of the pair $(\PP_T, \wP_T)$. The cumulant-generating function at $\alpha\in\rr$,
\[
e_T(\alpha)\defeq\log \EE\left(\e^{-\alpha \sigma_T}\right)=\log\sum_{\omega \in \supp\PP_T}
\PP_T(\omega)^{1-\alpha}\wP_T(\omega)^{\alpha}\eqdef S_{1-\alpha}(\PP_T|\wP_T),
\]
is the relative R\'enyi entropy. 
The next theorem summarizes the main results of
\cite{BJPP1} combined with refinements obtained in~\cite{CJPS-2017}.

\begin{theorem}\label{thm-int}
The following results hold.
\ben
\item \label{it:mean_ep_conv}The (possibly infinite) limit
\beq{eq:defEPsigma}
	\ep(\cJ,\rho)\defeq\lim_{T\to \infty}\frac1T \EE(\sigma_T)
\eeq
exists and is non-negative. We call it the \emph{mean entropy production rate} of $(\cJ,\rho)$. See~{\upshape\cite[Theorem~2.1]{BJPP1}} and~{\upshape\cite[Section~6.2]{CJPS-2017}}.
\item\label{it:as_ep_conv} The limit
\[
\overline{\sigma}(\omega)\defeq\lim_{T\to\infty}\frac1T\sigma_T(\omega)
\]
exists $\PP$-almost surely and satisfies $\overline{\sigma}\circ\phi=\overline{\sigma}$. Moreover,  $\EE(\overline{\sigma})=\ep(\cJ,\rho)$ and 
\[
\lim_{T\to\infty}\EE\left(\left|\frac1T\sigma_T-\overline{\sigma}\right|\right)=0
\]
holds whenever $\ep(\cJ,\rho)<\infty$.
The number $\overline{\sigma}(\omega)$ is the \emph{entropy production rate} of  $(\cJ,\rho)$ along the trajectory $\omega$. See~{\upshape\cite[Theorem~2.1]{BJPP1}}.

\item\label{it:ergoconvsigma} Suppose that the  dynamical system $(\Omega,\PP,\phi)$ is ergodic.\footnote{A sufficient condition for
ergodicity is that the completely positive map $\Phi=\sum_{a\in\cA}\Phi_a$ is irreducible.} Then  $\overline{\sigma}=\ep(\cJ,\rho)$ holds $\PP$-almost surely and
$\PP=\wP\iff \ep(\cJ,\rho)=0$.  See~{\upshape\cite[Proposition~2.2]{BJPP1}}.
\item\label{it:steinexpo}  For any $\epsilon\in{]}0,1{[}$, let
\[
s_T(\epsilon)\defeq\min\{\wP_T(\cT)\mid\cT\subset \Omega_T, \PP_T(\cT^c)\leq \epsilon\},
\]
where $\cT^c=\Omega_T\setminus\cT$. The numbers
\[
\underline{s}(\epsilon)\defeq\liminf_{T\to\infty}\tfrac1T \log s_T(\epsilon)\quad\text{and}\quad\overline{s}(\epsilon)\defeq\limsup_{T\to\infty}\tfrac1T\log s_T(\epsilon)
\]
are called the Stein's error exponents of the pair $(\PP, \wP)$. If  $(\Omega,\PP,\phi)$ is ergodic, then  for any $\epsilon\in{]}0,1{[}$
\[
\underline{s}(\epsilon)=\overline{s}(\epsilon)=-\ep(\cJ,\rho).
\]
See~{\upshape\cite[Theorem~2.3]{BJPP1}}.
\een

\medskip\noindent
Suppose now that Assumption~\assref{(C)} holds. We then have the following:

\medskip
\ben\setcounter{enumi}{4}
\item \label{it:differentiability_e_alpha} For all $\alpha \in \rr$, the (possibly infinite) limit
\beq{eq:defealpha}
e(\alpha)\defeq\lim_{T\to\infty}\tfrac1T e_T(\alpha)\in {]}{-}\infty, \infty]	
\eeq
exists.  We call it the \emph{entropic pressure} of $(\cJ,\rho)$. The  function $e$ is convex, satisfies $e(0)=e(1)=0$, and the symmetry
\beq{es-s}	
e(\alpha)=e(1-\alpha), \qquad \alpha \in \rr,
\eeq
holds.  Moreover, the function $e$ is  non-positive on $[0,1]$, non-negative on $\rr \setminus [0,1]$, differentiable on $]0,1[$, and\footnote{$\partial^\mp$ denotes the left/right derivative.}
\beq{eq:leftrightder}
(\partial^-e)(1)=-(\partial^+e)(0)=\ep(\cJ,\rho).
\eeq
See~{\upshape\cite[Theorem~2.4, Theorem~2.5]{BJPP1}} and~{\upshape\cite[Theorem~2.8]{CJPS-2017}}.

\item \label{it:AssDdifferentiability_e_alpha} If  Assumption~\assref{(C)} holds with  $\tau = 0$, then $e$ is finite and differentiable on $\rr$.  See~{\upshape\cite[Theorem~2.8]{BJPP1}}.
\item\label{it:LDPsigmaT} Under the laws $\PP$, the sequence of random variables $\left(\frac1 T\sigma_T(\omega)\right)_{T\in\nn^\ast}$   satisfies the LDP with a convex rate
function $I\colon\rr \to [0,\infty]$ in the sense that for any  Borel set  $S\subset\rr$,\footnote{$\mathrm{int}(S)/\mathrm{cl}(S)$ denotes the interior/closure of the set $S$.}
\begin{align}
-\inf_{s\in {\rm int}(S)}I(s) &\leq \liminf_{T\to\infty}\frac1T\log\PP\left(\left\{\omega\in\Omega\,\bigg|\,
\frac1T\sigma_T(\omega)\in S\right\}\right)\nonumber\\[-4pt]
&\label{weak-ldp-k}\\
&\leq 
\limsup_{T\to\infty}\frac1T\log\PP\left(\left\{\omega\in\Omega\,\bigg|\,
\frac1T\sigma_T(\omega)\in S \right\}\right)\leq
-\inf_{s\in {\rm cl}(S)} I(s).\nonumber
\end{align}
The rate function $I$ is the
Fenchel--Legendre transform of the function $\alpha \mapsto e(-\alpha)$, \ie
\[
I(s)\defeq\sup_{\alpha \in \rr}(\alpha s-e(-\alpha)),  \qquad s\in \rr,
\]
and it satisfies the Gallavotti--Cohen symmetry
\beq{gc-sym}
I(-s)=I(s) +s, \qquad s\in \rr.
\eeq
If $e$ is finite in a neighborhood of the origin, then $I$ is a good\footnote{We recall that $I$ is a rate function if it is non-negative, lower semicontinuous, and not everywhere infinite. We call $I$ a {\em good} rate function if, in addition, it has compact level sets.} rate function. See~{\upshape\cite[Theorem~2.11]{BJPP1}} and~{\upshape\cite[Theorem~2.8]{CJPS-2017}}.

\item \label{it:chernoff}
The Chernoff error exponents of the pair $(\PP, \wP)$ are defined as
\[
\underline{c}\defeq\liminf_{T\to\infty}\tfrac1T\log c_T\quad\text{and}\quad
\overline{c}\defeq\limsup_{T\to\infty}\tfrac1T\log c_T,
\]
where
\[
c_T\defeq\frac14\left(2-\sum_{\omega\in\Omega_T}\left|\PP_T(\omega)-\wP_T(\omega)\right|\right).
\]
One then has
\beq{che-ex}
\ubar c=\bar c=\min_{\alpha \in [0,1]}e(\alpha) = e\left(\frac{1}{2}\right).
\eeq
 See~{\upshape\cite[Theorem~2.12]{BJPP1}}.

\item\label{it:mainthoeffding} The Hoeffding error exponents of the pair $(\PP, \wP)$ are defined, for all $s\geq 0$, as
\begin{align*}
\overline{h}(s)\defeq&\inf_{(\cT_T)}\left\{\limsup_{T\to\infty}\tfrac1T\log\wP_T(\cT_T)\Big|\limsup_{T\to\infty}\tfrac1T\log\PP_T(\cT_T^c)<-s\right\},\\
\underline{h}(s)\defeq&\inf_{(\cT_T)}\left\{\liminf_{T\to\infty}\tfrac1T\log\wP_T(\cT_T)\Big|\limsup_{T\to\infty}\tfrac1T\log\PP_T(\cT_T^c)<-s\right\},\\
h(s)\defeq&\inf_{(\cT_T)}\left\{\lim_{T\to\infty}\tfrac1T\log\wP_T(\cT_T)\Big|\limsup_{T\to\infty}\tfrac1T\log\PP_T(\cT_T^c)<-s\right\},
\end{align*}
where the infimum is taken over all sequences of tests such that $\cT_T\subset \Omega_T$ and the last one is, moreover, restricted to the sequence of tests such that $\lim_{T\to\infty}\tfrac1T\log\PP_T(\cT_T^c)$ exists. One then has\footnote{Note that the formula given in~\cite{BJPP1} is different, but coincides with this one since $e(\alpha) = e(1-\alpha)$. The expression given here is convenient in view of the generalizations we discuss at the end of this section.} 
\[
\underline{h}(s)=\overline{h}(s)=h(s)=\inf_{\alpha\in]0,1]}\frac{(1-\alpha)s+e(\alpha)}{\alpha}.
\]
 See~{\upshape\cite[Theorem~2.13]{BJPP1}}.

\een
\end{theorem}

We refer the reader to~\cite{BJPP1} for  a discussion of the results stated in Theorem~\ref{thm-int} in the context of hypothesis testing and the emergence of the arrow of time in repeated quantum measurement processes.

The setting of this section and practically  all results of Theorem~\ref{thm-int}  extend to arbitrary pairs of instruments $(\cJ, \rho)$, $(\widehat {\cJ}, 
\widehat \rho)$ defined on possibly different  Hilbert spaces $\cH$ and $\widehat \cH$. In fact, the theory of~\cite{CJPS-2017} was already formulated for general pairs of invariant measures $(\PP, \wP)$ on $\Omega$ which are not necessarily related by OR (\ie no relation of the kind \eqref{eq:PphatTR} is assumed). We conclude this section with a discussion  of these generalizations.

Consider  two arbitrary instruments  $(\cJ,\rho)$ and $(\wcJ,\wrho)$, with $\cJ=(\Phi_a)_{a\in \cA}$ and $\wcJ = (\widehat \Phi_a)_{a\in \cA}$, both satisfying Assumption~\assref{(A)}. Instead of Assumption~\assref{(B)}, we shall assume the following weaker condition:\begin{assumption}{(B0)}
$\supp\,\PP_T\subset\supp\,\wP_T$ for all $T\in\nn^\ast$.
\end{assumption}

This assumption ensures that the  random variable  $\sigma_T$, given by~\eqref{eq:defsigmaT}, is well defined $\PP_T$-almost everywhere.  
Note that  when $\PP$ and $\wP$ are related by OR, \ie when~\eqref{eq:PphatTR} holds, then $\supp\,\PP_T$ and $\supp\,\wP_T$ have the same cardinality, and in that case Assumptions~\assref{(B0)} and~\assref{(B)} are equivalent.

In this more general setting, we replace the notation $\ep(\cJ, \rho)$ with
\beq{eq:newnotationEP}
\ep(\PP,\wP)\defeq\lim_{T\to \infty}\frac1T \EE(\sigma_T),
\eeq
since now the right-hand side depends on both measures. 

The results of 
Theorem~\ref{thm-int}  are adapted to this more general  setup as follows. 
\begin{itemize}
	\item Parts~\ref{it:mean_ep_conv} and~\ref{it:as_ep_conv} remain valid (with the change of notation~\eqref{eq:newnotationEP}). The observation that $\ep(\PP,\wP)$ is non-negative can be strengthened: we actually have $\ep(\PP,\wP)\geq -e(1)$ (which can be strictly positive, see \eqref{eq:eof1} below).
	\item Part~\ref{it:ergoconvsigma} needs to be reformulated as the following two statements:  
\begin{itemize}
\item If the dynamical system $(\Omega,\PP,\phi)$ is ergodic, then  $\overline{\sigma}=\ep(\PP,\wP)$ holds $\PP$-almost surely. This follows from an obvious modification of the proof of~\cite[Theorem~2.1]{BJPP1}. 
\item  If the dynamical system $(\Omega,\wP,\phi)$ is ergodic, then $\PP=\wP\iff \ep(\PP,\wP)=0$. Indeed, the argument in the proof of~\cite[Proposition~2.2]{BJPP1} only gives $\PP \ll \wP$ here, and in order to conclude that $\PP =\wP$, one needs $(\Omega,\wP,\phi)$ to be ergodic.\footnote{There are pairs of instruments such that $(\Omega,\PP,\phi)$ is ergodic and $\ep(\PP,\wP)=0$, yet $\PP\neq \wP$ and $\ep(\wP,\PP)>0$, even when $\supp\,\PP_T = \supp\,\wP_T$ for all $T\in\nn^\ast$. This happens for example if $\PP$, $\QQ$ are two distinct, fully supported ergodic measures, and $\wP = \frac 12 \PP + \frac 12 \QQ$.}
\end{itemize}

\item  Part~\ref{it:steinexpo} remains true up to the change of notation~\eqref{eq:newnotationEP}. It should be noted that in the definition of the exponents, one cannot interchange $\PP$ and $\wP$ in general.
\item Part~\ref{it:differentiability_e_alpha} is changed as follows. For all $\alpha \in \rr$, the (possibly infinite) limit
\eqref{eq:defealpha} exists. The  function $\alpha\mapsto e(\alpha)$ is convex, satisfies $e(0)=0$, and  is differentiable on $]0,1[$. However,
\beq{eq:eof1}
e(1) = \lim_{T\to\infty}\frac 1 T \log \sum_{\omega \in \supp \PP_T} \wP_T(\omega)
=\lim_{T\to\infty} \frac 1 T   \log \wP_T(\supp \PP_T)
\eeq
only ensures the inequality $e(1)\le0$. In particular, the symmetry~\eqref{es-s} does not necessarily hold. In order to adapt the proof of~\cite[Theorem~2.5]{BJPP1}, one has, in the notation therein, to replace the expression $f(\QQ^{(\alpha)})$ with $f(\QQ) - \alpha\lim_{T\to\infty} \frac 1 T \QQ[\sigma_T]$ in the variational principle (the two expressions coincide if $\PP$ and $\wP$ are related by OR).
\item Part~\ref{it:AssDdifferentiability_e_alpha} remains unchanged.
\item All the results in Part~\ref{it:LDPsigmaT} except the symmetry~\eqref{gc-sym} hold in the present setup (this is proved in greater generality in~\cite{CJPS-2017}).
\item All the conclusions of Part~\ref{it:chernoff} except for the last equality in~\eqref{che-ex} remain true.
\item Finally, all the assertions in Part~\ref{it:mainthoeffding} remain unchanged. 
\end{itemize}

The modifications to the proofs are minor (and some of them are already provided in~\cite{CJPS-2017}). The  hypothesis testing  discussion in 
\cite{BJPP1} is easily adapted to the above more general setting. 

\subsection{Von Neumann instruments}
\label{sec-vn-instruments}
Measurements of the kind discussed in Section~\ref{sec-direct-measurements} are called {\sl projective measurements.} To describe repeated projective measurements, one has to incorporate the unitary propagators
$U_<$ and $U_>$ describing the pre- and post-measurement time-evolutions. The corresponding instrument can be written as $\cJ=(\Phi_a)_{a\in\cA}$ with
\[
\Phi_a[X]\defeq U_<^\ast Q_aU_>^\ast XU_>Q_aU_<,
\]
$(Q_a)_{a\in\cA}$ being a partition of unity on the Hilbert space $\cH$. Setting $P_a=U_>Q_aU_>^\ast$ and $U=U_>U_<$, this can be rewritten as
\[
\Phi_a[X]=U^\ast P_aXP_aU,
\]
$(P_a)_{a\in\cA}$ being again a partition of unity and $U$ a unitary on $\cH$. We shall say that such a $\cJ$ is a {\sl von~Neumann instrument.} Since $\Phi_a^\ast(\rho)=P_a U\rho U^\ast P_a$, the maximal entropy state $\rho=\one/d$ always satisfies~\assref{(A)}. The unravelling $\PP$ turns out to be a Markov process (see Remark~\ref{rem:VonneumannMarkov}).
 Given an alphabet involution $\theta$, the induced OR $(\widehat\cJ,\widehat\rho)$ is again a von~Neumann instrument
\[
\widehat{\Phi}_a[X]=\widehat U^\ast\widehat P_aX\widehat P_a\widehat U,\qquad\widehat{\rho}=\rho,
\]
with $\widehat U=U^\ast$ and $\widehat P_a=U^\ast P_{\theta(a)}U$.
We note that if there exists an anti-unitary involution $\Theta$ on $\cH$ such that $\Theta U\Theta^\ast=U^\ast$ and
$\Theta P_a\Theta^\ast=P_{\theta(a)}$, then the unravelings satisfy $\wP=\PP$, so that $\ep(\cJ,\rho)=0$ and the entropic pressure $e$ vanishes identically.

\medskip
In \cite{ABL}, the discussion on the relationship between the arrow of time and the projection postulate is based on  sequential von~Neumann measurements of different observables. Assuming the sequence of observables is periodic of period $N$, the corresponding alphabet is $\mathcal A=\mathcal A_1\times\dotsb\times \mathcal A_N$ with each $\mathcal A_n$ being a finite alphabet labelling the outcome of one observable, and the instrument $\cJ=\{\Phi_a\}_{a\in\mathcal A}$ is
\[
\Phi_{(a_1,\dotsc,a_N)}[X]\defeq P_{a_N}^N\dotsb P_{a_1}^1 X P_{a_1}^1\dotsb P_{a_N}^N,
\]
where each $(P_{a_n}^n)_{a_n\in \cA_n}$ is a partition of unity by orthogonal projectors. Since 
\[\Phi^*_a[\rho]=P_{a_1}^1\dotsb P_{a_N}^NXP_{a_N}^N\dotsb P_{a_1}^1,\] the maximal entropy state $\rho=\one/d$ satisfies~\assref{(A)}.
Then, setting $\theta(a_1,\dotsc,a_N)=(a_N,\dotsc,a_1)$ makes the canonical OR instrument $\widehat{\cJ}$ equal to $\cJ$. Hence, $\PP=\wP$, so that $\ep(\cJ,\rho)=0$ and the entropic pressure $e$ vanishes identically. These observations relate our setting to one of the  results of \cite{ABL}, namely that the distribution of sequences of outcomes of von~Neumann measurements is time-reversal invariant.

\subsection{Probe measurements}
\label{sec-probes}

Ultimately, all measurements to be considered in the forthcoming examples will be of the kind described in Section~\ref{sec-direct-measurements}. However, from a physical perspective, the distinguished subclass of them describing indirect measurements is central to the  interpretation of repeated quantum measurement processes. More specifically, we will be interested in one-time and two-time probe measurements. 

Recall that our starting quantum system, which we denote by $\cS$ here, is described by  a finite-dimensional Hilbert space $\cH$ and a state $\rho$. 
In addition to $\cS$ we are given a countable collection $(\cP_t)_{t\in\nn^\ast}$ of independent, identical quantum probes, each of which is described by a finite-dimensional Hilbert space $\cH_p$ and a state $\rho_p$. In one-time measurement processes, at time $t=0$, $\cS$ and the probe ${\cP}_1$ are coupled and interact for  a unit of time. At the end of the interaction period,  a von~Neumann measurement of a specified probe  observable is performed and  the state of $\cS$ is updated according to the outcome of this measurement. Then the procedure is  repeated, with  $\cS$ in the updated state coupled to a probe ${\cP}_{2}$. Two-time measurement processes are similar, except that the probe observable subject to measurement is strongly related to the probe state $\rho_p$, this observable being measured before and after the interaction period. 
The details are as follows.

\subsubsection{One-time measurements}
\label{sec-one-time}
The Hilbert space of the coupled system is 
$\cH\otimes \cH_p$ and its initial state is $\rho\otimes \rho_p$. The evolution of the  coupled system over a unit time interval is described by a 
unitary operator $U\in\cB( \cH\otimes\cH_p)$. Let $(P_a)_{a\in\cA}$ be a partition
of unity on $\cH_p$ associated to a probe observable, and set\footnote{$\tr_{\cH_p}$ denotes the partial trace over $\cH_p$.}
\beq{ancila}
\Phi_a [X]\defeq\tr_{\cH_p}\left(U^\ast(X \otimes P_a)U(\one \otimes \rho_p)\right).  
\eeq
One easily checks that $\cJ=(\Phi_a)_{a\in\cA}$ is a quantum instrument. By Stinespring's dilation theorem~\cite[Theorem~3.6]{takesaki1979-1}, this construction has a converse: for any instrument  $(\Phi_a)_{a\in\cA}$ on $\cH$, one can find $\cH_p$, $\rho_p$, $U$ and a partition of unity $(P_a)_{a\in\cA}$ such that~\eqref{ancila} holds. Returning to measurements,  $\tr(\Phi_{\omega_1}^\ast[\rho])$ is the  probability that after the first interaction, the von~Neumann measurement of the probe observable yields the outcome $\omega_1$. After this measurement, the system $\cS$  is in the reduced state $\rho_{\omega_1}=\Phi_{\omega_1}^\ast[\rho]/\tr(\Phi_{\omega_1}^\ast[\rho])$. Repeating the measurement process with the system $\cS$ being  in  the state $\rho_{\omega_1}$, one derives that the probability of observing the sequence 
of outcomes $(\omega_1, \omega_2)$ is $\tr((\Phi_{\omega_2}^\ast\circ \Phi_{\omega_1}^\ast)[\rho])$. After the two measurements, the system $\cS$ is in the state $\rho_{\omega_1\omega_2}= (\Phi_{\omega_2}^\ast\circ \Phi_{\omega_1}^\ast)[\rho]/\tr((\Phi_{\omega_2}^\ast\circ \Phi_{\omega_1}^\ast)[\rho])$. Continuing in this way, one arrives at the expression \eqref{eq:defPttr} for $\PP_T$; see~\cite{BJPP1} for additional information and references. 


\subsubsection{Two-time measurements}
\label{sec-two-time}
In terms of the partition of unity $(P_l)_{l\in\cL}$, let the probe state be given by 
\beq{spec-ttm}
\rho_p\defeq\sum_{l\in\cL} \pi_l P_l,
\eeq
where the eigenvalues $\pi_l$  are strictly positive but not necessarily distinct.\footnote{$(P_l)_{l\in\cL}$ is the family of eigenprojections of an observable commuting with $\rho_p$, not necessarily the ones of $\rho_p$ itself. In particular, we do not assume the projections $P_l$ to be rank one.} The map $\theta(l,l')\defeq(l',l)$ defines an involution of the
alphabet $\cA=\cL\times\cL$. 
 Recall that  $U\in\cB(\cH\otimes \cH_p)$ 
denotes the unitary propagator over a unit interval of time. For $a=(l,l')\in\cA$, we set
\beq{twotimeviolin}
\Phi_a[X]\defeq{\rm tr}_{\cH_p}\left((U^\ast(X\otimes P_{l'})U)
(\one \otimes \pi_{l} P_{l})\right).
\eeq
The instrument  $\cJ\defeq(\Phi_a)_{a\in\cA}$ describes a  two-time indirect measurement (the probe state is measured before and after the interaction). In the following, $\PP$ denotes the unraveling of $(\cJ,\rho)$ for some invariant state $\rho$. For more information about the two-time measurement protocol see~\cite{HJPR2,HJPR1}.

Since the particular mathematical structure of repeated two-time measurements was not discussed in~\cite{BJPP1} and will play a role in the sequel, we elaborate on this point.

We will assume that the system is time-reversal invariant 
in the following sense: there exist anti-unitary involutions $\Theta\colon\cH\rightarrow \cH$ and $\Theta_p\colon\cH_p
\rightarrow \cH_p$ such that, for all $l\in\cL$, 
\beq{tri-tri}
\Theta \rho\Theta=\rho, \qquad 
\Theta_p P_l \Theta_p=P_l, \qquad 
(\Theta \otimes \Theta_p )U (\Theta \otimes \Theta_p)=U^\ast.
\eeq
For  $a=(l,l')\in\cA$, let $\Delta {\mathfrak S}(a)\defeq\log \pi_{l}- \log\pi_{l'}$. For $(\omega_1,\ldots\omega_T)\in\Omega_T$, let
\[
\Delta {\mathfrak S}_T(\omega_1, \dots, \omega_T)\defeq\sum_{t=1}^T\Delta {\mathfrak S}(\omega_t).
\]
Note that 
\beq{flu-again}
\Delta{\mathfrak S}_T=\sum_{t=0}^{T-1}\Delta {\mathfrak S}\circ \phi^t,
\eeq
where, for $\omega\in \Omega$, we have set $\Delta {\mathfrak S}(\omega)\defeq\Delta{\mathfrak S}(\omega_1)$. 

For $\alpha\in\rr$, we define  $\Phi(\alpha)\colon\cB(\cH)\rightarrow\cB(\cH)$ by
\[
\Phi(\alpha)[X]\defeq\sum_{a\in\cA}\e^{-\alpha \Delta {\mathfrak S}(a)}\Phi_a[X].
\]
This deformation of $\Phi=\Phi(0)$ is a completely positive map. It is irreducible  iff $\Phi$ is. In our setting,
$\Phi(\alpha)$ will play the role of the transfer operator.

\begin{theorem}\label{thm-two-time} 
Suppose that~\assref{(A)} holds. Then:
\ben
\item\label{it:2tone} For all $T\in\nn^\ast$ and all $\omega=(\omega_1,\ldots,\omega_T)\in\Omega_T$,
\[
{\rm tr} (\rho(\Phi_{\omega_1}\circ \cdots\circ \Phi_{\omega_T})[\one])
=\e^{\Delta {\mathfrak S}_T(\omega)}
{\rm tr} (\Phi_{\theta(\omega_T)}\circ \cdots \circ \Phi_{\theta(\omega_1)}[\rho]).
\]
\item\label{it:2ttwo}  Assumption~\assref{(B)} holds. If   $\Phi$ is irreducible, then~\assref{(C)} also holds and Theorem~\ref{thm-int} applies.

\item\label{it:2tthree} Let $r_+\defeq\max\sp(\rho)$, $r_-\defeq\min \sp(\rho)$. Then for all $T\in\nn^\ast$ and 
$\omega\in {\rm supp}\,\PP_T$, 
\[
\big|\sigma_{T}(\omega)-\Delta {\mathfrak S}_T(\omega) \big|\leq \log\frac{r_+}{r_-}.
\]

\item\label{it:2tfour}
\[
\ep(\cJ,\rho)=\int_\Omega \Delta {\mathfrak S}\,\d\PP
=\sum_{a\in\cA}\Delta{\mathfrak S}(a)\tr (\rho \Phi_a[\one]).
\]

\item\label{it:2tfive} 
\[
\ep(\cJ,\rho)= S(\rho\otimes \rho_p|U^\ast(\rho\otimes \rho_p)U),
\]
where $S(\mu_1|\mu_2)=\tr(\mu_1(\log \mu_1-\log\mu_2))$ denotes  the quantum relative entropy\footnote{All the  entropic notions we use are reviewed in~\cite[Section~2.1]{BJPP1}.
We recall that  $S(\mu_1|\mu_2)\geq 0$ and that the equality holds iff $\mu_1=\mu_2$.} of a pair  of density matrices $(\mu_1,\mu_2)$. In particular, 
$\ep(\cJ,\rho)=0$ if and only if  $U^\ast(\rho\otimes \rho_p)U=\rho\otimes \rho_p$. 

\item\label{it:2tsix} For all $\alpha \in \rr$, the limit 
\beq{cor-lim}
e(\alpha)=\lim_{T\rightarrow \infty}\frac{1}{T}
\log \left(\int_\Omega \e^{-\alpha \Delta {\mathfrak S}_T}\d\PP\right)
\eeq
exists and $e(\alpha)=\log r(\alpha)$, where $r(\alpha)$ is the spectral radius of $\Phi(\alpha)$. 
The function $\rr \ni\alpha \mapsto r(\alpha)$ is locally a branch of a multivalued analytic function with at worst algebraic  singularities.

\item\label{it:2tsev}  For $\alpha\not\in [0,1]$, 
\[
e(\alpha)\leq \min (|\alpha|, |1-\alpha|)\frac{r_+}{r_-}.
\]
For $\alpha \in [0,1]$, 
\[
e(\alpha)\geq -\max(\alpha, 1-\alpha)\frac{r_+}{r_-}.
\]
\een

\medskip
\noindent In the remaining statements we assume that $\Phi$ is irreducible.

\medskip 
\ben\setcounter{enumi}{7}
\item\label{it:2teight} $\ep(\cJ,\rho)=0$ if and only if $\tr (\rho \Phi_a[\one])=\tr(\rho\Phi_{\theta(a)}[\one])$ for all $a\in \cA$.
\item\label{it:2tnine} The function $\rr \ni \alpha \mapsto e(\alpha)$ is real analytic.
\item\label{it:2tten} The sequence of random variables $\left(\frac1T\Delta {\mathfrak S}_T\right)_{T\in\nn^\ast}$  under the laws $\PP_T$ satisfies the  LDP  with the rate function 
\[ 
I(s)=\sup_{\alpha \in \rr}( s\alpha - e(-\alpha)), \qquad s\in \rr.
\]
\item\label{it:2telev} The random variables  $(\Delta {\mathfrak S}_T)_{T\in\nn^\ast}$ satisfy the  central limit theorem. More precisely,
as $T\to\infty$, the random variable
\[
\frac{{\Delta {\mathfrak  S}_T-T\ep(\cJ,\rho)}}{{\sqrt{T}}}
\]
converges in law to a centered Gaussian with variance $e^{\prime\prime}(0)$.
\een
\end{theorem}
\proof The proof  is  based on well-known arguments and is relatively simple; see~\cite{JPW-2014,HJPR1,HJPR2}. We  sketch the argument for the reader's convenience and later reference.

One checks that for $a\in\cA$ and $X\in\cB(\cH)$,  
\beq{elem-1}
\Theta \Phi_{a}[\Theta X\Theta] \Theta = \e^{\Delta {\mathfrak S}(a)}\Phi_{\theta(a)}^\ast[X].
\eeq
This yields~\ref{it:2tone}. 

Part~\ref{it:2tone}  immediately gives that~\assref{(B)} holds. If $r_{\pm}$ are as in~\ref{it:2tthree}, the inequalities $r_-\one \leq \rho\leq r_+\one$ give that 
\beq{flu}
\frac{r_-}{r_+}\wP_T(\omega_1, \dots, \omega_T)\leq {\rm tr} (\Phi_{\theta(\omega_T)}\circ \cdots \circ \Phi_{\theta(\omega_1)}[\rho])\leq \frac{r_+}{r_-}\wP_T(\omega_1, \dots, \omega_T).
\eeq
If $\Phi$ is irreducible then, by Proposition~\ref{Irreprop}, there exist $C>0$ and $\tau\ge0$ such that for all
$\omega, \nu\in \Omega_{\rm fin}$ one can find $\xi\in \Omega_{\rm fin}$ satisfying  $0\leq |\xi|\leq \tau$, so that
\[
\PP([\omega\xi\nu])\geq
C\PP([\omega])\PP([\nu]).
\]
It then follows from~\eqref{flu} and~\ref{it:2tone} that  for the  same $\xi\in \Omega_{\rm fin}$,
 \[
 \wP([\omega\xi\nu])\geq
C\left(\frac{r_-}{r_+}\right)^3 \wP([\omega])\wP([\nu]),
\]
and~\assref{(C)} holds. This proves~\ref{it:2ttwo}.

Part~\ref{it:2tthree} follows from~\ref{it:2tone} and~\eqref{flu}. 

Part~\ref{it:2tfour} follows from~\ref{it:2tthree} and~\eqref{flu-again}.

To prove~\ref{it:2tfive}, we first note that 
\begin{align*}
\ep(\cJ,\rho)&=\sum_{a\in\cA}\Delta{\mathfrak S}(a)\tr (\rho \Phi_a[\one])\\[1mm]
&=
\sum_{l,l'\in\cL}(\log \pi_l -\log \pi_{l'}){\rm tr}\left((\rho\otimes \one)(U^\ast(\one\otimes P_{l'})U)
(\one \otimes \pi_l P_l)\right)\\[1mm]
&=\tr(\rho_p\log\rho_p)- \tr((\rho\otimes \rho_p)U^\ast(\one \otimes \log \rho_p)U).
\end{align*}
On the other hand, 
\[
\tr((\rho\otimes \rho_p)U^\ast(\log \rho \otimes \one)U)=\tr (\rho\Phi[\log\rho])=
\tr (\Phi^\ast[\rho]\log\rho)=\tr(\rho\log\rho),
\]
and so 
\begin{align*}
\ep(\cJ,\rho)&=\tr(\rho_p\log\rho_p)- \tr((\rho\otimes \rho_p)U^\ast(\one \otimes \log \rho_p)U) \\[2mm]
&\quad  +\tr(\rho\log\rho)-\tr((\rho\otimes \rho_p)U^\ast(\log \rho \otimes \one)U)\\[2mm]
&=S(\rho\otimes \rho_p|U^\ast(\rho\otimes \rho_p)U).
\end{align*}

The proof of~\ref{it:2tsix} starts with the identity
\beq{flu-2}
{\rm tr}(\rho \Phi^T(\alpha)[\one])=\int_\Omega \e^{-\alpha\Delta {\mathfrak S}_T}\d\PP,
\eeq
which yields  the bound
\[
\int_\Omega \e^{-\alpha \Delta {\mathfrak S}_T}\d \PP\leq \|\Phi^T(\alpha)\|.
\]
Gelfand's formula for the spectral radius gives
\beq{cor-1}
\limsup_{T\rightarrow \infty}\frac{1}{T}
\log \left(\int_\Omega \e^{-\alpha \Delta {\mathfrak S}_T}\d\PP\right)\leq 
\lim_{T\rightarrow \infty}\log \|\Phi^T(\alpha)\|^{1/T}=\log r(\alpha).
\eeq
The Perron--Frobenius theory for positive maps gives that there exists $0 \leq X(\alpha)\leq \one$, $X(\alpha)\not=0$, such that $\Phi(\alpha)X(\alpha)=r(\alpha) X(\alpha)$; see for example \cite[Theorem~2.5]{EHK}. The identity~\eqref{flu-2} then yields
\[
\int_\Omega \e^{-\alpha \Delta {\mathfrak S}_T}\d \PP \geq r(\alpha)^T\tr (\rho X(\alpha)),
\]
and so 
\beq{cor-2}
\liminf_{T\rightarrow \infty}\frac{1}{T}
\log \left(\int_\Omega \e^{-\alpha \Delta {\mathfrak S}_T}\d\PP\right)\geq\log r(\alpha).
\eeq
Relations~\eqref{cor-1} and~\eqref{cor-2} yield that the limit~\eqref{cor-lim} exists and that $e(\alpha)=\log r(\alpha)$. Since $r(\alpha)$ is an eigenvalue of $\Phi(\alpha)$ and the map $\cc \ni \alpha \mapsto \Phi(\alpha)$ is entire analytic, the stated regularity property  of  the map $\alpha \mapsto r(\alpha)$ follows from analytic perturbation theory~\cite[Section~II.1]{Katobook}.

\ref{it:2tsev} follows from~\ref{it:2tsix} and the symmetry $e(\alpha)=e(1-\alpha)$. 

The direction $\Rightarrow$ in~\ref{it:2teight} follows from Theorem~\ref{thm-int}~\ref{it:ergoconvsigma}. The direction $\Leftarrow$ follows 
from Part~\ref{it:2tfour} and does not require the assumption that $\Phi$ is irreducible.

If $\Phi$ is irreducible, then so is $\Phi(\alpha)$ for all $\alpha\in \rr$. This observation and the  Perron--Frobenius theory~\cite[Theorem~2.4]{EHK}  give that $r(\alpha)$ is a simple eigenvalue of $\Phi(\alpha)$ for all $\alpha\in \rr$. The analytic perturbation theory 
\cite[Section~II.1]{Katobook} yields 
that the map $\alpha \mapsto r(\alpha)$ is real analytic, and  Part~\ref{it:2tnine} follows from the identity $e(\alpha)=\log r(\alpha)$.

Part~\ref{it:2tten} is an immediate consequence of ~\ref{it:2tsev} and the G\"artner--Ellis theorem~\cite[Theorem~2.3.6]{DemboZeitouni}.

The central limit theorem stated in~\ref{it:2telev}  follows from the result of Bryc~\cite{bryc93}; the details are the same as in the proof of~\cite[Corollary~3.3]{JPW-2014}.\qed

\medskip\noindent
\begin{remark} It is well known that the time-reversal invariant two-time quantum measurement protocol induces a rich and deep mathematical structure; see~\cite{JOPP,JPW-2014}  for a discussion of this topic, references, and additional information. For the model discussed in this section, this structure emerges from the two elementary  identities \eqref{elem-1} and~\eqref{flu-2}, which allow for a simple proof of Theorem~\ref{thm-int}.
\end{remark} 

\begin{remark}  One of the special aspects of the two-time measurement protocol is the mathematical simplicity of $\sigma_T$, captured in Part~\ref{it:2tthree}, contrasting sharply the possible complexity of $\PP$.
 \end{remark} 
 
\begin{remark}  Continuing with the previous remark,  in contrast to Theorem~\ref{thm-int}, the statement and the proof of Theorem~\ref{thm-two-time} are deeply linked to the specific form of the reversal transformation. 
 The resulting $\wP$   is natural  in the context of the main theme of~\cite{BJPP1}: the emergence of an arrow of time in repeated quantum measurements. However, if $\PP$ and $\wP$ are unravelings of two unrelated two-time measurement instruments with the same alphabet, the structure behind the statement and the proof of Theorem~\ref{thm-two-time} is broken.
\end{remark} 
   
\begin{remark}\label{rem:marginals}  Identifying $\Omega\defeq(\cL\times\cL)^{\nn^\ast}$ with $\cL^{\nn^\ast}\times\cL^{\nn^\ast}$, the marginal of $\PP$ with the respect to the second factor is the unraveling of the instrument $((\Phi_l)_{l\in\cL}, \rho)$, where 
 \[
 \Phi_l[X]\defeq\tr_{\cH_p}\left(U^\ast(X\otimes P_l)U( \one \otimes \rho_p)\right).
 \]
This instrument corresponds to the one-time probe measurement associated to $(P_l)_{l\in\cL}$ with the special feature~\eqref{spec-ttm}. The marginal of $\PP$ with respect to the first factor is the Bernoulli measure on $\cL^{\nn^\ast}$  associated to the mass function $\cL\ni l\mapsto p(l)\defeq\pi_l\,\tr(P_l)$.
\end{remark} 

We now discuss several special cases of  the  two-time measurement process  that  are of particular physical and mathematical interest.



\paragraph{Case 1: Thermal probes.} 

This case corresponds to the choice $\rho_p\defeq\e^{-\beta H_p}/Z$, where $H_p$ is the Hamiltonian of the probe, $Z$ is a normalization constant and 
$\beta>0$ is the inverse temperature. Let 
\[
H_p\defeq\sum_{\varepsilon\in\sp(H_p)}\varepsilon P_\varepsilon
\]
 be the spectral resolution of $H_p$. 
 Setting $\cL\defeq\sp(H_p)$ and $\pi_\varepsilon\defeq\e^{-\beta\varepsilon-\log Z}$ gives the representation~\eqref{spec-ttm}. Note that
in this case the  $\pi_\varepsilon$ are distinct, and that the two-time measurement process corresponds to the measurement of the energy of the probe, before and after the interaction.

Note that  if $a=(\varepsilon,\varepsilon')$, then 
$\Delta {\mathfrak S}(a)=\beta(\varepsilon'-\varepsilon)$ is the entropy variation of  the probe due to the measurement process. 
The function $\Delta {\mathfrak S}_T$ describes the entropy variation in the probe subsystem in the time interval $[0, T]$. It follows that 
Part~\ref{it:2tthree} of Theorem~\ref{thm-two-time} relates the information-theoretic entropy production $\sigma_T$ to the physical entropy variation of the probes described by $\Delta{\mathfrak S}_T$. Thus, the information-theoretic/hypothesis testing interpretation of Theorem~\ref{thm-int}, in terms of the emergence of the quantum arrow of time, is directly related to the thermodynamic interpretation that stems from Theorem~\ref{thm-two-time}. 
For example, $\ep(\cJ,\rho)$ is the Stein error exponent in the hypothesis testing of the quantum arrow of time and
coincides with the expected entropy variation per unit time in the probes. For more information about this link see 
\cite{JOPP,JOPS}.

\paragraph{Case 2: Random thermal probes.}\label{par-tworantherm}
We start with a collection of $K\geq 2$ probes, each of which is in thermal equilibrium at inverse temperature $\beta_k$, $k\in\lbr1,K\rbr$. The two-time measurement process  has the additional feature that, at each time step $t=0, 1,\ldots$, the system $\cS$ is coupled to a randomly chosen probe, the $k$-th probe being chosen with  probability $w_k>0$. Thus, 
\[
\cH_p\defeq\bigoplus_{k=1}^K \cH_k, \qquad \rho_p\defeq\bigoplus_{k=1}^K w_k\rho_k, \qquad \rho_k\defeq\frac{\e^{-\beta_k H_k}}{Z_k},
\qquad\cL\defeq\coprod_{k=1}^K\sp(H_k),
\]
where $\coprod$ stands for the disjoint union, $H_k$ is the Hamiltonian of the $k$-th probe and $Z_k$ a normalization constant. Denoting the spectral decomposition of $H_k$ by
\beq{vacations}
H_k=\sum_{\varepsilon\in\sp(H_k)}\varepsilon P_{k,\varepsilon}
\eeq
and, for $\varepsilon\in\sp(H_k)$,  setting $\pi_\varepsilon\defeq w_k\e^{-\beta_k\varepsilon-\log Z_k}$
and $P_\varepsilon\defeq0 \oplus \cdots \oplus P_{k,\varepsilon}\oplus\cdots \oplus 0$,
the representation~\eqref{spec-ttm} holds. Writing 
\[
\cH\otimes \cH_p=\bigoplus_{k=1}^K \cH\otimes \cH_k,
\]
we  assume that  the unitary propagator over the interaction period has the form $U=\bigoplus_{k=1}^K U_k$, where $U_k\colon\cH\otimes \cH_k \rightarrow\cH\otimes \cH_k$, and we identify the alphabet  $\cA$ with $\coprod_{k=1}^K\sp(H_k)\times\sp(H_k)$.

If $(\Phi_a^{(k)})_{a\in\sp(H_k)\times\sp(H_k)}$ denotes the instrument describing the thermal probe two-time measurement with  $\cS$ coupled 
only to the $k$-th probe subsystem, as in Case~1 above, it is immediate that the random thermal probes two-time measurement process is described by the instrument 
\beq{rp-1}
\cJ\defeq\{w_k\Phi_a^{(k)}\mid k\in \lbr 1, K\rbr,a\in\sp(H_k)\times\sp(H_k)\},
\eeq
and that for all $\alpha$, 
\beq{rp-2}
 \Phi(\alpha)=\sum_{k=1}^K w_k \Phi^{(k)}(\alpha).
 \eeq

\paragraph{Case 3: Multi-thermal probes.} 
In this setting  the probe  consists of $K\geq 2$ independent sub-probes, each of which is in thermal 
equilibrium at inverse temperature $\beta_k>0$, $k\in\lbr1,K\rbr$. Hence,
\[
\cH_p\defeq\bigotimes_{k=1}^K \cH_k, \qquad \rho_p\defeq\bigotimes_{k=1}^K \rho_k, \qquad \rho_k\defeq\frac{\e^{-\beta_k H_k}}{Z_k},
\qquad
\cL\defeq\bigtimes_{k=1}^K\sp(H_k),
\]
$H_k$ denoting the Hamiltonian of the $k$-th probe and $Z_k$ a normalization constant. Given the spectral representation~\eqref{vacations}, \eqref{spec-ttm} holds with
\[
\pi_\varepsilon=\e^{-\sum_{k=1}^K\beta_k\varepsilon_k+\log Z_k},
\qquad 
P_\varepsilon=\bigotimes_{k=1}^K P_{k,\varepsilon_k},
\]
for $\varepsilon=(\varepsilon_1,\ldots,\varepsilon_K)\in\cL$.
Note that the $\pi_\varepsilon$'s are not necessarily distinct.

\begin{remark}
Case~3  corresponds to  the well-known setting of 
 open quantum  systems in non-equilibrium statistical mechanics. However, in the context of repeated quantum measurements, 
 Case~2 is both physically more natural and mathematically simpler to analyze. As we shall see on the example  of the spin instruments, relations~\eqref{rp-1} and~\eqref{rp-2} considerably facilitate the computation of the mean entropy production rate 
 $\ep(\cJ,\rho)$ and of the entropic pressure $e(\alpha)$. The related computations are much more complicated in Case~3. For these reasons we shall  illustrate Case~3 only on the example of the XXZ-spin instrument with $K=2$.
\end{remark}

\begin{remark}
	For a related use of the formalism of Case 2, see \cite{bougron_linear_2020}.
\end{remark}

\subsection{On the role of the thermodynamic formalism}
\label{sec-thermo}

In this section we further discuss Theorem~\ref{thm-int} in view of the  examples we shall study.

The classical Ruelle--Walters  thermodynamic  formalism~\cite{Ru1,Waltersbook}  provides  a conceptual framework and powerful technical  tools for the study of the questions addressed in Theorem~\ref{thm-int}. However, the
application of these tools requires certain regularity assumptions that are natural in the mathematical theory of classical spin systems, and which may or may not hold  for a given repeated quantum measurement process.

The following definition is basic. Let $\QQ\in\cP_\phi(\Omega)$. For $\omega\in\Omega$ and $T\in\nn^\ast$ let\footnote{Note that if $\QQ$ is the unraveling of some quantum instrument, then this definition is consistent with the notation introduced in Section~\ref{sec-setup}.}
\[
\QQ_T(\omega)\defeq\QQ([\omega_1\cdots\omega_T]).
\]
Consider the subshift $(\supp \QQ, \phi)$. The measure $\QQ$  is called weak Gibbs if there exist a continuous function $F\colon\supp \QQ\rightarrow \rr$ and  positive constants $(C_{T})_{T\in\nn^\ast}$ satisfying 
$\lim_{T\rightarrow \infty}\frac{1}{T}\log C_T=0$,   such that for all $T\in\nn^\ast$ and all $\omega \in \supp \QQ$, 
\beq{w-gibbs}
C_T^{-1} \e^{-S_TF(\omega)}\leq \QQ_T(\omega)\leq C_T\e^{-S_T F(\omega)},
\eeq
where $S_T F\defeq\sum_{t=0}^{T-1} F\circ\phi^t$.  In this context, we shall refer to $F$ as a {\em potential} for $\QQ$. If the constants $C_T$ can be chosen independent of $T$, then the measure $\QQ$ is called Gibbs.
The following criteria are useful. 
\begin{theorem}\label{noe-thm}
\hspace{0em} %
\ben 
\item\label{it:noeone} Suppose that for some $\QQ\in\cP_\phi(\Omega)$ there exist positive constants $(D_T)_{T\in\nn^\ast}$ satisfying 
$\lim_{T\rightarrow \infty}\frac{1}{T}\log D_T=0$, and such that, for all $T, S\in\nn^\ast$ and all $\omega\in\supp\QQ$,
\[
D_T^{-1}\QQ_T(\omega)\QQ_{S}(\phi^T(\omega))\leq  \QQ_{T+S}(\omega)\leq 
 D_T\QQ_T(\omega)\QQ_{S}(\phi^T(\omega)).
\]
Then $\QQ$ is weak Gibbs. 
\item\label{it:noetwo} If $\QQ\in\cP_\phi(\Omega)$ is weak Gibbs, then
\[
\lim_{T\rightarrow \infty}\frac 1 T \sup_{S\in [1,T-1]}\sup_{ \omega\in\supp\QQ}\left|\log  \frac{\QQ_T(\omega)}{\QQ_{S}(\omega)
\QQ_{T-S}(\phi^S(\omega))}\right|=0.
\]
\een
\end{theorem}
\proof Part~\ref{it:noeone} follows from \cite[Theorem 1.2]{cuneo_asympt_2020} (see also Corollary 4.2 therein for a derivation in the case where $\supp\QQ = \Omega$). To prove~\ref{it:noetwo}, note that  w.l.o.g.\;we may assume that the constants $C_T$ in~\eqref{w-gibbs} are  non-decreasing. Then the obvious estimates 
\[
C_T^{-3}\leq (C_TC_{T-S}C_S)^{-1}\leq \frac{\QQ_T(\omega)}{\QQ_{S}(\omega)
\QQ_{T-S}(\phi^S(\omega))}\leq C_T C_{T-S}C_S\leq C_T^3
\]
yield the statement.\qed

Consider an instrument $(\cJ,\rho)$ satisfying Assumption~\assref{(A)} and let $\PP$ be its unraveling. If $\PP$ is weak Gibbs, then most of the  conclusions of Theorem~\ref{thm-int} are standard results, as we now briefly discuss.

Let $\Omega^+\defeq \supp \PP=\supp \wP$. First, when $\PP$ is weak Gibbs, so is $\wP$. Indeed, a potential $\widehat F$ for
$\wP$ can be constructed as follows. For all $\omega\in \Omega^+$, let 
$$
F^{(k)}(\omega)=F^{(k)}(\omega_{\lbr 1,k\rbr})
\defeq\sup\{F(\eta)\mid \eta\in \Omega^+, \eta_{\lbr 1,k\rbr} = \omega_{\lbr 1,k\rbr}\},
$$
where $F$ is a potential for $\PP$. Then $F^{(k)}$ converges uniformly to $F$ on $\Omega^+$, and we can define a sequence of potentials $(\widehat F^{(k)})_{k\in \nn^\ast}$ by $\widehat F^{(k)}(\omega) = \widehat F^{(k)}(\omega_{\lbr 1,k\rbr})\defeq F^{(k)}(\theta(\omega_k), \theta(\omega_{k-1}), \dots, \theta(\omega_1))$ for all $\omega\in \Omega_+$. It is then not hard to prove that
$$
\lim_{k\to\infty}\limsup_{T\to\infty}\frac 1T \sup_{\omega\in \supp \wP}|\log \wP_T(\omega_{\lbr 1,T\rbr})+S_T\widehat F^{(k)}(\omega)| = 0,
$$ and thus \cite[Theorem 1.2]{cuneo_asympt_2020} provides a potential $\widehat F$ with respect to which $\wP$ is weak Gibbs.

We then have, assuming $\PP$ is weak Gibbs: 

\begin{itemize}
\item[\bf (a)] \ref{it:mean_ep_conv}--\ref{it:as_ep_conv} follow from the ergodic theorem. 
 
\item[\bf (b)] Regarding~\ref{it:differentiability_e_alpha}, the Ruelle--Walters thermodynamic formalism of weak Gibbs measures gives that  the limit~\eqref{eq:defealpha}  exists and is finite. The formalism, of course, gives much more.  The variational principle 
\[
e(\alpha)=\sup_{\QQ\in\cP_\phi(\Omega^+)}\left( h_\phi(\QQ) +\int_{\Omega^+} \left(\alpha F + (1-\alpha) \widehat F\right)
\d\QQ\right)
\]
holds, where $h_{\phi}(\QQ)$ is the Kolmogorov--Sinai entropy of $\QQ$. The set 
\[
\cP_{\rm eq}(\alpha)\defeq\left\{\QQ\in\cP_\phi(\Omega^+)\,\biggm|\, 
e(\alpha)= h_\phi(\QQ) +\int_{\Omega^+} \left(\alpha F + (1-\alpha) \widehat F\right)
\d\QQ\right\}
\]
is a non-empty,  convex,  compact subset of $\cP_\phi(\Omega^+)$.   It is a Choquet simplex and a face of $\cP_\phi(\Omega^+)$.   The extreme points of $\cP_{\rm eq}(\alpha)$ are $\phi$-ergodic. If $\cP_{\rm eq}(\alpha)$ is a singleton, then $e$ is differentiable at $\alpha$. 
Whether $\cP_{\rm eq}(\alpha)$ is a singleton or not cannot be resolved with having only information that 
$\PP$ is weak Gibbs with an unknown potential; see ${\bf (b')}$ below. 

\item[\bf (c)] The Level III LDP  holds on $(\Omega^+, \phi)$ w.r.t.\;$\PP$ with the rate function 
\[
\II(\QQ)\defeq\begin{cases}
 \int_{\Omega^+} F \d\QQ - h_\phi(\QQ)&\text{if }\QQ\in\cP_\phi(\Omega^+);\\[2pt]
 \infty&\text{otherwise,}
\end{cases}
\]
see~\cite{comman1,pfister1} and~\cite[Appendix~A.3]{CJPS-2017} (note that Assumption~\assref{(C)} implies the various specification conditions therein). The Level III LDP and the contraction principle yield the LDP for the entropy production observable stated in Part~\ref{it:LDPsigmaT}.

\item[\bf (d)] The proof of Parts~\ref{it:LDPsigmaT}--\ref{it:mainthoeffding} follows the standard route that relies only on the validity of the LDP for the sequence  $\left(\frac1 T\sigma_T(\omega)\right)_{T\in\nn^\ast}$  under the laws $\PP_T$; see~\cite{JOPS}.
\end{itemize}

In conclusion, whenever the unraveling $\PP$ is weak Gibbs, all the statements of Theorem~\ref{thm-int} follow from well-known results via the classical thermodynamic formalism, except for Parts~\ref{it:ergoconvsigma}, \ref{it:steinexpo}, and   the differentiability properties of $e$ stated at the end of Part~\ref{it:differentiability_e_alpha} and in Part~\ref{it:AssDdifferentiability_e_alpha}. Although the proofs of Parts~\ref{it:ergoconvsigma} and~\ref{it:steinexpo} are by no means hard, they are not related to the thermodynamic formalism.
We will further comment on differentiability issues later in this section.

As we have already mentioned, the unraveling $\PP$ of a quantum instrument may or may not be weak Gibbs. In practically all examples that we will consider 
in the main body of the  paper the unraveling  will not be weak Gibbs. The interest of Theorem~\ref{thm-int}  is that it  applies to such cases. We emphasize that Theorem~\ref{thm-int} combines results obtained in \cite{BJPP1} and \cite{CJPS-2017}, which provide two very different and complementary routes to the LDP for the entropy production.

The methods of~\cite{BJPP1} were based on a non-additive thermodynamic formalism that appears well suited for the study of questions addressed in 
Theorem~\ref{thm-int} in the cases where standard thermodynamic formalism fails. More precisely:

\begin{itemize}
\item[\bf (a')] Parts~\ref{it:mean_ep_conv}--\ref{it:as_ep_conv} follow from Kingman's sub-additive ergodic theorem.  

\item[\bf (b')]  Parts~\ref{it:differentiability_e_alpha} and~\ref{it:AssDdifferentiability_e_alpha} follow from the non-additive thermodynamic formalism. Assumptions~\assref{(A)}  and~\assref{(B)}  suffice
to develop this formalism for $\alpha \in [0,1]$, \ie to prove a suitable variant of the variational principle and to define $\cP_{\rm eq}(\alpha)$ which has the same properties as in~{\bf (b)} above. If in addition~\assref{(C)} holds, one also proves that $\cP_{\rm eq}(\alpha)$ is a singleton for $\alpha \in {]}0,1[$ and hence that $e$ is differentiable on $]0,1[$. In comparison to~{\bf (b)}, the proof that $\cP_{\rm eq}(\alpha)$  is singleton for $\alpha \in {]}0,1[$ uses the decoupling  properties~\eqref{dec-i} and~\eqref{dec-int} in an essential way. Under the Assumption of 
Part~\ref{it:AssDdifferentiability_e_alpha} these results are global, \ie hold for  all $\alpha\in\rr$. 

\item[\bf (c')]  The G\"artner--Ellis theorem and  the differentiability of $e$ on $]0,1[$ yield a local version of the LDP stated in Part~\ref{it:LDPsigmaT}: the relation~\eqref{weak-ldp-k} holds  for any $S\subset [(\partial^{-}e)(0),(\partial^+e)(1)]$.

\item[\bf (d')] The local LDP described in ${\bf (c^\prime)}$ suffices for the proof of Parts~\ref{it:chernoff}--\ref{it:mainthoeffding}; 
see~\cite{JOPS}.
\end{itemize}

As illustrated by the rotational instrument discussed in Sections~\ref{sec-rotinst} and~\ref{sec-ex-ro}, the methods of~\cite{BJPP1} cannot be used to derive  the global LDP stated in  Theorem~\ref{thm-int}~\ref{it:LDPsigmaT}. 

The work~\cite{CJPS-2017} concerns the LDP for so-called selectively decoupled measures that are considerably more general than unravelings of repeated quantum measurement processes considered here. The methods of~\cite{CJPS-2017} bypass the use of thermodynamic formalism and are centered around the so-called Ruelle--Lanford functions. In the context of unravelings, under Assumption 
\assref{(C)}, \cite{CJPS-2017}  provides a proof of the Level III LDP for $\PP$ and the global LDP stated in Theorem~\ref{thm-int}~\ref{it:LDPsigmaT}. We point out that in contrast to~{\bf (c)} above, Part~\ref{it:LDPsigmaT} cannot be deduced from the Level III LDP and the contraction principle, and a separate argument is required; see~\cite{CJPS-2017} for a discussion and additional information.

\subsection{Miscellaneous remarks}
\label{sec-remarks}

\paragraph{1}\label{rempar1}  As we have already mentioned, the research program initiated in~\cite{BJPP1} can be broadly described as the study of the statistical properties of quantum-mechanical probabilistic rules carried out in the specific setting of repeated quantum measurement processes. Such a  program could have been formulated in 1932, the year of publication of von~Neumann's classic~\cite{vonNeumann}.
 Needless to say,  there is an enormous  body of  literature on the subject, although the dynamical system/sub-additive thermodynamic formalism perspective developed in~\cite{BJPP1} and here appear to be new; 
see the introduction in~\cite{BJPP1} for references and additional information. The program will continue with additional works~\cite{Dima1,DetBal}, dealing with the statistical mechanics of repeated quantum measurements and a study  of the quantum detailed balance condition. 

We would like to complement the general discussion in~\cite{BJPP1} with the following  remark. Our own interest in the subject stems from  the advent of  experimental methods in cavity and circuit QED, and in particular the experimental breakthroughs  of the Haroche--Raimond  and Wineland groups~\cite{HarocheN,WinelandN,HarocheRaimondB}. These advances made it possible to sample the classical stochastic process of the unravelings $\PP$ for  a wide range  of quantum instruments. However, 
there is an essential difference in the questions we  study in comparison with  those directly related to the experimental setups. While the experiments are focused  on the tracking and manipulation of the  states of the system $\cS$,  our focus is on the unravelings.  We neither study nor gain any information about $\cS$. Rather, we study the possible mathematical and physical  richness of the statistics of repeated quantum measurements with focus on the mathematical complexity of the unravelings $\PP$. 

There is a large body of literature on the mathematical theory of quantum measurements. The monographs \cite{daviesbook,kraus1983states,holevo2003statistical} are classical; see also \cite{ballesteros2020appearance,ballesteros2018non,BFFS,ballesteros2019perturbation,bbqnd,BaBeBe2,BaBeTi2016,bbtjumprate,BFPP,bernardin2018spiking, ozawa1984quantum,maassenpurification,maassenergodic,barchiellibelavkin,barchielligregoratti09,barchielliholevo,belavkin89,bouten} for some standard results and recent developments.

\paragraph{2} In this work, among the various notions of Gibbsianness introduced in  the literature, we emphasize the weak Gibbs viewpoint. Although the notion of weak Gibbs measure was implicit in  classical works on thermodynamic formalism of spin systems, its definition was formalized only in 2002 by Yuri~\cite{yuri}. The weak Gibbs notions are  particularly well suited as a characterization of the thermodynamic regularity of the unraveling of quantum instruments, and in this context we will continue their study in~\cite{Dima1}.

\paragraph{3}\label{rempar2} Returning to the generalization of Theorem~\ref{thm-int} discussed at the end of Section~\ref{sec-setup}, assume that $\wP$ is the uniform measure on $\Omega$, \ie that  $\wP_T(\omega) = |\cA|^{-T}$ for all $\omega \in \Omega$ and $T\in\nn$. Such a measure is obtained, for example, by the instrument $\widehat \Phi_a[\one] =\widehat \Phi_a^\ast[\one] = |\cA|^{-1} \one$ for all $a\in \cA$ and the choice  $\wrho = d^{-1}\one$.  Assumption~\assref{(B0)} obviously holds. As was already noted in~\cite[Remark~2.11]{CJPS-2017}, in this case 
\beq{future-stat}
e(\alpha)=\lim_{T\to\infty}\frac1T \log\left(\sum_{\omega\in\supp\PP_T}\PP_T^{1-\alpha}(\omega)\right)
-\alpha\log|\cA|.
\eeq
Note that the first term under the above  limit sign is the R\'enyi entropy per time step $T^{-1}S_{1-\alpha}(\PP_T)$. By declaring 
\[
{\cal E}_T(\omega_1, \dots, \omega_T)\defeq-\log \PP_T(\omega_1, \dots, \omega_T)
\]
the energy of the configuration $(\omega_1, \dots, \omega_T)$ and $\beta=1-\alpha$ the inverse temperature, the limit~\eqref{future-stat} turns into 
the pressure of  a ``spin system'' defined by the family  of finite-volume Hamiltonians $({\cal E}_T)_{T\in\nn^\ast}$. The resulting statistical mechanics, with  examples that complement those considered in this paper, is developed in~\cite{Dima1}.  

\paragraph{4}\label{rempar3} We use the opportunity to correct a minor inaccuracy in~\cite{BJPP1}: according to~\cite[Proposition~2.5]{JOPS}, the statement of~\cite[Theorem~2.10~(2)]{BJPP1} lacks a non-triviality condition, namely that {\sl the entropic pressure $e(\alpha$) should not vanish identically for $\alpha\in]0,1[$.} This omission propagates to~\cite[Theorem~2.12~(2)]{BJPP1}. Fortunately, the two omissions do not affect the other results of~\cite{BJPP1}. Indeed, as already mentioned, $\PP$ is $\phi$-ergodic thanks to Assumption~\assref{(C)} while the vanishing of the entropic pressure implies that of the mean entropy production rate. Hence, it follows from~\cite[Theorem~2.2~(3)]{BJPP1} that $\wP=\PP$ from which one easily concludes that the Stein and Hoeffding exponents vanish. Thus, in cases where the entropic pressure vanishes identically on the interval $]0,1[$, the conclusions of Theorem~2.12~(3) and~2.13 in~\cite{BJPP1} still hold, despite the failure of~\cite[Theorem~2.10~(2) and Theorem~2.12~(2)]{BJPP1}.

\section{Examples}
\label{sec-examples}


In this section we state our main results.
In Section~\ref{sec-PMP} we introduce the structural class of the positive matrix product (PMP) measures and  instruments. 
All our examples, except for the rotational instrument discussed in Section~\ref{sec-rotinst}, belong to this category. 
In Section~\ref{sec-basic-ex} we discuss basic examples --- Bernoulli instruments and Markov chain instruments. For those  instruments  Theorem~\ref{thm-int} 
is a textbook result. We describe them because of their importance and because they will reappear as special cases  in the analysis  of  more complex examples.  The  Keep--Switch instrument, our first novel example, is discussed in Section~\ref{sec-int-ks}. The spin instruments are discussed in Sections~\ref{spin-intro} and~\ref{sec-nevnnevxx}. In these two sections 
the results are stated in the abbreviated form, and the complete formulations are presented in Section~\ref{spin-main} where the proofs are given. The far-reaching relation  between the PMP measures and hidden Markov models is discussed in 
Section~\ref{sec-hmcp}.  Section~\ref{sec-rotinst} is devoted to the rotational instrument. 

The proofs are given in Sections~\ref{sec:flip--keep}, \ref{spin-main}, and~\ref{sec-ex-ro}. The spin-instruments proofs, 
given in Section~\ref{spin-main}, involve in part  tedious straightforward linear algebra computations 
that we will largely  omit. Finally, for the reader's convenience, we collect in Appendix~\ref{sec-cont-fra} some basic results about continued fractions that are used in the analysis of the rotational instrument.

\subsection{Positive matrix product measures and instruments}
\label{sec-PMP}

Let $(M_a)_{a\in\cA}$ be a  collection of $d\times d$ matrices with non-negative entries such that  
$M\defeq\sum_{a\in\cA} M_a$ satisfies  $M{\bf 1}={\bf 1}$, where\footnote{We denote the transpose of an arbitrary matrix $A$, by $A^\mathsf{T}$.}  ${\bf 1}\defeq[1\ \cdots\ 1]^{\mathsf T}$.  We write $M_a\eqdef[m_{ij}(a)]_{1\leq i, j\leq d}$. Let  ${\bf p}\defeq[p_1\ \cdots\  p_d]$ be a probability vector such that  ${\bf p}M={\bf p}$, and for $\omega\in\Omega_T$, set 
\[
\PP_T(\omega)\defeq{\bf p}M_{\omega_1}\cdots M_{\omega_T}{\bf 1}.
\]
There is a unique probability measure $\PP\in\cP_\phi(\Omega)$ such that
\[
\PP([\omega])=\PP_T(\omega)
\]
for all $T\in\nn^\ast$ and all $\omega\in\Omega_T$.

\begin{definition}\label{def:PMP}
We shall call $\PP$ the positive matrix product (abbreviated PMP) measure generated by $((M_a)_{a\in\cA}, {\bf p})$. 
\end{definition}

The class of PMP  measures is identical to the class of  hidden Markov models. This point is both of conceptual and technical importance, and we shall discuss it in Section~\ref{sec-hmcp}.

If $\PP$ is  a PMP measure, we  shall  call any element of $\fJ_\PP$ a PMP instrument. 
The following construction shows that $\fJ_\PP$ is non-empty and exhibits  its  canonical element.  Let $\cH$ be a Hilbert space of dimension $d$ and $(v_i)_{1\leq i \leq d}$  one of its orthonormal bases. 
Given $u,v\in\cH$ we use Dirac's notation $|v\rangle\langle u|$ for the linear map $w\mapsto\langle u|w\rangle v$.
Let $\cJ\defeq(\Phi_a)_{a\in\cA}$ and $\rho$ be the quantum instrument and the density matrix on $\cH$ defined by
\beq{eq:canonicalPMPinstr}
\Phi_a[X]\defeq\sum_{i,j=1}^dm_{ij}(a) \langle v_j| X v_j\rangle |v_i\rangle \langle v_i|,\qquad
\rho\defeq\sum_{i=1}^d p_i |v_i\rangle \langle v_i|.
\eeq
A simple computation gives that $\PP$ is the unraveling of $(\cJ,\rho)$. Note that the $d$-dimensional subspace of $\cB(\cH)$ consisting of all operators with a diagonal matrix representation w.r.t.\;the basis $(v_i)_{1\leq i \leq d}$ contains the range of each map $\Phi_a$ and $\Phi_a^\ast$, as well as the density matrix $\rho$.

In the remaining part of this section, we assume that all entries in ${\bf p}$ are strictly positive, which ensures that Assumption~\assref{(A)} holds.
This construction of an  instrument canonically associated to a PMP measure gives more. Invoking~\cite[Theorem~2.1]{JPW-2014}, one easily shows that the map
$\Phi\defeq\sum_{a\in\cA}\Phi_a$ is irreducible iff the right-stochastic matrix $M$ is irreducible.\footnote{$M$ is called irreducible whenever, for any index pair $ij$, there is a power $M^n$ with non-vanishing $ij$ entry.} Given an involution $\theta$ on
$\cA$, the OR measure $\wP$ is the PMP measure generated by 
$((\widehat M_a)_{a\in\cA}, {\bf p})$, where $\widehat M_a\defeq D^{-1}M_{\theta(a)}^{\mathsf T} D$, $D$ being the diagonal matrix with entries $p_1,\ldots,p_d$. Note that  the OR instrument  $(\wcJ, \wrho)$ constructed in~\eqref{eq:canonical_OR_instrument} is given by
\[
\widehat \Phi_a[X] =\sum_{i,j=1}^d\widehat m_{ij}(a) \langle v_j| X v_j\rangle |v_i\rangle \langle v_i|,\qquad\wrho=\rho,
\]
and coincides with the instrument canonically associated to the PMP measure $\wP$.

\begin{lemma}\label{PMP-props}
Let $\PP$ be the PMP measure generated by $((M_a)_{a\in\cA},\mathbf{p})$, and denote by $(\cJ,\rho)$ the instrument canonically associated to it.
\ben 
\item\label{it:PMPtwo} If $m_{ij}(a)=0 \Leftrightarrow m_{ji}(\theta(a))=0$ for any index pair $ij$ and all $a\in\cA$, then Assumption~\assref{(B)}  holds.
\item\label{it:PMPthree}  If the matrix $\sum_{a\in\cA}M_a\otimes \widehat M_a$ is irreducible, then Assumption~\assref{(C)} holds.
\item\label{it:PMPfour} If all entries in $M_a$ are positive entries for all $a\in\cA$, then Assumption~\assref{(C)} holds with $\tau=0$, and $\PP$ is weak Gibbs.
\een
\end{lemma}

\proof 
Part~\ref{it:PMPtwo} is easily deduced from the fact that, 
since all entries in ${\bf p}$ are strictly positive, 
${\bf p}M_{a_1}\cdots M_{a_T}{\bf 1}=0$ iff
$m_{i_1i_2}(a_1)m_{i_2i_3}(a_2)\cdots m_{i_Ti_{T+1}}(a_T)=0$ for all indices $i_1,i_2,\ldots,i_{T+1}$. To prove Part~\ref{it:PMPthree}, observe that $(\Phi_a\otimes\widehat{\Phi}_a)_{a\in\cA}$ is the instrument canonically associated to the PMP measure generated by $((M_a\otimes\widehat{M}_a)_{a\in\cA},\mathbf{p}\otimes\mathbf{p})$. The result then follows from Remark~\ref{rem:4}. Part~\ref{it:PMPfour} is a variation of~\cite[Theorem~2.1]{feng02}. Namely, with
\[
C\defeq\min_{a\in\cA}\min_{i,j}\frac{m_{ij}(a)}{p_j\sum_k m_{ik}(a)},
\]
one has $m_{ij}(a)\ge C \sum_k m_{ik}(a)p_j$ for all $a\in\cA$ and any pair of indices $ij$. We can conclude that
\beq{covid20}
\PP_{T+S}(\omega)\ge C\PP_T(\omega)\PP_S\circ\phi^T(\omega)
\eeq
holds for all $\omega\in\Omega$ and $T,S\in\nn^\ast$. A similar estimate clearly holds for the OR measure $\wP$, which 
yields Assumption~\assref{(C)} with $\tau=0$. Together with the upper-decoupling property~\eqref{dec-i}, \eqref{covid20}
allows us to invoke Theorem~\ref{noe-thm}~\ref{it:noeone} to conclude that $\PP$ is weak Gibbs.\qed
 
 Note that  if~\assref{(B)} holds, then the limit~\eqref{eq:defealpha}  is finite  for all $\alpha$.  

Our final remark concerns  unravelings of two-time measurement processes.  If  such  an unraveling  is a PMP measure generated by $( (M_a)_{a\in\cA}, {\bf p})$, then
\beq{tbus-l}
\ep(\cJ,\rho)=\sum_{a\in\cA} \Delta {\mathfrak  S}(a)\mathbf{p}M_a\mathbf{1},\qquad e(\alpha)=\log r(\alpha),
\eeq
where $r(\alpha)$ is the spectral radius of the matrix
 \[
M(\alpha)=\sum_{a\in\cA}\e^{-\alpha \Delta {\mathfrak S}(a)}M_a.
\]

\subsubsection{Basic examples}
\label{sec-basic-ex}


\paragraph{Bernoulli instruments.} 
Let $Q$ be a probability mass function on the alphabet $\cA$ and let $\PP$ be the Bernoulli measure generated by $Q$, \ie the unique measure on $\Omega=\cA^{\nn^\ast}$ such that, for all $T\in\nn^\ast$ and all $\omega\in\cA^T$,
\[
\PP\left([\omega]\right)=Q(\omega_1)\cdots Q(\omega_T).
\]
Any element of $\fJ_\PP$ will be called a Bernoulli instrument generated  by $Q$. $\PP$ is a PMP measure generated by 
$((M_a)_{a\in\cA}, {\bf p})$, where, for any $d\geq 1$, one can take $M_a\defeq Q(a)\one$,  and 
where  ${\bf p}$ is  an arbitrary probability vector in $\rr^d$ with strictly positive entries. Note that, given some involution $\theta$ of $\cA$, the OR 
$\wP$ is the Bernoulli measure generated by ${\widehat Q}\defeq Q\circ\theta$  and that Assumptions~\assref{(A)} 
and~\assref{(C)} obviously hold. Assumption~\assref{(B)} holds iff 
$Q(a)=0 \Longleftrightarrow\widehat{Q}(a)=0$.   Note also that  $M\defeq\sum_{a\in\cA} M_a=\one$, and so  if $d>1$, $M$ is not irreducible.

For a Bernoulli instrument we have 
\[
\ep(\cJ,\rho)=S(Q|\widehat Q)=\sum_{a\in\cA}Q(a) \log \frac{Q(a)}{{\widehat Q}(a)},
\]
and
\[ 
e(\alpha)=S_{1-\alpha}(Q\mid\widehat Q)=\log \left( \sum_{a\in\cA}Q(a)^{1-\alpha} {\widehat Q}(a)^{\alpha}\right).
\]
Note in particular that $\ep(\cJ,\rho)>0$ iff $Q\not=\widehat Q$.

Examples of Bernoulli instruments are given in Remark~\ref{rem:KSber}, Theorems~\ref{xy-ot-on}, \ref{xx-one-int}~\ref{it:otone}+\ref{it:otsev}+\ref{it:otnine}, \ref{thm:X00tp}~\ref{it:ttone}+\ref{it:ttsix}, \ref{thm:X00random}~\ref{it:rmone}+\ref{it:rmsix}, \ref{thm:XXZot}, \ref{thm-xy-tp}~\ref{it:thsix},  \ref{thm-wbde}~\ref{it:rteight}; see also Theorem~\ref{thm-xxz-multi}~\ref{it:mtone}, and the ``trivial case'' in Section~\ref{sec-tris}.

\paragraph{Markov chain instruments.} 
Let $P=[p_{xy}]_{x, y\in\cA}$ be a right-stochastic matrix on $\rr^\cA$ and ${\bf p}=[p_x]_{x\in\cA}$  a probability 
vector with strictly positive entries satisfying ${\bf p}P={\bf p}$. The Markov measure $\PP$ generated by $(P, {\bf p})$
is the unique measure on $\Omega=\cA^{\nn^\ast}$ such that, for all $T\in\nn^\ast$ and all $\omega\in\cA^T$,
\beq{eq:MarkovM}
\PP\left([\omega]\right)
=p_{\omega_1}p_{\omega_1\omega_2}\cdots p_{\omega_{T-1}\omega_T}.
\eeq
One easily verifies that $\PP$ is the PMP measure generated by $((M_a)_{a\in\cA}, {\bf p})$, where $M_a$ has the entries
\[
m_{xy}(a)\defeq p_{xy}\delta_{xa},
\]
$\delta$ denoting the Kronecker symbol.\footnote{$m_{xy}(a) = p_{xy}\delta_{ya}$ is an equally valid choice.} Obviously,  $\sum_{a\in\cA}M_a =P$. The canonically associated instrument in \eqref{eq:canonicalPMPinstr} takes the form
\beq{eq:canonicalMarkov}
\Phi_a[X] =\sum_{x\in \cA}p_{ax} \langle v_x| X v_x\rangle |v_a\rangle \langle v_a|,\qquad
\rho=\sum_{x\in \cA} p_x |v_x\rangle \langle v_x|.
\eeq

Given an alphabet involution $\theta$, the OR measure $\wP$ is also Markov and is generated by $(\widehat{P}, \widehat {\bf p})$, where 
\[
\widehat p_x =p_{\theta(x)}, \qquad \widehat p_{xy}=\frac{p_{\theta(y)}}{p_{\theta(x)}}p_{\theta(y)\theta(x)}.
\]
\begin{theorem}\label{ex-markov-c}
Suppose that $P$ is irreducible and denote by $(\cJ,\rho)$ the instrument canonically associated to the PMP measure $\PP$. Then: 
\ben
\item Assumptions~\assref{(A)}  and~\assref{(C)} hold. Assumption~\assref{(B)} holds iff, for all $x,y\in\cA$,
\[
p_{xy}=0 \,\,\Longleftrightarrow\,\, p_{\theta(y)\theta(x)}=0.
\]
\item 
\[ 
\ep(\cJ,\rho)=\sum_{\substack{x, y \in\cA\\ p_{xy}>0}} p_x\log \frac{p_{xy}}{\widehat p_{xy}}.
\]
In particular, $\ep(\cJ,\rho)=0$ iff, for all $x,y\in\cA$, 
\[
 p_x p_{xy}=p_{\theta(y)}p_{\theta(y)\theta(x)}.
 \]
\item For $\alpha\in \rr$, the matrix\footnote{We use the convention $0/0=0$.} 
\[
P(\alpha)=\left[p_{xy}^{1-\alpha}\widehat p_{xy}^{\,\alpha}\right]_{x,y\in\cA}
\]
is irreducible and $e(\alpha)=\log r(\alpha)$, where $r(\alpha)$ is spectral radius  of $P(\alpha)$. The map 
$\rr \ni \alpha \mapsto e(\alpha)$ is real analytic. 
\een
\end{theorem}
\begin{remark}\label{rem:VonneumannMarkov}
Physically important Markov chain instruments  are described in Theorems~\ref{thm-xy-tp}~\ref{it:thsev} and~\ref{thm-wbde}~\ref{it:rtnin}. 
A special class of Markov chain instruments is provided by von~Neumann instruments with rank-one projections. Indeed, with the notation of Section~\ref{sec-vn-instruments}, setting $p_{xy}=\tr(P_yUP_xU^\ast)$ clearly yields a doubly stochastic matrix $P=[p_{xy}]_{x,y\in\cA}$, and one easily checks that the Markov measure generated by $(P,{\bf 1})$ coincides with the unraveling of the associated von~Neumann instrument. In fact, the matrix $P$ obtained here is unistochastic (in dimension strictly larger than $2$, such matrices form a proper subset of the set of doubly stochastic matrices). Conversely, for any unistochastic matrix $P$, one can find a von~Neumann instrument whose unraveling is the Markov process generated by $(P,{\bf 1})$.
\end{remark}

\begin{remark}
We mention without giving any details that one can, in the same way, describe multistep Markov chains (see for example \cite{Dima1, fannes1992functions}), \ie measures of the form 
\[\PP_T(\omega_1, \dots, \omega_T)\defeq	p_{\omega_1\cdots\omega_k}  \prod_{j=0}^{T-k-1}p_{\omega_{j+1} \cdots \omega_{j+k}; \omega_{j+k+1}}\]
 under some standard invariance condition, where $k\geq 1$ (the case $k=1$ corresponds to the usual Markov chains discussed above). $\PP$ is then a PMP measure, as one sees by choosing $M_a$ acting on $\rr^{d}$, with $d\defeq |\cA|^k$, and ${\bf p}$ as 
\[
M_a\defeq\sum_{y_1, \dots, y_k\in\cA}p_{ay_1\cdots y_{k-1};y_k}|v_{ay_1\cdots y_{k-1}}\rangle \langle v_{y_1\cdots y_k}|, \qquad {\bf p}\defeq\sum_{x_{1},\, \dots, x_k\in \cA} p_{x_1\cdots x_k}v_{x_1\cdots x_k},
\]
where $(v_{x_1\cdots x_k})_{x_1, \dots, x_k\in \cA}$ is an orthonormal basis of $\rr^d$. \end{remark}


\subsubsection{The Keep--Switch PMP instrument} 
\label{sec-int-ks}

This example, studied in Section~\ref{sec:flip--keep}, is the instrument with alphabet $\cA\defeq\{K, S\}$ canonically associated to the PMP measure $\PP$ generated by $((M_K,M_S),{\bf p})$, where
\beq{ksdef}
M_K\defeq\begin{bmatrix}q_1 &0\\0&q_2\end{bmatrix}, \qquad 
M_S\defeq\begin{bmatrix}0 &r_1\\r_2&0\end{bmatrix}, \qquad 
{\bf p}\defeq(r_1+r_2)^{-1}\begin{bmatrix}r_2&r_1\end{bmatrix},
\eeq
$q_1,q_2$ being two parameters in the interval ${]}0,1{[}$ and $r_1\defeq1-q_1$, $r_2\defeq1-q_2$. Let $\theta$ be the unique non-trivial involution of $\cA$, \ie $\theta(K)=S$. We will discuss the probabilistic interpretation of the measure $\PP$ in Section~\ref{sec-hmcp}.

\begin{remark}\label{rem:KSexchange}
Exchanging $q_1$ and $q_2$ (and hence $r_1$ and $r_2$) does not change the measure $\PP$.
\end{remark}

\begin{remark}\label{rem:KSber}
In the case  $q_1=q_2$, $\PP$ is the Bernoulli measure
generated by the mass function $Q(K)=q_1$. The case $q_1=1-q_2$ has a long history in the literature; see  Section~\ref{sec-hmcp}. To the best of our knowledge, the cases $q_1\not=1-q_2$, 
$q_1\not=q_2$,  have not been studied before. 
\end{remark}

\begin{remark}\label{rem:epzero}  We have $\PP = \wP$ iff $q_1 = q_2 = \frac 12$. The $\Leftarrow$ implication is obvious, and the $\Rightarrow$ implication follows, for example, by writing the two identities $\PP_1(K) = \PP_1(S)$ and $\PP_3(KKK) = \PP_3(SSS)$ in terms of $q_1, q_2$ and solving the resulting system of equations.
\end{remark}

\begin{remark} Replacing $\theta$ by the identity would lead to the OR measure satisfying $\wP=\PP$, see Proposition~\ref{lem:reversibilityKS}. Thus, the central role here is played by the choice of a non-trivial $\theta$, and time-reversal in itself plays no role.
\end{remark}

To avoid uninteresting cases, we assume throughout that $q_1\not=q_2$. Our main result is: 

\begin{theorem}
\hspace{0em}
\label{mainthm-KS}
\ben
\item\label{KS-i}  Assumptions~\assref{(A)}, \assref{(B)} and~\assref{(C)} hold. In particular, Theorem~\ref{thm-int} applies.
\item\label{KS-ii} The measure $\PP$ is  not weak Gibbs. 
\item\label{KS-iii} Let $\gamma =\frac{1}{2}\log \frac{q_1}{q_2}$ and $\eta=\frac12\log\frac{q_1q_2}{r_1r_2}$. Then,
\[
\ep(\cJ,\rho)=\frac{(r_2-r_1)\gamma+\left(r_1-4r_1r_2+r_2\right) \eta}{r_1+r_2}>0.
\]
\item\label{KS-iv}  The  function $\alpha\mapsto e(\alpha)$ is differentiable and strictly convex. It is real analytic on $\rr\setminus\{0,1\}$,  but fails to be twice differentiable at $\alpha\in\{0,1\}$. Moreover,
\beq{eq:eppjump}
(\partial^+e')(0)-(\partial^-e')(0)=(\partial^-e')(1)-(\partial^+e')(1)
=\frac{4r_1r_2}{(q_1+q_2)(r_1+r_2)}\gamma^2.
\eeq
\item\label{KS-v} The random variables  $(\sigma_T)_{T\in\nn^\ast}$ satisfy a non-Gaussian central limit theorem: as $T\to\infty$,
\beq{eq:CLTKSPMP}
 \frac{{\sigma_T-T\ep(\cJ,\rho)}}{{\sqrt{T}}}
 \eeq
converges in law towards $Z_1 - |Z_2|$, where $Z_1$ and $Z_2$ are independent, centered normal random variables of variance given by
\begin{align*}
{\rm Var}(Z_1)&=\frac{4r_1r_2 }{(r_1+r_2)^3}\left(
(q_1+q_2)\gamma^2+4(q_1-q_2)\gamma\eta+4(q_1r_2^2+q_2r_1^2)\eta^2\right)>0,\\
{\rm Var}(Z_2) &=\frac{4r_1r_2}{(q_1+q_2)(r_1+r_2)}\gamma^2>0.
\end{align*}
\item\label{KS-vi} The  fluctuation--dissipation relation fails (see Section~\ref{sec:FDR} for a more precise statement).
\een
\end{theorem}
 
The proof of Theorem~\ref{mainthm-KS} is given in Section~\ref{sec:flip--keep}. We remark that Part~\ref{KS-i} is easily established. Indeed,  \assref{(A)} is immediate and~\assref{(B)} follows from the observation that $\supp \PP_T=\supp\wP_T=\Omega_T$ for all $T\in\nn^\ast$. To prove~\assref{(C)}, one shows that the matrix $M_K \otimes M_S + M_S \otimes M_K$ is irreducible on $\rr^2\otimes \rr^2$ by verifying that all the entries of the matrix 
$(\one + M_K \otimes M_S + M_S \otimes M_K)^2$ are strictly positive. Thus, we will start the proof of Theorem~\ref{mainthm-KS} in Section~\ref{sec:flip--keep} with Part~\ref{KS-ii}. The proof gives an explicit expression for $e(\alpha)$; 
see~\eqref{eq:KSfinalealpha} and~\eqref{eq:defQlambda}.

We finish with the following remark. In addition to the canonical PMP instrument $((\Phi_K, \Phi_S), \rho)$ defined in~\eqref{eq:canonicalPMPinstr}, $\fJ_\PP$ contains two instruments that are physically more natural. 

The first one is the instrument $((\Psi_K, \Psi_S), \rho)$  on $\cH\defeq\cc^2$  defined as
\begin{align*}
\Psi_K[X]&\defeq\begin{bmatrix}\cos\varphi_1&0\\0&\cos\varphi_2\end{bmatrix}X\begin{bmatrix}\cos\varphi_1&0\\0&\cos\varphi_2\end{bmatrix},\\[4pt]
\Psi_S[X]&\defeq\begin{bmatrix}0&\sin\varphi_1\\ \sin\varphi_2&0\end{bmatrix}X\begin{bmatrix}0&\sin\varphi_2\\ \sin\varphi_1&0\end{bmatrix},
\end{align*}
and
\[
\rho\defeq(r_1+r_2)^{-1}\begin{bmatrix} r_2&0\\0&r_1\end{bmatrix},
\]
where  $\varphi_1, \varphi_2$ are such that $q_1=\cos^2\varphi_1$, $q_2=\cos^2\varphi_2$.
One easily checks that $((\Psi_K, \Psi_S), \rho)\in\fJ_\PP$. The canonical OR instrument is $(\widehat{\Psi}_K,\widehat{\Psi}_S)=(\Psi_S,\Psi_K)$. 

The second  one is the  X00-spin instrument with $\rho_p$ a pure state; see Section~\ref{sec-nevnnevxx} and in particular 
Parts~\ref{part:X00-KSp} and~\ref{part:X00-KSm} of Theorem~\ref{xx-one-int}.

\subsubsection{XXZ-spin instruments} 
\label{spin-intro}
With the notations of Section~\ref{sec-probes},  the setting of the probe measurements that leads to the XXZ-spin instruments is the following.
The Hilbert spaces are  $\cH=\cH_p\defeq\cc^2$. We denote by  $\sigma_x, \sigma_y, \sigma_z$ the usual Pauli matrices. 
The model depends on the parameters $\epsilon>0$, $\omega>0$, $\lambda>0$,  $\mu\in \rr$, $t>0$, and $\eta \in {]}{-}1/2, 1/2[$. The Hamiltonians of the system $\cS$ and the probes are 
\beq{eq-xxz-ham}
 H_\cS\defeq\frac{\omega}{2}\sigma_z, \qquad H_p\defeq\frac{\epsilon}{2}\sigma_z,
 \eeq
and the interaction between $\cS$ and a single probe is described by 
\beq{eq-xxz-v}
V\defeq\frac\lambda2(\sigma_x\otimes\sigma_x+\sigma_y\otimes\sigma_y) +\frac\mu2\sigma_z\otimes\sigma_z.
\eeq
The full Hamiltonian is 
\beq{xy-ham}
H\defeq H_\cS\otimes \one + \one \otimes H_p + V,
\eeq
and the corresponding propagator over a period $t$, $U\defeq\e^{-\i t H}$, will be computed in  Section~\ref{xy-main}. The state of the probes is 
\beq{pr-st-xy}
\rho_p\eqdef\begin{bmatrix} \frac{1}{2}-\eta &0\\0 & \frac{1}{2}+\eta\end{bmatrix}=Z_p^{-1}\e^{-\beta_pH_p},
\eeq
where, in the cases of thermal probes, $\eta$ is linked to the inverse temperature $\beta_p\in\rr$ through the relation $\eta=\frac12\tanh\frac{\beta_p\epsilon}{2}$. Finally, $\cL\defeq\{-,+\}$, and the partition of unity
\beq{eq:Ppm}
 P_+\defeq\begin{bmatrix} 1 &0\\0 & 0\end{bmatrix}, \qquad 
P_-\defeq\begin{bmatrix} 0 &0\\0 & 1\end{bmatrix}
\eeq
is associated to measurements  of the probe energy $H_p$.
The time-reversal invariance~\eqref{tri-tri} holds with $\Theta$ and $\Theta_p$ acting as complex conjugation. This setting will be used for the one-time measurement protocol (with $\cA=\cL$) as well as for two-time measurements with thermal, random thermal, and multi-thermal probes (with $\cA=\cL\times\cL$).

\paragraph{One-time measurements.} 
The instrument $\cJ=(\Phi_-,\Phi_+)$  is described in detail in Section~\ref{sec-xxz-one}. One shows that $\Phi=\Phi_-+\Phi_+$ is irreducible and that $\rho=\rho_p$ is the unique density matrix for which $(\cJ, \rho)$ satisfies Assumption~\assref{(A)}. Moreover, one has a complete description of the unraveling $\PP$ of this instrument.
\begin{theorem}\label{xy-ot-on}
 The unraveling $\PP$ is the Bernoulli measure generated by the mass function $Q(\pm)=\frac12\mp\eta$. In particular, 
$\PP$  does not depend on $\epsilon$, $\omega$, $\lambda$, $\mu$, and $t$.  
\end{theorem}

\paragraph{Two-time measurements with a thermal probe.}\label{sec:xxz-results}
The instrument $\cJ=(\Phi_{++},\Phi_{+-},\Phi_{-+},\Phi_{--})$ is described in Section~\ref{sec-xy-tttp}. We set 
\beq{Eq-sDef}
\delta\defeq\sqrt{\left(\frac{\epsilon-\omega}2\right)^2+\lambda^2},\qquad
s\defeq\left(\lambda\frac{\sin(\delta t)}{\delta}\right)^2.
\eeq
Applying  Theorems~\ref{thm-two-time} and~\ref{noe-thm}, the following result will be proved in Section~\ref{sec-xy-tttp}.
\begin{theorem} \label{thm-xxz-twotime}
\hspace{0em}
\ben
\item $\rho=\rho_p$ is the unique density matrix for which $(\cJ,\rho)$ satisfies Assumption~\assref{(A)}.
\item $\ep(\cJ,\rho)=0$ and $e\equiv 0$. 
\item The unraveling $\PP$ of $(\cJ,\rho)$ is a PMP measure. 
\item Suppose that $s\in{]}0,1{[}$. Then $\PP$ is not weak Gibbs for any $\beta_p\in\rr$.
\een
\end{theorem}
The surprising aspect of this result is that although $\PP$ is not a weak Gibbs measure, its two marginals
are Bernoulli measures. This follows from Remark~\ref{rem:marginals} and Theorem~\ref{xy-ot-on}.

For additional information see Theorem~\ref{thm-xy-tp}.

\paragraph{Two-time measurements with random thermal probes.}
We now consider the case of XXZ-spin interactions with $K$ random thermal probes labeled by $k\in\lbr1,K\rbr$. At each step the $k$-th probe, which is in thermal equilibrium at inverse temperature $\beta_k$, is selected with probability $w_k$. The corresponding instrument $\cJ$ is described in Section~\ref{sec-xy-rtb}, where more details can be found. In particular, we shall prove that there exists a unique density matrix $\rho$ such that $(\cJ,\rho)$ satisfies Assumption~\assref{(A)}. The following theorem only gives the main properties of this instrument.

\begin{theorem}\label{thm-xxz-ran}
\hspace{0em}
\ben
\item
\[
\ep(\cJ,\rho)=\frac{s}{2}\sum_{k,l=1}^Kw_kw_l
\frac{((\beta_k-\beta_l)\epsilon/2)\,\sinh((\beta_k-\beta_l)\epsilon/2)}
{\cosh((\beta_k+\beta_l)\epsilon/2)+\cosh((\beta_k-\beta_l)\epsilon/2)},
\]
where $s$ is given by~\eqref{Eq-sDef}.
\item $\ep(\cJ,\rho)=0$ if and only if $s=0$ or   $\beta_1=\beta_2=\dots=\beta_K$.
\item The entropic pressure is given by
\[
e(\alpha)=\log\left(1+\frac{s}{2}\left(\sqrt{1-\Delta(\alpha)}-1\right)\right),
\]
where
\[
\Delta(\alpha)\defeq
\sum_{k,l=1}^Kw_kw_l\frac{\sinh(\alpha(\beta_k-\beta_l)\epsilon/2)\sinh((1-\alpha)(\beta_k-\beta_l)\epsilon/2)}
{\cosh(\beta_k\epsilon/2)\cosh(\beta_l\epsilon/2)}
\]
is an entire analytic function such that $\Delta(\alpha)<1$ for all $\alpha\in\rr$.
\item The unraveling $\PP$ of $(\cJ,\rho)$ is a PMP measure. 
\item Suppose that $s\in{]}0,1{[}$. Then $\PP$ is not weak Gibbs for any $\beta_1,\dots, \beta_K\in\rr$.
\een
\end{theorem}

\paragraph{Two-time measurements with multi-thermal probes.}
This is  a computationally involved  example, and we will consider only the  case $K=2$. The probe Hilbert space 
is $\cH_p\defeq\cc^2\otimes \cc^2$, the Hamiltonian of each sub-probe is $H_k\defeq\tfrac\epsilon2\sigma_z$, and 
\[ 
H_p\defeq\frac\epsilon2\left( \sigma_z\otimes \one +\one \otimes \sigma_z\right).
\]
The interaction between $\cS$ and the probes is described by 
\[
\begin{split}
 V\defeq\frac\lambda2(\sigma_x\otimes \sigma_x\otimes \one + 
\sigma_x\otimes \one\otimes \sigma_x&+ \sigma_y \otimes \sigma_y\otimes \one +
\sigma_y\otimes\one\otimes \sigma_y) \\[1mm]
+\frac\mu2(\sigma_z \otimes \sigma_z\otimes \one &+ \sigma_z\otimes \one\otimes \sigma_z),
\end{split}
\]
with coupling constants $\lambda>0$ and $\mu\in\rr$. The full Hamiltonian is 
\[ 
H\defeq H_\cS\otimes \one + \one \otimes H_p +V,
\]
and $U\defeq\e^{-\i t H}$, where $t>0$. The matrix $U$ and the resulting quantum  instrument $\cJ$ are computed in Section~\ref{sec-tris}, where we also carry out the  complete analysis of the model. Here we state the results in the special case $\epsilon=\omega>0$,  $\mu=0$. 
\begin{theorem}
\label{thm-xxz-multi}
\hspace{0em}
\ben
\item\label{it:mtone} If $\lambda t\in\frac\pi2\nn^\ast$, then any state $\rho$ satisfies $\Phi^\ast[\rho]=\rho$ and the unraveling $\PP$ of the instrument $(\cJ,\rho)$ is a convex combination of two Bernoulli measures.
\een

\medskip
\noindent In the remaining statements we assume that $\lambda t\not\in\frac\pi2\nn^\ast$.

\medskip 
\ben\setcounter{enumi}{1}
\item\label{it:mttwo} There is a unique state $\rho$ such that $(\cJ,\rho)$ satisfies Assumption~\assref{(A)}, and all the conclusions of Theorem~\ref{thm-two-time} hold.
\item\label{it:mtthree}
\[
\ep(\cJ,\rho)=2s
\frac{\left((\beta_1-\beta_2)\epsilon/2\right)\sinh\left((\beta_1-\beta_2)\epsilon/2\right)}
{\cosh\left((\beta_1-\beta_2)\epsilon/2\right)+\cosh\left((\beta_1+\beta_2)\epsilon/2\right)}.
\]
Obviously, $\ep(\cJ,\rho)=0$ if and only if $\beta_1=\beta_2$.
\item\label{it:mtfour}The entropic pressure is given by
\[
e(\alpha)=2\log\left(1+s\left(\sqrt{1-\Delta(\alpha)}-1\right)\right),
\]
where
\[
\Delta(\alpha)\defeq
\frac{\sinh\left(\alpha(\beta_1-\beta_2)\epsilon/2\right)\sinh\left((1-\alpha)(\beta_1-\beta_2)\epsilon/2\right)}
{\cosh\left(\beta_1\epsilon/2\right)\cosh\left(\beta_2\epsilon/2\right)}
\]
is an entire analytic function such that $\Delta(\alpha)<1$ for all $\alpha\in\rr$.
\item\label{it:mtfive} The unraveling $\PP$ is a PMP measure.
\item\label{it:mtsix} $\PP$ is not weak Gibbs for any $\beta_1>0,\beta_2>0$.
\een
\end{theorem}

\subsubsection{X00-spin instruments} 
\label{sec-nevnnevxx}
The models to be considered in  this section are similar to the previous XXZ-spin instruments, except for the system--probe 
interaction which is now given by
\[
V\defeq\frac\lambda2 \sigma_x\otimes \sigma_x,
\]
and the fact that we also allow for the values $\eta=\pm 1/2$ in the probe state~\eqref{pr-st-xy}. The total Hamiltonian is
\beq{xx-ham}
H\defeq H_\cS\otimes \one + \one \otimes H_p + V,
\eeq
with $H_\cS$ and $H_p$ given by~\eqref{eq-xxz-ham}. The explicit form of the propagator $U\defeq\e^{-\i t H}$ will be given in Section~\ref{xx-main}.

\paragraph{One time measurements.}
The properties of the instrument $\cJ=(\Phi_-,\Phi_+)$ describing one-time measurements of the X00-spin interaction, as well as that of its unraveling $\PP$ are given in the following theorem. We set
\beq{eq-upm} 
s_\pm\defeq\left(\lambda\frac{\sin\left(\frac t2\sqrt{\lambda^2+(\omega\pm\epsilon)^2}\right)}
{\sqrt{\lambda^2+(\omega\pm\epsilon)^2}}
\right)^2.
\eeq

\begin{theorem}\label{th:xx-M}\label{xx-one-int}
\hspace{0em}
\ben
\item\label{it:otone} If $s_-=s_+=0$, then any density matrix $\rho$ satisfies $\Phi^\ast[\rho]=\rho$ and the unraveling $\PP$ of the associated instrument $(\cJ,\rho)$ is the Bernoulli measure generated by the mass function $Q(\pm)=1/2\mp\eta$. 
\een

\medskip\noindent
In the following, we assume that $s_-+ s_+ >0$ and  set
\beq{str-3}
p\defeq\frac12+\eta\frac{s_+-s_-}{s_++s_-}.
\eeq

\medskip
\ben\setcounter{enumi}{1}
 \item\label{it:ottwo} There is a unique density matrix,
\beq{str-2}
\rho\defeq\begin{bmatrix} p &0\\0 & 1-p \end{bmatrix},
\eeq
such that the instrument $(\cJ,\rho)$ satisfies Assumption~\assref{(A)}. The time-reversal invariance~\eqref{tri-tri} obviously holds.
\item\label{it:otthree} The unraveling $\PP$ of the instrument $(\cJ, \rho)$  is a PMP measure.
\item\label{it:otfour} If $s_-=1$ and $s_+=0$, then 
\[
\PP=\left(\frac{1}{2}-\eta\right)\delta_+ + \left(\frac{1}{2}+\eta\right)\delta_-,
\]
where $\delta_{\pm}$ is the Dirac  measure at $(\pm,\pm,\dots)$. 
\een

\medskip\noindent
In the remaining statements we assume that $s_{\pm}\in{]}0,1{[}$.

\medskip
\ben\setcounter{enumi}{4}
\item\label{part:X00-KSp}  If  $\eta=-\frac12$, then $\PP$ is the Keep--Switch PMP measure with $(K,S)=(+,-)$ and $(r_1,r_2)=(s_+,s_-)$.
\item\label{part:X00-KSm}  If $\eta=\frac12$, then $\PP$ is the Keep--Switch PMP measure with $(K,S)=(-,+)$ and $(r_1,r_2)=(s_-,s_+)$.
\item\label{it:otsev} If $\eta=0$, then $\PP$ is the Bernoulli measure generated by the mass function $Q(\pm)=1/2$.
\een

\medskip\noindent
In the remaining statements we assume in addition that $|\eta|\in{]}0, 1/2{[}$.

\medskip
\ben\setcounter{enumi}{7}
\item\label{it:oteight}  $\PP$ is a weak Gibbs measure and the entropic pressure $e$ is a differentiable function on $\rr$. 
\item\label{it:otnine} If $\theta(+)=-$, then  $\ep(\cJ,\rho)=0$ iff $s_-=s_+=1/2$. The latter condition implies that $\PP$ is the Bernoulli measure generated by the mass function $Q(\pm)=1/2$.
\item\label{it:otten} If  $\theta(+)=+$, then  $\ep(\cJ,\rho)= 0$.
\een
\end{theorem}

\begin{remark} Assumption~\assref{(A)} always holds. Assumptions~\assref{(B)} and~\assref{(C)} hold apart from the trivial boundary cases $(s_-,s_+)\in\{(0,0),(1,0)\}$.
\end{remark} 
 
 \begin{remark} With regard to~\ref{part:X00-KSp} and~\ref{part:X00-KSm}, we emphasize that $\rho_p$ given by~\eqref{pr-st-xy} is a pure state iff  $\eta=\pm 1/2$. One easily sees that as the parameters $\omega,\epsilon,\lambda$ and $t$ vary in ${]}0,\infty{[}$, the pair $(s_-,s_+)$ can take any value in $[0,1]\times[0,1{[}$. Thus,  if $\eta=\pm 1/2$, then for any Keep--Switch PMP measure $\PP$ one can find parameters $\omega,\epsilon,\lambda$ and $t$ such that the unraveling of the corresponding X00-spin instrument is equal to $\PP$. The failure of the fluctuation--dissipation relation for the Keep--Switch PMP instruments translates to its failure for the X00-spin instruments with $\eta=\pm\frac12$. \end{remark} 
 
\paragraph{Two-time measurements with a thermal probe.}
The associated instrument $\cJ$ is described in Section~\ref{sec-xx-ttm}, where we prove the following theorem.
\begin{theorem}\label{thm:X00tp}
\hspace{0em}
\ben
\item\label{it:ttone} If $s_-=s_+=0$, then any diagonal density matrix $\rho>0$ satisfies Assumption~\assref{(A)} and the unraveling  of $(\cJ,\rho)$ is the  Bernoulli measure on $\{{++},{--}\}^{\nn^\ast}$  generated by the mass function
$Q(\pm\pm)=\e^{\mp\beta\epsilon/2}/2\cosh(\beta \epsilon/2)$.
\een

\medskip\noindent
In the following, we assume that $s_-+ s_+ >0$.

\medskip
\ben\setcounter{enumi}{1}
\item\label{it:tttwo} The map $\Phi$ is irreducible and the unique density matrix $\rho$ such that $(\cJ,\rho)$ satisfies Assumption~\assref{(A)} is given by~\eqref{str-3} and~\eqref{str-2} with $\eta\defeq\frac12\tanh(\beta\epsilon/2)$.
\item\label{it:ttthree} The unraveling $\PP$ of the instrument $(\cJ,\rho)$ is a PMP measure. 
\item\label{it:ttfour} 
\[
\ep(\cJ,\rho)=\frac{2s_+s_-}{s_++s_-}\beta\epsilon\tanh(\beta\epsilon/2),
\]
and in particular $\ep(\cJ,\rho)>0$ iff $s_\pm$ are both non-vanishing.
\item\label{it:ttfive} The entropic pressure is given by
\[
e(\alpha)=\log\left(1+\frac{s_++s_-}2\left(\sqrt{1-\Delta(\alpha)}-1
\right)\right),
\]
where
\[
\Delta(\alpha)\defeq\frac{4s_+s_-}{(s_++s_-)^2}
\frac{\sinh(\alpha\beta\epsilon)\sinh((1-\alpha)\beta\epsilon)}{\cosh^2(\beta\epsilon/2)}
\]
is an entire analytic function such that $\Delta(\alpha)<1$ for all $\alpha\in\rr$.
\item\label{it:ttsix} If $s_+=s_-=1/2$, then $\PP$ is Bernoulli. In the opposite cases  $\PP$ is not weak Gibbs for any $\beta >0$.
\een
\end{theorem}

\begin{remark} As noted in~\cite{HJPR1}, the invariant state $\rho$ is not the Gibbs state at inverse temperature $\beta$ and Bohr frequency $\epsilon$ unless $s_-=0$, that is,  unless $t\sqrt{\lambda^2+(\omega-\epsilon)^2}\in2\pi\nn^\ast$.
\end{remark}

\begin{remark} This strict positivity of  entropy production stated in Part~\ref{it:ttfour} has  been previously observed~\cite{HJPR1,HJPR2}.
\end{remark}

For additional information see Section~\ref{sec-xx-ttm}.

\paragraph{Two-time measurements with random thermal probes.}

The corresponding instrument $\cJ$ is described in Section~\ref{sec-xx-rttm}. The main result is as follows.

\begin{theorem}\label{thm:X00random}
\hspace{0em}
\ben
\item\label{it:rmone} If $s_-=s_+=0$, then any diagonal density matrix $\rho>0$ satisfies Assumption~\assref{(A)} and the unraveling  of $(\cJ,\rho)$ is the  Bernoulli measure on $(\lbr1,K\rbr\times\{{++},{--}\})^{\nn^\ast}$  generated by the mass function $Q(k{\pm\pm})=w_k\e^{\mp\beta_k\epsilon/2}/2\cosh(\beta_k \epsilon/2)$.
\een

\medskip\noindent
In the following, we assume that $s_-+ s_+ >0$.

\medskip
\ben\setcounter{enumi}{1}
\item\label{it:rmtwo} The map $\Phi$ is irreducible and the unique density matrix $\rho$ such that $(\cJ,\rho)$ satisfies Assumption~\assref{(A)} is given by~\eqref{str-3} and~\eqref{str-2} with 
\beq{equ:X00eta2}
\eta\defeq\frac12\sum_{k=1}^Kw_k\tanh(\beta_k\epsilon/2).
\eeq
\item\label{it:rmthree} The unraveling $\PP$ of the instrument $(\cJ,\rho)$ is a PMP measure. 
\item \label{it:rmfour}
\begin{align*}
\ep(\cJ,\rho)&=\frac{s_+^2+s_-^2}{s_++s_-}\sum_{k,l=1}^Kw_kw_l\frac{((\beta_k-\beta_l)\epsilon/2)\sinh((\beta_k-\beta_l)\epsilon/2)}{\cosh((\beta_k-\beta_l)\epsilon/2)+\cosh((\beta_k+\beta_l)\epsilon/2)}\\
&\quad +\frac{2s_+s_-}{s_++s_-}\sum_{k,l=1}^Kw_kw_l\frac{((\beta_k+\beta_l)\epsilon/2)\sinh((\beta_k+\beta_l)\epsilon/2)}{\cosh((\beta_k-\beta_l)\epsilon/2)+\cosh((\beta_k+\beta_l)\epsilon/2)}.
\end{align*}
In particular $\ep(\cJ,\rho)>0$ if $s_+s_->0$ or if there is a pair of indices such that $w_kw_l>0$ and $\beta_k\not=\beta_l$.
\item\label{it:rmfive} The entropic pressure is given by
\[
e(\alpha)=\log\left(1+\frac{s_++s_-}2\left(\sqrt{1-\Delta(\alpha)}-1\right)\right),
\]
where
\begin{align*}
\Delta(\alpha)\defeq\sum_{k,l=1}^Kw_kw_l
\biggl(
&\frac{s_+^2+s_-^2}{(s_++s_-)^2}\frac{\sinh(\alpha(\beta_k-\beta_l)\epsilon/2)\sinh((1-\alpha)(\beta_k-\beta_l)\epsilon/2)}{\cosh(\beta_k\epsilon/2)\cosh(\beta_l\epsilon/2)}\\
+&\frac{2s_+s_-}{(s_++s_-)^2}\frac{\sinh(\alpha(\beta_k+\beta_l)\epsilon/2)\sinh((1-\alpha)(\beta_k+\beta_l)\epsilon/2)}{\cosh(\beta_k\epsilon/2)\cosh(\beta_l\epsilon/2)}
\biggr)
\end{align*}
is an entire analytic function such that $\Delta(\alpha)<1$ for all $\alpha\in\rr$.
\item\label{it:rmsix} If $s_+=s_-=1/2$, then $\PP$ is Bernoulli. In the opposite cases  $\PP$ is not weak Gibbs for any $\beta_1,\ldots,\beta_K$.
\een
\end{theorem}
For additional information see Section~\ref{sec-xx-rttm}.


\subsection{The hidden Markov model perspective}
\label{sec-hmcp}
 We show here that the class of PMP measures introduced in Section~\ref{sec-PMP} coincides with the standard class of {\sl Hidden Markov models.} For completeness, we start by introducing a third (equivalent) class of measures: {\sl Function Markov measures.}
 
We denote by $\cL$ an auxiliary finite alphabet, and by $\Xi\defeq\cL^{\nn^\ast}$ the associated path space, equipped with a stationary Markov measure $\QQ\in\cP_\phi(\Xi)$.  Let $f\colon\cL\to\cA$ be onto and define the map
$F\colon\Xi\to\Omega$ by $F\colon(\xi_t)_{t\in\nn^\ast}\mapsto(f(\xi_t))_{t\in\nn^\ast}$.

\begin{definition}\label{def:FM}
The measure $\PP\defeq\QQ\circ F^{-1}\in\cP_\phi(\Omega)$ is the Function Markov (FM) measure generated by the pair $(\QQ,f)$.
\end{definition}
Obviously, the same $\PP$ may be generated by many distinct pairs $(\QQ,f)$.
\begin{remark}\label{rem:gibbsFM}
 If $\supp\QQ=\Xi$, then $\PP$ is Gibbs for a H\"older continuous potential (see, e.g., \cite[Sections~2--3]{Verbitskiy2011}). However, necessary and sufficient conditions on the pair $(\QQ,f)$ for the induced FM measure $\PP$ to be Gibbs or weak Gibbs are unknown.
\end{remark}

It is sometimes convenient to define  $\PP$ as the law of the process  $(f(\xi_t,\ldots,\xi_{t+k-1}))_{t\in\nn^\ast}$, where $f\colon\cL^k\to\cA$ is an onto map. This case reduces to Definition~\ref{def:FM} by considering the stationary Markov Chain ${\cal X}_t=(\xi_t,\ldots,\xi_{t+k-1})$ with state space $\cL^k$.  

Let $R\eqdef[R_{la}]_{l\in\cL,a\in\cA}$ be a right-stochastic matrix. For $(\omega_1,\ldots,\omega_T)\in\cA^T$, set
\[
\PP_T(\omega_1,\ldots,\omega_T)\defeq\sum_{(\xi_1,\ldots,\xi_T)\in\cL^T}
\QQ([\xi_1\cdots\xi_T])\prod_{t=1}^T R_{\xi_t\omega_t}.
\]
$\PP_T$ induces a probability measure on $\cA^T$ and there is a unique probability measure $\PP\in\cP_\phi(\Omega)$
such that $\PP([\omega_1\cdots\omega_T])=\PP_T(\omega_1, \dots, \omega_T)$ for all $T\in\nn^\ast$ and $(\omega_1,\ldots,\omega_T)\in\Omega_T$.

\begin{definition}\label{def:HM}
The measure $\PP$ is the Hidden Markov (HM) measure generated by the pair $(\QQ, R)$.
\end{definition}

Of course, here again, different pairs $(\QQ,R)$ may generate the same $\PP$. It follows from the proof of the next proposition and Remark~\ref{rem:gibbsFM} that $\PP$ is Gibbs with a H\"older continuous potential provided $\supp\QQ=\Xi$ and the entries of the matrix  $R$ are strictly positive. Sufficient and necessary conditions for $\PP$ to be Gibbs or weak Gibbs are not known; see 
 \cite{berghout_regularity_2021} for some recent results in this direction.

The following result is basic.
\begin{proposition}\label{fm-hm}
 Let $\PP\in\cP_\phi(\Omega)$. The following statements are equivalent.
\ben
\item $\PP$ is an FM measure.
\item $\PP$ is an HM measure.
\item $\PP$ is a PMP measure. 
\een
\end{proposition}
\proof 
(i) $\Rightarrow$ (ii). Let the FM measure $\PP\in\cP_\phi(\Omega)$ be generated by the pair $(\QQ,f)$, and set $R_{la}\defeq\delta_{f(l),a}$. The matrix $R\defeq[R_{la}]_{(l,a)\in\cL\times\cA}$ is right-stochastic and $\PP$ is the HM measure generated by $(\QQ,R)$.

\medskip\noindent
(ii) $\Rightarrow$ (i). Suppose that the HM measure $\PP\in\cP_\phi(\Omega)$ is generated by the pair $(\QQ, R)$, the Markov measure $\QQ\in\cP_\phi(\Xi)$ being itself generated by $(Q,{\bf q})$. Setting $\cX\defeq\cL\times\cA$, one easily checks that the pair $(P,{\bf p})$, where ${\bf p}\defeq[q_{l}R_{la}]_{(l,a)\in\cX}$ and $P\defeq[q_{ll'}R_{l'a'}]_{(l,a),(l',a')\in\cX}$, generates a Markov measure $\mathbb{L}\in\cP_\phi(\cX^{\nn^\ast})$.
Defining $f\colon\cX\to\cA$ by $f(l,a)\defeq a$, one concludes that $\PP$ is the FM measure generated by the pair $(\mathbb{L},f)$.

\medskip\noindent
(ii) $\Rightarrow$ (iii).  Let $\PP\in\cP_\phi(\Omega)$ be the HM measure generated by $(\QQ, R)$, with the Markov measure $\QQ\in\cP_\phi(\Xi)$ generated by $(Q,{\bf q})$. For $a\in\cA$, define the matrix $M_a\defeq[q_{ll'}R_{l'a}]_{l,l'\in\cL}$. Then, $\PP$ is the PMP measure generated by $((M_a)_{a\in\cA}, {\bf q})$. 

\medskip\noindent
(iii) $\Rightarrow$ (ii). Suppose that $\PP\in\cP_\phi(\Omega)$ is the PMP measure generated  by $((M_a)_{a\in\cA}, {\bf p})$, where $M_a\eqdef[m_{ij}(a)]_{i,j\in\lbr1,d\rbr}$ and ${\bf p}\eqdef[p_i]_{i\in\lbr1,d\rbr}$.  Let $\cL\defeq\lbr1,d\,\rbr\times\cA$ and consider the Markov measure $\QQ\in\cP_\phi(\Xi)$ generated by $({\bf q},Q)$ with
\[
q_{(i,a)}\defeq\sum_{h=1}^dp_hm_{hi}(a),\qquad q_{(i,a)(j,b)}\defeq m_{ij}(b).
\]
Then $\PP$ is the HM measure generated by  $(\QQ, R)$, where $R_{(i,b)a}\defeq\delta_{ab}$.\qed

The equivalence (i) $\Leftrightarrow$ (ii) goes back to the seminal work~\cite{Petrie1} where the FM/HM measures were introduced. Following~\cite{Petrie1}, the subject developed rapidly with applications extending to ecology, automatic speech recognition, communications and information theory, econometrics, biology, to mention  some of them. The review article~\cite{Ephraim1} is an excellent introduction to the subject from the statistical/information theoretic perspective. For the dynamical system perspective, see the collection of papers in~\cite{Marcus_petersen_weissman_2011}. 

Although mathematically elementary, we are not aware of the equivalence (ii) $\Leftrightarrow$ (iii) appearing previously in the literature. The likely explanation for this  is that the particular  probabilistic intuition behind the FM/HM construction  is invisible in the PMP picture.\footnote{We mention here that the link between HM and matrix products was used in~\cite{Jacquet1} for the study of the Kolmogorov--Sinai entropy.}
On the other hand, from the perspective of our work, it is precisely the PMP representation that plays the central role. The reason for this is the particular quantum mechanical interpretation of  the PMP representation  which in turn is invisible in the FM/HM picture.  From the foundational perspective, this interpretation complements the probabilistic FM/HM construction. We will now argue that it is also useful from the technical perspective.

In the proof of the implication (ii) $\Rightarrow$ (iii) one associates to the generating pair $(\QQ, R)$ 
of the HM measure  $\PP$ the  PMP generating pair $((M_a)_{a\in\cA}, {\bf p})$ with the simple identification 
\[
m_{ll'}(a)=\QQ(\xi_2=l'\mid\xi_1=l)R_{l'a}.
\]
The subadditive thermodynamic formalism of~\cite{BJPP1} and the Lanford--Ruelle 
LDP theory of~\cite{CJPS-2017} then lead to information about the statistics of $\PP$ that is of independent interest and might be difficult to prove with other approaches. For example, in~\cite{CJPS-2017} it is proven that if the matrix $M$ is irreducible, then the Level III LDP holds for $\PP$.
The results of~\cite{BJPP1} give a very general criterion  for validity of binary hypothesis testing for a pair $(\PP, \wP)$ of HM measures, as described at the end of Section~\ref{sec-setup}.  We hope that these  results will  be of further theoretical use and will find applications in  the fields where the HM/FM measure modeling plays an important role.

Turning to the examples, the FM picture of the Keep--Switch PMP measure $\PP$ introduced in Section~\ref{sec-int-ks} arises as follows. Consider the stationary Markov chain $(\xi_t)_{t\in\nn^\ast}$ on the state space $\cL\defeq\{+,-\}$ generated by the pair $(P,{\bf p})$ with
\beq{eq:MatrixM}
P\defeq\begin{bmatrix}q_1 &r_1\\r_2&q_2\end{bmatrix}, \qquad 
{\bf p}\defeq(r_1+r_2)^{-1}\begin{bmatrix}r_2&r_1\end{bmatrix}.
\eeq
Then, in the FM picture,  $\PP$ is the law of $(f(\xi_t,\xi_{t+1}))_{t\in\nn^\ast}$, where 
\beq{eq:deffKSpm}
f(\xi,\xi')\defeq\begin{cases}
K&\text{if }\xi=\xi';\\
S&\text{otherwise},
\end{cases}
\eeq
see Section~\ref{sec:hiddenMarkovKS}. In the particular case $q_1=1-q_2$, $q_1\not=1/2$, 
the FM measure $\PP$ is sometimes called  the Blackwell--Furstenberg--Walters--van den Berg (BFWB) measure and has appeared independently in probability theory, dynamical systems, and statistical physics~\cite{Blackwell-1957,Walters-1986,Verbitskiy-2016}.
The interest in this example is the non-Gibbsian character of $\PP$, which was typically examined 
on the Dobrushin--Lanford--Ruelle level: the limsup of conditional probabilities 
\[
{\cal C}(\omega)\defeq\limsup_{T\rightarrow \infty} \frac{\PP_T(\omega_1, \dots, \omega_T)}{\PP_{T-1}(\omega_2,\dots, \omega_T)}
\]
is discontinuous {\em everywhere} on $\Omega$, see~\cite{LOMAVE-1998,Verbitskiy2011}. Due to this fact, in the literature the BFWB measure is a canonical  toy example illustrating  the possible non-Gibbsian  character of the HM/FM measures. To the best of our knowledge, the case $q_1\not=1-q_2$ has not been considered previously. 

Theorem~\ref{mainthm-KS}  can be viewed as an extension of the BFWB example  adapted to the topics studied in this paper. Here it is also important to note that the  proof of Theorem~\ref{mainthm-KS}, given in Section~\ref{sec:flip--keep}, is technically centered 
around the FM representation of $\PP$. But we also emphasize that  Parts~\ref{part:X00-KSp}--\ref{part:X00-KSm} of Theorem~\ref{th:xx-M}\footnote{In this context, see also Remark~\ref{rem:rtp}.} shed a different light on the 
Theorem~\ref{mainthm-KS} and in particular the BFWB example. Firstly, the probe measurements leading to $\PP$ are realized in a physically natural  way,  singling out its theoretical  and experimental relevance. Secondly, the generalization 
to $q_1\not=1-q_2$ that goes beyond the BFWB  example is not only of mathematical interest, but is also motivated by the fact that the parameters of the X00-spin instrument would be unnaturally restricted if the values of $q_1$ and $q_2$ were not independent of each other.

In view of other possible applications and in the spirit of the discussion at the end of Section~\ref{sec-setup}, it is of interest to consider the  case 
where $\PP$ and $\wP$ are two Keep--Switch PMP measures with unrelated parameters (\ie no time reversal or involution $\theta$ is involved here). More precisely, let $\PP$ and $\wP$ be the Keep--Switch PMP measures respectively generated by the matrices and probability vectors
\begin{align*}
M_K\defeq\begin{bmatrix}q_1 &0\\0&q_2\end{bmatrix}, \qquad 
M_S\defeq\begin{bmatrix}0 &r_1\\r_2&0\end{bmatrix}, \qquad 
{\bf p}\defeq(r_1+r_2)^{-1}\begin{bmatrix}r_2&r_1\end{bmatrix},\\
\widehat M_K\defeq\begin{bmatrix} \widehat q_1 &0\\0&\widehat q_2\end{bmatrix}, \qquad 
\widehat M_S\defeq\begin{bmatrix}0 &\widehat r_1\\ \widehat r_2&0\end{bmatrix}, \qquad 
\widehat{\bf p}\defeq(\widehat r_1+\widehat r_2)^{-1}\begin{bmatrix}\widehat r_2&\widehat r_1\end{bmatrix},
\end{align*}
where $0 < q_1, q_2, \widehat q_1, \widehat q_2 < 1$ and $r_i\defeq1-q_i$, $\widehat r_i\defeq1-\widehat q_i$, $i=1,2$. The canonically associated instruments satisfy Assumptions~\assref{(A)}, \assref{(B)} and~\assref{(C)}, and the generalization of Theorem~\ref{thm-int} formulated at the end of Section~\ref{sec-setup} applies to the pair $(\PP, \wP)$. We complement it with the following result, in which the numbers
\begin{align}
\gamma&\defeq\frac 12\log\frac {q_1}{q_2},\quad&\quad
\chi&\defeq\frac{1}{2}\left(\left|\log\frac{q_1}{q_2}\right|
-\left|\log\frac{\widehat q_1}{\widehat q_2}\right|\right),\label{eq:xichigamma}\\ 
\eta&\defeq\frac 12\log \frac{q_1q_2\widehat r_1\widehat r_2}{r_1r_2\widehat q_1 \widehat q_2},
\quad&\quad
\delta&\defeq\frac 12\log\frac{r_1r_2}{\widehat r_1 \widehat r_2},\label{eq:etadelta}
\end{align}
will play a particular role. Unlike in Theorem~\ref{mainthm-KS}, we do not assume here that $\gamma\neq0$.

\medskip 
\begin{theorem}\label{KS-new}
\hspace{0em}
\ben
\item\label{point:iieppairks} We have
\[
\ep(\PP, \wP) =\delta+\frac {r_1-2r_1r_2+r_2}{r_1+r_2}\eta+\frac {|r_1-r_2|}{r_1+r_2}\chi.
\]
\item\label{point:part2ep0} The following four statements are equivalent: (a) $\ep(\PP, \wP)=0$, (b) $\PP =\wP$, (c) $(q_1,q_2)=(\widehat q_1,\widehat q_2)$ or  $(q_1,q_2)=(\widehat q_2,\widehat q_1)$, (d) $\eta = \chi = 0$.
\een

\medskip\noindent
We assume that $\ep(\PP, \wP)>0$ in the remaining statements.

\medskip
\ben\setcounter{enumi}{2}
\item\label{point:iiipartanalytic}  The  function $\alpha\mapsto e(\alpha)$ is differentiable and strictly convex. If $\chi=0$, then 
$e$ is real analytic. If $\chi\not=0$, then $e$ is analytic on 
$\rr\setminus\{|\gamma|/\chi\}$, but not twice differentiable at $|\gamma|/\chi$.
\item\label{point:pairKSCLT1} If $\chi=0$ or $\gamma\not=0$, then the  random variables  $(\sigma_T)_{T\in\nn^\ast}$ satisfy the central limit theorem\footnote{Here and below, $\Rightarrow $ denotes convergence in law.}
\[
\frac{{\sigma_T-T\ep(\PP, \wP)}}{{\sqrt{T}}}\,\,\Rightarrow \ Z,
\]
where $Z$ is a centered normal random variable of variance 
\[
{\rm Var}(Z)=\frac{4r_1r_2}{(r_1+r_2)^3}\left((q_1r_2^2+q_2r_1^2)\eta^2+   2 |r_2-r_1|\eta\chi + (q_1+q_2)\chi^2 \right)>0.
\]  
\item\label{point:pairKSCLT2} If $\chi \not=0$ and $\gamma=0$, then the random variables  $(\sigma_T)_{T\in\nn^\ast}$ satisfy the non-Gaussian central limit theorem
\[
\frac{{\sigma_T-T\ep(\PP, \wP)}}{{\sqrt{T}}}\,\,\Rightarrow \, Z_1 - |Z_2|,
\]
where $(Z_1,Z_2)$ is a pair of independent, centered normal random variables with 
\begin{gather*}
{\rm Var}(Z_1)=q_1r_1\eta^2,\qquad
{\rm Var}(Z_2)=  \frac{q_1}{r_1}\chi^2>0.
\end{gather*}
\een
\end{theorem}

\begin{remark} Unlike in Theorem~\ref{KS-new}, the measure $\wP$ discussed in context of Theorem~\ref{mainthm-KS}  {\em is not}  a Keep--Switch PMP measure, and the two results are not  directly related. However, on the technical level, the proof of Theorem~\ref{KS-new} is only a slight 
modification of the proof of Theorem~\ref{mainthm-KS}. We will sketch it  in Section~\ref{sec-KS-new}.
\end{remark}
\begin{remark} The proof gives an explicit expression for $e(\alpha)$; see  \eqref{eq:formuleealphapairks}.

\end{remark}
 
\begin{remark}
Note that $\gamma=0$ iff $q_2=q_1$, that is, iff $\PP$ is a Bernoulli measure. Moreover, $\chi=0$ 
iff $q_1/q_2=\widehat q_1/\widehat q_2$ or $q_1/q_2=\widehat q_2/\widehat q_1$. Thus, the non-standard central limit theorem 
stated in Part~\ref{point:pairKSCLT2} holds iff $\PP$ is Bernoulli and $\wP$ is not. Note also that, unsurprisingly, this coincides with the situation where $e$ is not twice differentiable at $\alpha = 0$.
\end{remark}

\begin{remark} In the case $\widehat q_1=\widehat q_2=1/2$, $\wP$ is the symmetric Bernoulli measure on $\Omega$ (recall Remark~\ref{rem:KSber}), and Theorem~\ref{KS-new} yields information about the specific 
entropy and the specific R\'enyi entropy of the Keep--Switch PMP measure $\PP$. More precisely,
\[
S(\PP)\defeq\lim_{T\rightarrow \infty}\frac{1}{T}S(\PP_T)=\frac{r_1S_2+r_2S_1}{r_1+r_2},
\]
where $S_i\defeq-q_i\log q_i-r_i\log r_i$ is the entropy of the measure on $\cA$ generated by $Q_i(K)=q_i$.
The limit defining the specific R\'enyi entropy (recall~\eqref{future-stat})
\[
r(\alpha)\defeq\lim_{T\rightarrow \infty} \frac{1}{T}S_\alpha (\PP) = \lim_{T\rightarrow \infty} \frac{1}{T}\log\left[\sum_{\omega\in\supp\,\PP_T} (\PP_T(\omega))^\alpha\right] = e(1 - \alpha)+(1-\alpha)\log 2
\]
exists and the function $\alpha \mapsto r(\alpha)$ is differentiable and strictly convex. If 
$q_1=q_2$, that is, if $\PP$ is Bernoulli, the function $r$ is real analytic, otherwise $r$ is real analytic on 
$\rr\setminus\{0\}$ but not twice differentiable at $0$. Let $S_T(\omega)\defeq-\log \PP_T(\omega)$. 
The Shannon--McMillan--Breiman $\PP$-a.s.\;convergence (recall that $\PP$ is $\phi$-ergodic),
\[
\lim_{T\rightarrow \infty}\frac{S_T(\omega)}{T}=h_\phi(\PP),
\]
is accompanied by the central limit theorem
\[
\frac{{S_T-Th_\phi(\PP)}}{{\sqrt{T}}}\,\,\Rightarrow  Z,
\]
where $Z$ is centered normal random variable with variance
\begin{align*}
{\rm Var}(Z)= \frac{4r_1r_2}{(r_1+r_2)^3}
\left((q_1+q_2)\gamma^2+2(r_2-r_1)\gamma\eta+(q_1r_2^2+q_2r_1^2)\eta^2\right),
\end{align*}  
and the LDP which holds with the rate function $I(s)\defeq\sup_{\alpha\in \rr}(s\alpha- r(-\alpha))$. 
\end{remark}

\begin{remark} There is abundant literature on entropies of hidden Markov models, part of which is devoted to the study of 
concrete examples; see~\cite{Marcus_petersen_weissman_2011}.  In this context Theorem~\ref{KS-new} gives detailed information for a class of examples which, to the best of our knowledge, have not been studied previously. 
\end{remark}

The unravelings of spin instruments also give novel classes of FM/HM measures that are not weak Gibbs; see Remark 2 after Theorem 4.4 in \cite{vanenter1993} for a related discussion. We will not go into details here since we shall return to the FM/HM/PMP connection in the continuation of this work~\cite{Dima1} where many additional examples are discussed.


\subsection{Rotational instruments}
\label{sec-rotinst}
Let $\cH\defeq\cc^2$ and  $\cA\defeq\{0,1,2,3\}$. Let, moreover, $\Delta\in [0, 2[$ be a parameter. We define the rotational instrument $(\cJ,\rho)$ by 
$\cJ\defeq(\Phi_a)_{a\in\cA}$ and $\rho\defeq\frac{1}{2}\one$, where
\begin{align*}
\Phi_0[X]&\defeq\frac13R_{\Delta}XR_{-\Delta},&
\Phi_1[X]&\defeq\frac1{12}VXV^\ast,\\[4pt]
\Phi_2[X]&\defeq\frac16\tr(X)\one-\Phi_1[X],&
\Phi_3[X]&\defeq\frac16\tr(X)\one,
\end{align*}
with
\[
R_\Delta\defeq\begin{bmatrix}
\cos (\pi \Delta) & -\sin (\pi\Delta)\\[4pt]
\sin(\pi \Delta) &\cos(\pi  \Delta )
\end{bmatrix},\qquad
V\defeq\begin{bmatrix}
0&0\\[4pt]
1&0\end{bmatrix}.
\]
 We note that $R_\Delta$ is a rotation matrix and that $R_\Delta^\ast= R_{-\Delta}$. Since $\Phi_3$ is positivity improving, so is $\Phi\defeq\sum_{a\in \cA}\Phi_a$. In particular, $\Phi$ is irreducible.
The involution $\theta$ is defined by 
\[
\theta(0)\defeq 2, \quad
\theta(1)\defeq 1, \quad
\theta(2)\defeq 0, \quad 
\theta(3)\defeq 3.
\]

Denote by ${\cal I}$ the set of irrational numbers in $[0,2[$. In Section~\ref{sec-ex-ro}, we will prove
\begin{theorem}\label{rot-theorem}
\hspace{0em}
\ben 
\item\label{prt:rot-ass} For  all $\Delta\in{\cal I}$, Assumptions~\assref{(A)}, \assref{(B)}, and~\assref{(C)} hold.
\item\label{prt:rot-reg} For a set of $\Delta$'s in ${\cal I}$  of full measure, we have $e(\alpha)<\infty$ for all $\alpha\in\rr$.
\item\label{prt:rot-nost} For a dense set of $\Delta$'s in ${\cal I}$, we have  $e(\alpha)=+\infty$ for all $\alpha \notin [0,1]$, and 
\[
(\partial^{-}e)(1)=-(\partial^+e)(0)=\ep(\cJ,\rho)<\infty.
\]
\item\label{prt:rot-st} For a dense set of $\Delta$'s in $ {\cal I}$, we have  $e(\alpha)=+\infty$ for all $\alpha \not\in [0,1]$ and 
\[
(\partial^-e)(1)=-(\partial^+e)(0)=\ep(\cJ,\rho)=\infty.
\]
\een
\end{theorem}

\begin{remark} By Theorem~\ref{thm-int}~\ref{it:differentiability_e_alpha}, $e$ is finite and differentiable on $]0,1[$. In Part~\ref{prt:rot-reg} the function $e$ is finite on 
$\rr$, but its differentiability properties outside $]0,1[$ are not known.
\end{remark}

\begin{remark} Obviously, in Parts~\ref{prt:rot-nost} and~\ref{prt:rot-st} the unraveling $\PP$ is not PMP\footnote{For PMP unravelings $e(\alpha) <\infty$ for all $\alpha\in\rr$.}  and is not  weak Gibbs. The surprising aspect  of these two cases is the extent of this failure, captured by the singularities of  $e$ at $\alpha=0,1$. 
\end{remark}

\begin{remark} We conjecture that for all  $\Delta$ in ${\cal I}$ the unraveling $\PP$ is not  PMP and not weak Gibbs. We will return 
to this point in~\cite{Dima1}.
\end{remark}

\begin{remark} With our choice of  $\Phi_0$ and $\Phi_1$ one easily computes that
\[
\PP([10^T1])=(288)^{-1}3^{-T}\sin^2(T\pi\Delta).
\] 
The quantity $\sin^2(T\pi \Delta)$ plays the central role in the proof of Theorem~\ref{rot-theorem} and in the intuition behind 
the result.  Depending on the number-theoretic properties of $\Delta$,  $\sin^2(T\pi \Delta)$ can be extremely small yet non-zero for some large values of $T$. The maps $\Phi_2$ and $\Phi_3$, on the other hand, play a secondary role and are  chosen  for easy verification of 
Assumptions~\assref{(A)}, \assref{(B)}, and~\assref{(C)}.
\end{remark}

\begin{remark} As with the Keep--Switch instrument, the specific choice of $\theta$ here is crucial, as pure time-reversal does not produce the above singularities. To see this, consider the involution
\[
\Upsilon[X]\defeq UXU^\ast,\qquad U\defeq\begin{bmatrix}0&1\\1&0\end{bmatrix},
\]
and observe that, for any $a\in\cA$, $\Upsilon\circ\Phi_a=\Phi^\ast_a\circ\Upsilon$. Since $\Upsilon[\one]=\one$ and $\tr\circ\Upsilon=\tr$, one has
\begin{align*}
\PP([\omega_1\cdots\omega_T])
&=\frac12\tr\left(\Upsilon\circ\Phi_{\omega_1}\circ\cdots\circ\Phi_{\omega_T}[\one]\right)
=\frac12\tr\left(\Phi^\ast_{\omega_1}\circ\cdots\circ\Phi^\ast_{\omega_T}\circ\Upsilon[\one]\right)\\
&=\frac12\tr\left(\Phi_{\omega_T}\circ\cdots\circ\Phi_{\omega_1}[\one]\right)
=\PP([\omega_T\cdots\omega_1]),
\end{align*}
for any $T\in\nn^\ast$ and $\omega\in\Omega_T$.
\end{remark}

\label{sec-proofs}
\section{Keep--Switch instruments}\label{sec:flip--keep}
Sections~\ref{sec-KS-i}--\ref{sec:FDR} are devoted to the proof of Theorem~\ref{mainthm-KS}. Theorem~\ref{KS-new} is proved in Section~\ref{sec-KS-new}. We shall freely use some standard  definitions and results of Large Deviation theory. The classical references for those are~\cite{DemboZeitouni,EllisBook}.

In the proof of Theorem~\ref{mainthm-KS}, recalling Remark~\ref{rem:KSexchange} and our standing assumption $q_1\not=q_2$, we shall also assume that
\beq{equ:q1big}
q_1>q_2.
\eeq

\subsection{Proof of Theorem~\ref{mainthm-KS},  Part~\ref{KS-ii}}
\label{sec-KS-i}

We have to show that the PMP measure $\PP$ generated by $((M_K,M_S),{\bf p})$, as defined in~\eqref{ksdef}, is not weak Gibbs. For $T\in\nn^\ast$, a simple computation gives 
\[
\frac{\PP_{2T+1}({K^T}{S}{K^T})}{\PP_{T+1}({K^T}{S})\PP_T(K^T)}=2(r_1+r_2)\left((1+r_1)\left(\frac{q_1}{q_2}\right)^T+(1+r_2)\left(\frac{q_2}{q_1}\right)^T\right)^{-1}.
\]
Hence, 
\[
\lim_{T\to\infty}\frac1{2T+1}\log\left(\frac{\PP_{2T+1}({K^T}{S}{K^T})}{\PP_{T+1}({K^T}{S})\PP_T(K^T)}\right)=
\frac12\log\frac{q_2}{q_1}<0,
\]
and the statement follows from Theorem~\ref{noe-thm}~\ref{it:noetwo}.~

\subsection{The FM representation}\label{sec:hiddenMarkovKS}

We mentioned in Section~\ref{sec-hmcp} that the FM representation suggested by Proposition~\ref{fm-hm} turns out to be an efficient technical tool in the study of PMP or HM measures. We shall exemplify this statement in the proof of the remaining parts of Theorem~\ref{mainthm-KS}.

Let $\QQ$ be the Markov measure on $\Xi=\{+,-\}^{\nn^\ast}$ generated by the pair $(P,{\bf p})$ given by~\eqref{eq:MatrixM}.
The reader should  keep in mind that $+$ stands for $+1$ and $-$ for $-1$, and recall that according to the notation introduced in Section~\ref{sec-setup}, $\Xi_T=\{+,-\}^T$ for every integer $T$. For $a\in\{+,-\}$, we define $\overline{a}=-a$ and we extend this map to arbitrary (finite or infinite) sequences of elements of $\{+,-\}$ in the obvious way.

By the definition of $\QQ$ (recall~\eqref{eq:MarkovM}), for any $T\in\nn^\ast$ and $\xi\in\Xi_{T+1}$, we have
\beq{eq:Qbnupinu}
\QQ([\xi])=p_{\xi_1} p_{\xi_1\xi_2}p_{\xi_2\xi_3}\cdots p_{\xi_T \xi_{T+1}}.
\eeq
Denoting by $n_{ab}(\xi)$ the number of transitions from $a$ to $b$ in $\xi$, \ie the number of indices $t\in \lbr1,T\rbr$ such that $(\xi_t, \xi_{t+1}) = (a,b)$,  we can rewrite
\beq{eq:QNUnij}
\QQ([\xi])=p_{\xi_1} q_1^{n_{{++}}(\xi)}q_2^{n_{{--}}(\xi)}r_1^{n_{{+-}}(\xi)}r_2^{n_{{-+}}(\xi)}.
\eeq
Let $f:\Xi_2 \to \cA$ be as in \eqref{eq:deffKSpm}. We let then $F\colon\Xi_{T+1} \to \Omega_T$ be given by
\[
F(\xi)\defeq(f(\xi_1, \xi_2), f(\xi_2, \xi_3), \dots, f(\xi_T, \xi_{T+1})),
\]
and with a slight abuse of notation, we also denote by $F$ its obvious extension to infinite sequences. These maps are 2-to-1: It is immediate that $F(\overline\xi)=F(\xi)$ for any $\xi$. Moreover, for any finite or infinite sequence $\omega$, one has $F^{-1}(\{\omega\})=\{\xi,\overline{\xi}\}$ with
\[
\xi_1\defeq+,\qquad\xi_t\defeq(-1)^{N_t(\omega)}, \qquad N_t(\omega)\defeq|\{s\in \lbr 1, t-1\rbr\mid \omega_s = S \}|,\qquad  t>1.
\]
\begin{proposition}
For all $T\in\nn^\ast$ and all $\omega\in\Omega_T$,
\beq{eq:PomegasumQnu}
\PP([\omega])=\QQ([\xi])+\QQ([\overline\xi])
=p_{\xi_1} q_1^{n_{++}}q_2^{n_{--}}r_1^{n_{+-}}r_2^{n_{-+}}
+p_{\overline\xi_1}q_1^{n_{--}}q_2^{n_{++}}r_1^{n_{-+}}r_2^{n_{+-}},
\eeq
where $\xi\in F^{-1}(\{\omega\})$ and  $n_{ab}=n_{ab}(\xi)$. It follows that
\[
\PP = \QQ\circ F^{-1}.
\]
\end{proposition}

\proof We denote here the canonical basis of $\rr^2$ by $(e_+, e_-)$ and by $[p_{ab}]_{a,b\in\{+,-\}}$ the entries of the matrix $P$ defined in~\eqref{eq:MatrixM}. One verifies that $e_a^{\mathsf{T}} M_{f(a,b)} = p_{ab}e_{b}^{\mathsf{T}}$ for $a,b\in\{+,-\}$, and by iterating we obtain that, for any $\omega,\xi$ as in the statement,
\begin{align*}
\PP([\omega])&={\bf p}M_{\omega_1}\cdots M_{\omega_T}{\bf 1}
=\left(p_{\xi_1}e_{\xi_1}^{\mathsf{T}}+p_{\overline\xi_1}e_{\overline\xi_1}^{\mathsf{T}}\right) M_{\omega_1}\cdots M_{\omega_T}{\bf 1}\\
&=p_{\xi_1}e_{\xi_1}^{\mathsf{T}}M_{f(\xi_1,\xi_2)}\cdots M_{f(\xi_T,\xi_{T+1})}{\bf 1}
+p_{\overline\xi_1}e_{\overline\xi_1}^{\mathsf{T}}M_{f(\overline\xi_1,\overline\xi_2)}\cdots M_{f(\overline\xi_T,\overline\xi_{T+1})}{\bf 1}\\
&=p_{\xi_1} p_{\xi_1\xi_2}p_{\xi_2\xi_3}\cdots p_{\xi_{T}\xi_{T+1}}
+p_{\overline\xi_1} p_{\overline\xi_1\overline\xi_2}p_{\overline\xi_2\overline\xi_3}\cdots p_{\overline\xi_{T}\overline\xi_{T+1}},
\end{align*}
which establishes the first equality in~\eqref{eq:PomegasumQnu}. 
 The second equality follows from~\eqref{eq:QNUnij} and the relation $n_{ab}(\overline\xi)=n_{\overline a\,\overline b}(\xi)$.\qed

As a first application of the obtained FM representation, we prove

\begin{proposition}\label{lem:reversibilityKS}
The measure $\PP$ satisfies $\PP([\omega_1\cdots\omega_T])=\PP([\omega_T\cdots\omega_1])$ for any $T\in\nn^\ast$ and any $(\omega_1,\ldots,\omega_T)\in\Omega_T$.
\end{proposition}
\proof Observing that $(\omega_1,\ldots,\omega_T)=F(\xi_1,\ldots,\xi_{T+1})$ iff
$F(\xi_{T+1},\ldots,\xi_1)=(\omega_T,\ldots,\omega_1)$ and that, as any stationary two-state Markov chain, $\QQ$ is reversible in the sense that $\QQ([\xi_1\cdots\xi_T])=\QQ([\xi_T\cdots\xi_1])$, the statement immediately follows from the first equality in~\eqref{eq:PomegasumQnu}.\qed

\subsection{The FM entropy production}\label{sec:FMEP}
Set 
\beq{equ:hsigdef}
\widehat \sigma_T\defeq\sigma_T \circ F.
\eeq
Obviously, the law of  $\sigma_T$ under $\PP$ is the same as the law of $\widehat \sigma_T$ under $\QQ$. Much of our analysis 
is focused on $\widehat \sigma_T$. In this section we study the statistics of a family $([U_T\ \ V_T\ \ W_T])_{T\in\nn^\ast}$ of $\rr^3$-valued random vectors on $(\Xi,\QQ)$, and we establish the  asymptotic form
\beq{eq:sigmaexprasympt}
\widehat \sigma_T \sim \eta\,U_T+\gamma(|V_T|-|W_T|),
\eeq
where
\beq{eq:defetaxi}
\gamma\defeq\12\log\frac{q_1}{q_2}>0,\qquad\eta\defeq\12\log\frac{q_1q_2}{r_1r_2}.
\eeq
As we shall see, the ``pathologies'' in Theorem ~\ref{mainthm-KS} are the consequences of the absolute values appearing in~\eqref{eq:sigmaexprasympt}.

We start with some computations leading to an  expression of $\widehat \sigma_T$ that allows for the identification of the family $([U_T\ \ V_T\ \ W_T])_{T\in\nn^\ast}$.

 Since $\sigma_T$ depends only on $\omega_1\cdots\omega_T$, $\widehat \sigma_T$ depends only on  $\xi_1\cdots\xi_{T+1}$. For $\xi\in\Xi_{T+1}$, we rewrite~\eqref{eq:PomegasumQnu} as
\beq{eq:PfnuFM}
\PP([F(\xi)]) =2\frac{(q_1q_2)^{(n_{++}+n_{--})/2}(r_1r_2)^{(n_{+-}+n_{-+}+1)/2}}{r_1+r_2}
\cosh\left(\gamma(n_{++}-n_{--})+\delta(\xi) \right),
\eeq
where
\[
\delta(\xi)\defeq\frac12\left((n_{+-}-n_{-+})\log\frac{r_1}{r_2}+\log\frac{p_{\xi_1}}{p_{\overline\xi_1}}\right).
\]
Since $2|n_{-+}-n_{+-}|=|\xi_{T+1}-\xi_1|\leq2$, we have a bound
\beq{equ:deltabound}
|\delta(\xi)|\leq C,
\eeq
for some constant $C$ depending only on $q_1, q_2$.\footnote{Recall that $n_{ab} = n_{ab}(\xi)$.}

It follows from Proposition~\ref{lem:reversibilityKS} that for $\omega\in\Omega$,
\[
\sigma_T(\omega)=\log\frac{\PP([\omega_1\cdots\omega_T])}{\PP([\theta(\omega_1)\cdots\theta(\omega_T)])}.
\]

For $T\in\nn^\ast$, let $\psi\colon\Xi_T\to\Xi_T$ be given by 
\[
\psi(\xi_1\xi_2\xi_3\xi_4\cdots\xi_T)\defeq\overline\xi_1\xi_2\overline\xi_3\xi_4\cdots
\begin{cases}
\overline\xi_T&\text{if }T\text{ is odd;}\\
\xi_T&\text{otherwise.} 
\end{cases}
\]
It is immediate that $\psi(\overline\xi)=\overline{\psi(\xi)}$. Moreover, for $\xi \in \Xi_{T+1}$ and $\omega=F(\xi)\in \Omega_{T}$, we find
\[
F(\psi(\xi))=F(\psi(\overline\xi))= \theta(\omega_1)\theta(\omega_2)\cdots \theta(\omega_{T}),
\]
and hence
\[
\widehat\sigma_T(\xi)=\sigma_T(F( \xi))=\log\frac{\PP_T(F(\xi))}{\PP_T(F(\psi(\xi)))}.
\]
In order to compute $\PP_T(F(\psi(\xi)))$, we need the identifications
\begin{align*}
n_{++}(\psi(\xi))&=n_{-+}^o(\xi)+n_{+-}^e(\xi),& 
n_{--}(\psi(\xi))&=n_{+-}^o(\xi)+n_{-+}^e(\xi),\\
n_{+-}(\psi(\xi))&=n_{--}^o(\xi)+n_{++}^e(\xi),&
n_{-+}(\psi(\xi))&=n_{++}^o(\xi)+n_{--}^e(\xi),
\end{align*}
where $n_{ab}^{o/e}$ is the number of {\em odd/even} integers $t\in \lbr1,T\rbr$ such that $(\xi_t, \xi_{t+1})=(a,b)$. As a consequence,
\begin{align*}
n_{++}(\psi(\xi))+n_{--}(\psi(\xi))&=n_{+-}(\xi)+n_{-+}(\xi),\\
n_{+-}(\psi(\xi))+n_{-+}(\psi(\xi))&=n_{--}(\xi)+n_{++}(\xi).
\end{align*}
It follows from~\eqref{eq:PfnuFM} that, omitting again the argument $\xi$ of $n_{ab}$,
\[
\PP_T(F(\psi(\xi)))=2\frac{(q_1q_2)^{(n_{+-}+n_{-+})/2}(r_1r_2)^{(n_{++}+n_{--}+1)/2}}{r_1+r_2}
\cosh\left(\gamma(\Delta_{+-}-\Delta_{-+})+\delta(\psi(\xi))\right),
\]
with $\Delta_{ab}=n_{ab}^e-n_{ab}^o$.
This leads to the expression
\beq{eq:sigmathatnij}
\widehat\sigma_T(\xi) =\log\left(\left(\frac{q_1q_2}{r_1r_2}\right)^{(n_{++}+n_{--}-n_{+-}-n_{-+})/2}
\frac{\cosh\left(\gamma(n_{++}-n_{--})+\delta(\xi)\right)}
{\cosh\left(\gamma(\Delta_{+-}-\Delta_{-+})+\delta(\psi(\xi)) \right)}\right).
\eeq
We are  now ready to define the  $\rr^3$-valued random vectors 
$$
X_T\defeq[U_T\ \ V_T\ \ W_T]
$$ 
on $(\Xi,\QQ)$ by setting
\begin{align*}
U_T(\xi)&\defeq n_{++}(\xi_{[1,T+1]})+n_{--}(\xi_{[1,T+1]})-n_{+-}(\xi_{[1,T+1]})-n_{-+}(\xi_{[1,T+1]}),\\
V_T(\xi)&\defeq n_{++}(\xi_{[1,T+1]})-n_{--}(\xi_{[1,T+1]}),\\
W_T(\xi)&\defeq\Delta_{+-}(\xi_{[1,T+1]})-\Delta_{-+}(\xi_{[1,T+1]}).
\end{align*}
Recalling~\eqref{eq:defetaxi}, we have 
\[
\widehat\sigma_T=\eta\,U_T+\log\cosh\left(\gamma V_T+\delta(\xi_{[1,T+1]})\right)
-\log\cosh\left(\gamma W_T+\delta(\psi(\xi_{[1,T+1]}))\right).
\]
Using the bound~\eqref{equ:deltabound} and the inequalities $\frac 12 \e^{|x|} \leq  \cosh(x) \leq \e^{|x|}$, we conclude that
\begin{align}\label{eq:hatsigmaeta}
\left|\widehat\sigma_T-(\eta\,U_T+\gamma(|V_T|-|W_T|))\right|\leq C',
\end{align}
for some constant $C'$ depending only on $q_1,q_2$.

\begin{remark} Note that $U_T(\overline\xi)=U_T(\xi)$, $V_T(\overline\xi)=-V_T(\xi)$ and $W_T(\overline\xi)=-W_T(\xi)$. As a consequence, $V_T$ and $W_T$ cannot be expressed as a function of $\omega\in\Omega$, but $U_T$, $|V_T|$ and $|W_T|$ can. 
\end{remark}

We now turn to the study of  the  statistics  of the family $(X_T)_{T\in\nn^\ast}$. To this end,  for $\lambda\in\rr^3$ and $T\in\nn^\ast$, we set\footnote{Here $\lambda\cdot X$ denotes the Euclidean inner product on $\rr^3$.}
\[
q_T(\lambda)\defeq\frac1T\log\EE\left(\e^{\lambda\cdot X_T}\right). 
\]

\begin{proposition}\label{prop:conv_freeenergy_1D_Ising}
\hspace{0em}
\ben
\item\label{it:KSone} For all $\lambda=(\lambda_1,\lambda_2,\lambda_3)\in\rr^3$, 
\beq{eq:defQlambda}
Q(\lambda)\defeq\lim_{T\to\infty}q_T(\lambda) = \log\left(A_-(\lambda)+A_+(\lambda)\right),
\eeq
with
\[
A_\pm(\lambda)\defeq(q_1q_2r_1r_2)^{1/4}\left(
\e^{\pm(2\lambda_1+\eta)}+\e^{2\lambda_1+\eta}\sinh^2(\lambda_2+\gamma)
+\e^{-(2\lambda_1+\eta)}\sinh^2(\lambda_3)\right)^{1/2}.
\]
\item\label{it:KStwo} The function $Q$ is real analytic on $\rr^3$. Its gradient and its Hessian matrix at $\lambda=0$ are given by
\begin{gather}
\nabla Q(0)=\frac1{r_1+r_2}\begin{bmatrix}r_1-4r_1r_2+r_2& r_2-r_1&0\end{bmatrix},\label{eq:gradQ000}\\[4pt]
D^2Q(0)=\frac{4r_1r_2}{(r_1+r_2)^3}\begin{bmatrix}
4(q_1r_2^2+q_2r_1^2)&2(q_1-q_2)&0\\[2pt]
2(q_1-q_2)&q_1+q_2&0\\[2pt]
 0&0& \frac{(r_1+r_2)^2}{q_1+q_2}\end{bmatrix}.\label{eq:Hessian}
\end{gather}
\item\label{prt:ldpXT} The family $(T^{-1}X_T)_{T\in\nn^\ast}$ satisfies the LDP with a good convex rate function $I$, given by the Legendre transform of $Q$. In particular, the following weak law of large numbers holds: for any $\epsilon > 0$, there exist constants $C,\delta >0$ such that for all $T$,
\beq{eq:convergence_mean_1D_Ising}
\QQ\left(\left\{\xi\mid\|T^{-1}X_T(\xi)-\nabla Q(0)\|>\epsilon\right\}\right)\leq C\e^{-\delta T}.
\eeq
\item\label{prt:cltXT} The following central limit theorem holds: as $T\to\infty$, the random vector
\beq{eq:TCL_1D_Ising}
\frac{X_T-T\nabla Q(0)}{\sqrt{T}}
\eeq
converges in distribution to a centered Gaussian random vector in $\rr^3$ whose covariance matrix is given by $D^2Q(0)$.
\een
\end{proposition}

\proof
\ref{it:KSone} Consider the following deformations of the generating matrix $P$
\[
P_o(\lambda)\defeq\begin{bmatrix}\e^{\lambda_1 + \lambda_2}q_1 & \e^{-\lambda_1-\lambda_3}r_1\\
\e^{-\lambda_1+\lambda_3}r_2&\e^{\lambda_1-\lambda_2}q_2\end{bmatrix},
\qquad 
P_e(\lambda)\defeq\begin{bmatrix}\e^{\lambda_1 + \lambda_2}q_1& \e^{-\lambda_1+\lambda_3}r_1\\
\e^{-\lambda_1-\lambda_3}r_2&\e^{\lambda_1-\lambda_2}q_2\end{bmatrix},
\]
and set $P(\lambda)\defeq P_o(\lambda)P_e(\lambda)$. By~\eqref{eq:Qbnupinu} and the definition of $U_T, V_T, W_T$,  we have
\beq{eq:qTpiMmo}
q_T(\lambda)=\frac1T\log\left({\bf p}
P(\lambda)^{\left\lfloor\frac{T}2\right\rfloor}
P_o(\lambda)^{T-2\left\lfloor\frac{T}2\right\rfloor}{\bf1}\right),
\eeq
where $\left\lfloor\frac T 2\right\rfloor$ is the integer part of $\frac T 2$. The eigenvalues of $P(\lambda)$ are
\[
\kappa_\pm(\lambda)=\left(A_+(\lambda)\pm A_-(\lambda)\right)^2,
\]
with $A_\pm(\lambda)$ as in the statement. For any $\lambda\in\rr^3$, the Perron--Frobenius theorem implies that the spectral projection corresponding to the dominant eigenvalue $\kappa_+(\lambda)$ has strictly positive entries. It follows that, as $T\to\infty$ through odd/even integers,
\beq{eq:qTform}
{\bf p}P(\lambda)^{\left\lfloor\frac{T}2\right\rfloor}
P_o(\lambda)^{T-2\left\lfloor\frac{T}2\right\rfloor}{\bf1}
=\kappa_+(\lambda)^{T/2}\left(C_{{o}/{e}}(\lambda)+o(1)\right),
\eeq
where $C_{{o}/{e}}(\lambda)>0$. Thus, by~\eqref{eq:qTpiMmo}, the limit in~\eqref{eq:defQlambda} exists and equals $\log(\kappa_+(\lambda))/2$.

\medskip\noindent
\ref{it:KStwo} The functions $A_\pm$ are clearly real analytic on $\rr^3$. Since 
\[
A_-(\lambda)+A_+(\lambda)\ge 2(q_1q_2r_1r_2)^{1/4},
\]
the same is true of the function $Q$. The remaining statements follow from simple calculations.

\medskip\noindent
\ref{prt:ldpXT} Since  $Q$ is real analytic, the  G\"artner--Ellis theorem applies, and the stated LDP follows.  It is well known that  Part~\ref{it:KStwo}  yields~\eqref{eq:convergence_mean_1D_Ising}; see for example~\cite[Theorem II.6.3]{EllisBook}.

\medskip\noindent
\ref{prt:cltXT} Finally, in order to obtain the CLT, we observe that the remainder on the right-hand side of~\eqref{eq:qTform} is locally uniform in $\lambda\in\cc$. Therefore, there exist a complex neighborhood $0\in U\subset\cc$ and a number $T_0\geq 1$ such that $q_T$ has an analytic continuation on $U$ for each $T\geq T_0$ and
\beq{eq:unifboundTtouT}
\sup_{T\geq T_0}\sup_{\lambda\in U}|q_T(\lambda)|<\infty.
\eeq
It then follows from the version Bryc's theorem~\cite{bryc93} given in~\cite[Theorem~A.8]{JOPP} that
\[
\frac1{\sqrt{T}}\left(X_T-\EE(X_T)\right) 
\]
converges in distribution to a Gaussian random vector as in the statement. To obtain the CLT for~\eqref{eq:TCL_1D_Ising}, it remains to show that, for $T\to\infty$,\footnote{We note that~\eqref{eq:toshowsqrtQe} does not follow from~\eqref{eq:defQlambda} and~\eqref{eq:unifboundTtouT} alone. Some estimate on the speed of convergence is required.}
\beq{eq:toshowsqrtQe}
\frac1{\sqrt{T}}\left(T\nabla Q(0)-\EE(X_T)\right)
=\sqrt T \left(\nabla Q(0)-\nabla q_T(0)\right)\to 0.
\eeq
By~\eqref{eq:unifboundTtouT}, $Q$ admits an analytic continuation on $U$, and there is a neighborhood $0\in U'\subset U$ on which the analytic continuation of $q_T$ converges uniformly to $Q$ (see for example~\cite[Appendix~A.4]{JOPP}). Using again~\eqref{eq:qTform}, we obtain that 
\[
\sup_{T\geq T_0}\sup_{\lambda\in U'}
T|q_T(\lambda)-Q(\lambda)|<\infty,
\]
and then~\eqref{eq:toshowsqrtQe} follows from Cauchy's integral formula.\qed

\subsection{Proof of Theorem~\ref{mainthm-KS}, Parts~\ref{KS-iii} and~\ref{KS-iv}}
\label{sec-KS-iv}

These two parts of Theorem~\ref{mainthm-KS} concern the entropic pressure and the entropy production rate.
In view of~\eqref{equ:hsigdef} and~\eqref{eq:hatsigmaeta} the entropic pressure reads
\beq{eq:ealphautvtwt}
e(\alpha)=\lim_{T\to\infty}\frac1T\log\EE\left(\e^{ -\alpha\left(\eta\,U_T+\gamma(|V_T|-|W_T|)\right)}\right),
\eeq
where the expectation is with respect to the measure $\QQ$.

We first need a lemma to deal with the absolute values in the exponent above (we shall use the lemma once for each absolute value).

\begin{lemma}\label{lem:pressurevalabs}
Let $([X_T\ \ Y_T])_{T\in\nn^\ast}$ be a family of random vectors, with $[X_T\ \ Y_T]\in\rr^d\times\rr$. Assume that $(T^{-1}[X_T\ \ Y_T])_{T\in\nn^\ast}$ satisfies the LDP with some convex rate function $I\colon\rr^d\times\rr\to [0,+\infty]$. Then, $(T^{-1}[X_T\ \ |Y_T|])_{T\in\nn^\ast}$ also satisfies the LDP with the rate function $\overline I$ given by
\beq{eq:contractIvabs}
\overline I(x,y)\defeq\begin{cases}
\min(I(x,y), I(x,-y))&\text{if } y\geq 0;\\
+\infty&\text {if  }y<0.
\end{cases}
\eeq
Assume in addition that, for all $(\alpha,\beta)\in\rr^d\times\rr$, the limit
\[
q(\alpha,\beta)\defeq\lim_{T\to\infty}\frac1T\log\EE\left(\e^{\alpha\cdot X_T+\beta Y_T}\right)
\]
exists, is finite and satisfies $q(\alpha,\beta_0+\beta)=q(\alpha,\beta_0-\beta)$ for some $\beta_0\in \rr$. Then, the rate function $\overline I$ is the Legendre transform of the function
\[
\overline q(\alpha,\beta)
\defeq\lim_{T\to\infty}\frac1T\log\EE\left( \e^{\alpha\cdot X_T+\beta |Y_T|}\right)
=\begin{cases}
q(\alpha,\beta\sign(-\beta_0)) &\text {if }\beta\geq-|\beta_0|;\\
q(\alpha,\beta_0)&\text {otherwise,}
\end{cases}
\]
with the convention that $\sign(0)=1$. In particular, $\overline I$ is convex.
\end{lemma}
\proof
The LDP for $(T^{-1}[X_T\ \ |Y_T|])_{T\in\nn^\ast}$ with rate function $\overline I$ is a direct consequence of the contraction principle. Turning  to the second part of the lemma, since $q$ is finite everywhere, we have $q=I^\ast$ by Varadhan's theorem.\footnote{$f^\ast$ denotes the Legendre transform of $f$.} 
Then, since $I$ is convex by assumption, we have
\[
I(x,y)=q^\ast(x,y)=\beta_0y+h^\ast(x,y),
\]
where $h(\alpha,\beta)=q(\alpha,\beta_0+\beta)$. Since $h$ is even in its second argument, so is $h^\ast$, and~\eqref{eq:contractIvabs} reads
\beq{eq:reexprIbxy}
\overline I(x,y)=\begin{cases}
h^\ast(x,y)-|\beta_0|y&\text {if  }y\geq0;\\
+\infty&\text {if  }y<0.
\end{cases}
\eeq
$\overline I$ is obviously convex, since $h^\ast$ is. Next, since
\begin{align*}
\limsup_{T\to\infty}\frac1T\log\EE\left(\e^{\alpha\cdot X_T+\beta|Y_T|}\right)
&\leq\limsup_{T\to\infty}\frac1T\log\EE\left(\e^{\alpha\cdot X_T+\beta Y_T}
+\e^{\alpha\cdot X_T-\beta Y_T}\right)\\
&\leq\max(q(\alpha,\beta), q(\alpha,-\beta))<\infty,
\end{align*}
invoking again Varadhan's theorem, we obtain that the limit defining $\overline q$ exists, is finite everywhere, and satisfies $\overline q=\overline I^\ast$. Thus, by~\eqref{eq:reexprIbxy} we have
\beq{eq:supryp}
\overline q(\alpha,\beta)=\sup_{(x,y)\in\rr^d\times\rr_+} (\alpha\cdot x+\beta y
-\bar I(x,y))
=\sup_{(x,y)\in\rr^d\times\rr_+}(\alpha\cdot x+(\beta+|\beta_0|)y-h^\ast(x,y)).
\eeq
Since $h$ is convex, we have $h^{\ast\ast}=h$. We consider the following two cases:
\begin{itemize}
\item  If $\beta>-|\beta_0|$, then the supremum in~\eqref{eq:supryp} is actually a supremum over $y\in \rr$, so that
\begin{align*}
\overline q(\alpha,\beta)&=\sup_{(x,y)\in\rr^d\times\rr}(\alpha\cdot x+(\beta+|\beta_0|)y-  h^\ast(x,y))=h(\alpha,\beta+|\beta_0|)\\
&=h(\alpha,\beta\sign(-\beta_0)-\beta_0)=q(\alpha,\beta\sign(-\beta_0)).
\end{align*}
\item If $\beta\leq -|\beta_0|$, the supremum in~\eqref{eq:supryp} is reached at $y=0$ and we get
\[
\overline q(\alpha,\beta)=\sup_{x\in\rr^d}(\alpha\cdot x-h^\ast(x,0))=\sup_{(x,y)\in\rr^d\times\rr}  (\alpha\cdot x-h^\ast(x,y))=h(\alpha,0)=q(\alpha,\beta_0).
\]
\end{itemize}
This completes the proof.\qed

The next proposition identifies the entropic pressure $e(\alpha)$.

\begin{proposition}\label{prop:explicit_e_alpha}
For any $\alpha\in \rr$, one has
\beq{eq:KSfinalealpha}
e(\alpha)=Q(-\eta\alpha,-\gamma(\alpha\wedge1),\gamma(\alpha\vee0)).
\eeq
\end{proposition}
\proof We recall that, by  Proposition~\ref{prop:conv_freeenergy_1D_Ising}, the family $(T^{-1}X_T)_{T\in\nn^\ast}$ satisfies the LDP with the convex rate function $I=Q^\ast$, the cumulant-generating function $Q$ being given by~\eqref{eq:defQlambda}. It is obvious from this formula that for fixed $\lambda_1$, $\lambda_2$, the map $\lambda_3\mapsto Q(\lambda_1, \lambda_2, \lambda_3)$ is even and reaches its minimum at $\lambda_3=0$. Thus, we obtain from Lemma~\ref{lem:pressurevalabs} (with $\beta_0 = 0$) that the family $(T^{-1}[U_T\ \ V_T\ \ |W_T|])_{T\in\nn^\ast}$ satisfies the LDP with a convex rate function and that
\[
\overline Q(\lambda)
\defeq\lim_{T\to\infty}\frac1T\log\EE\left(\e^{\lambda_1U_T+\lambda_2V_T+\lambda_3|W_T|}\right)
=\begin{cases}
Q(\lambda_1,\lambda_2,\lambda_3)&\text{ if }\lambda_3\geq0;\\
Q(\lambda_1,\lambda_2,0)&\text{ if }\lambda_3<0.
\end{cases} 
\]
It follows from the last formula and~\eqref{eq:defQlambda} that $\bar Q(\lambda_1,-\gamma+\lambda_2,\lambda_3)=\bar Q(\lambda_1,-\gamma-\lambda_2,\lambda_3)$ for all $\lambda\in\rr^3$. Thus, invoking Lemma~\ref{lem:pressurevalabs} again, this time with $\beta_0=-\gamma$, we conclude that  $(T^{-1}[U_T\ \ |V_T|\ \ |W_T|])_{T\in\nn^\ast}$ satisfies the LDP with a convex rate function and that
\[
\overline{\overline Q}(\lambda)
\defeq\lim_{T\to\infty}\frac1T\log\EE\left(\e^{\lambda_1U_T+\lambda_2 |V_T|+\lambda_3|W_T|}\right)
=\begin{cases}
\overline Q(\lambda_1, \lambda_2,\lambda_3)&\text{if }\lambda_2\geq-\gamma;\\
\overline Q(\lambda_1, -\gamma,\lambda_3)&\text{if }\lambda_2<-\gamma.
\end{cases}
\]
From this and~\eqref{eq:ealphautvtwt}, we derive  that
\[
e(\alpha)=\overline{\overline Q} (-\eta\alpha,- \gamma\alpha,\gamma\alpha)
=\begin{cases}
Q(-\eta\alpha,-\gamma\alpha,0)&\text{if }\alpha<0;\\
Q(-\eta\alpha,-\gamma\alpha,\gamma\alpha)&\text{if }0\leq\alpha\leq 1;\\
Q(-\eta\alpha,-\gamma,\gamma\alpha)&\text{if }\alpha>1,
\end{cases}
\]
which is~\eqref{eq:KSfinalealpha}.\qed

We are now ready to complete the proofs of Parts~\ref{KS-iii} and~\ref{KS-iv} of Theorem~\ref{mainthm-KS}. The function $e$ is obviously real analytic on $\rr\setminus\{0,1\}$, since $Q$ is real analytic on $\rr^3$. Moreover, using~\eqref{eq:KSfinalealpha} and the expression~\eqref{eq:gradQ000} for $\nabla Q(0)$, we immediately see that
\[
(\partial^-e)(0)=\nabla Q(0)\begin{bmatrix}-\eta&-\gamma&0\end{bmatrix}^{\mathsf{T}}=\nabla Q(0)\begin{bmatrix}-\eta&-\gamma&\gamma\end{bmatrix}^{\mathsf{T}}=(\partial^+e)(0),
\]
since the last component of $\nabla Q(0)$ is zero. Thus, $e$ is differentiable at $\alpha = 0$, and hence also at $\alpha = 1$ by the symmetry~\eqref{es-s}. In particular, we have
\beq{eq:EPfromNabla}
\ep(\cJ,\rho)=-e'(0)=-\nabla Q(0)\begin{bmatrix}-\eta&-\gamma&0\end{bmatrix}^{\mathsf{T}},
\eeq
which gives the formula in~\ref{KS-iii}. Since our assumptions exclude the case where  $\PP =\wP$, Theorem~\ref{thm-int}~\ref{it:ergoconvsigma} guarantees that $\ep(\cJ,\rho) >0$.

By Theorem~\ref{thm-int}~\ref{it:differentiability_e_alpha}, $e$ is convex. To show that it is strictly convex, it suffices, in view of~\eqref{eq:KSfinalealpha}, to observe that $Q$ is strictly convex, and to recall that $\gamma>0$ by assumption.

Finally, writing $H=D^2Q(0)$ (recall~\eqref{eq:Hessian}), we have
\[
(\partial^+e')(0)-(\partial^-e')(0)=\begin{bmatrix}-\eta&-\gamma&\gamma\end{bmatrix}H\begin{bmatrix}-\eta&-\gamma&\gamma\end{bmatrix}^{\mathsf{T}}-\begin{bmatrix}-\eta&-\gamma&0\end{bmatrix}H\begin{bmatrix}-\eta&-\gamma&0\end{bmatrix}^{\mathsf{T}},
\]
which  yields~\eqref{eq:eppjump} and completes the proof of~\ref{KS-iv}.

\subsection{Proof of Theorem~\ref{mainthm-KS}, Part~\ref{KS-v}}
\label{sec:proofCLTsigmaKS}
To derive the anomalous central limit theorem satisfied by $(\sigma_T)_{T\in\nn^\ast}$,  we consider the random variables
\beq{eq:sigmaTmTep}
\frac{\widehat \sigma_T-T\ep(\cJ,\rho)}{\sqrt{T}}
\eeq
defined on $(\Xi,\QQ)$, which have, by construction, the same law as the random variables in the left-hand side of~\eqref{eq:CLTKSPMP} defined on $(\Omega,\PP)$. In view of~\eqref{eq:hatsigmaeta}, the random variables \eqref{eq:sigmaTmTep} and 
\[
\varsigma_T
\defeq\frac{\eta\,U_T+\gamma(|V_T|-|W_T|)-T\ep(\cJ,\rho)}{\sqrt{T}}
\]
have the same limiting law (if any). We now show that the same is true for  $\varsigma_T$ and 
\[
\varsigma_T'
\defeq\frac{\eta\,U_T+\gamma(V_T-|W_T|)-T\ep(\cJ,\rho)}{\sqrt{T}}.
\]
For this, it suffices to show that
\[
\lim_{T\to\infty}\QQ(\{|V_T|\neq V_T\})=0.
\] 
By Part~\ref{prt:ldpXT} of Proposition~\ref{prop:conv_freeenergy_1D_Ising} and Formula~\eqref{eq:gradQ000}, we have
\beq{eq:convWT}
\lim_{T\to\infty}\QQ\left(\left\{\left|\frac1T V_T-\frac{r_2-r_1}{r_1+r_2}\right|>\epsilon\right\}\right)=0,
\eeq
for any $\epsilon >0$. Setting $\epsilon=\frac {r_2-r_1}{r_1+r_2}>0$ (recall~\eqref{equ:q1big}), we find
\[
\QQ\left(\{|V_T|\neq V_T\}\right)=\QQ\left(\{T^{-1}V_T<0\}\right)
\leq\QQ\left(\left\{\left|T^{-1}V_T -\epsilon\right| >\epsilon \right\}\right),
\]
which converges to zero by~\eqref{eq:convWT}.\footnote{Note that, since the third component of $\nabla Q(0)$ vanishes, we cannot get rid of the absolute value of $W_T$ in the same way.}

Next, recalling the expression~\eqref{eq:EPfromNabla} for $\ep(\cJ,\rho)$, we can write
\[
\varsigma_T'=\frac{\eta (\,U_T - T(\partial_1Q)(0))
+\gamma (V_T-T(\partial_2Q)(0)-|W_T|)}{\sqrt{T}}.
\]
By Part~\ref{prt:cltXT} of Proposition~\ref{prop:conv_freeenergy_1D_Ising} and the continuous mapping theorem (applied to the function $(x,y,z)\mapsto\eta x+\gamma(y-|z|)$), we conclude that $\varsigma'_T$ converges in law to
\[
\eta X_1+\gamma(X_2-|X_3|),
\]
where $X\eqdef[X_1\ X_2\ X_3]$ is a centered Gaussian random vector of covariance matrix  $H\defeq D^2Q(0)$. The block-diagonal structure of $H$ displayed in~\eqref{eq:Hessian} implies that the third component of $X$ is independent of the first two. Thus, \eqref{eq:CLTKSPMP} holds with
\[
Z_1=\eta X_1+\gamma X_2,\qquad 	Z_2=\gamma X_3,
\]
which are independent, centered Gaussian random variables with variances given by
\[
{\rm Var}(Z_1)=\begin{bmatrix}\eta&\gamma&0\end{bmatrix}H\begin{bmatrix}\eta&\gamma&0\end{bmatrix}^{\mathsf{T}},\qquad
{\rm Var}(Z_2)=\begin{bmatrix}0&0&\gamma\end{bmatrix}H\begin{bmatrix}0&0&\gamma\end{bmatrix}^{\mathsf{T}},
\]
and Formula~\eqref{eq:Hessian} yields the claimed expressions. Since $\gamma\not=0$ by assumption,
the strict positivity of these variances follows from the fact that $H$ is positive definite. To see this, observe that the third diagonal element of $H$ as well as the trace and determinant (compute it!) of its upper $2\times2$ block are strictly positive. The proof of Part~\ref{KS-v} is complete.
\subsection{Proof of Theorem~\ref{mainthm-KS}, Part~\ref{KS-vi}}\label{sec:FDR}

In this subsection, we show that the infinite-time fluctuation--dissipation relation does not hold for the Keep--Switch instrument. More precisely, we show that suitably defined currents follow a central limit theorem at equilibrium (defined by  $\ep(\cJ,\rho)=0$) with covariance matrix $D_\infty$, while the associated linear response coefficients $L_\infty$ do not verify the fluctuation--dissipation relation $L_\infty=\frac12D_\infty$. The finite-time fluctuation--dissipation relation, however, is satisfied.

As mentioned in Remark~\ref{rem:epzero}, we have $\ep(\cJ,\rho)=0$ iff $q_1=r_1=q_2=r_2=\frac12$. We shall therefore parametrize the model by $\varepsilon=(\varepsilon_1,\varepsilon_2)\in{]}{-1/2},1/2{[}\times{]}{-1/2},1/2{[}$, with
\beq{eq:paramspijepsi}
q_1\defeq\frac12-\varepsilon_1,\quad q_2\defeq\frac12-\varepsilon_2,\quad r_1\defeq\frac12+\varepsilon_1,\quad r_2\defeq\frac12+\varepsilon_2
\eeq
(we release the constraint~\eqref{equ:q1big} in this section). Equilibrium thus corresponds to $\varepsilon=0$. We denote by $\PP^{(\varepsilon)}$ the Keep--Switch PMP measure corresponding to $\varepsilon$. We write in the same way $\EE^{(\varepsilon)}$, $\ep^{(\varepsilon)} = \ep^{(\varepsilon)}(\cJ^{(\varepsilon)},\rho^{(\varepsilon)})$ and $\sigma_T^{(\varepsilon)}$ for the corresponding quantities. Viewing $\varepsilon$ as a thermodynamic force, we define the corresponding current by
\[
J_{T}^{(\varepsilon)}\defeq\int_0^1\left(\nabla_\varepsilon\sigma_T\right)^{(\lambda\varepsilon)}\d\lambda,
\]
which we view as a column vector here.
With this definition, we have the {\em flux relation} (see~\cite[Definition~4.1]{JPRB-2011})
\beq{eq:fluxrelsigma}
\sigma_T^{(\varepsilon)}=\varepsilon\cdot J_{T}^{(\varepsilon)}.
\eeq
Let $\Theta_T\colon\Omega_T\to\Omega_T$ denote reversal, \ie 
\[
\Theta_T(\omega_1, \omega_2, \dots, \omega_T)
\defeq(\theta(\omega_T), \theta(\omega_{T-1}), \dots, \theta(\omega_1)).
\]
Since $\sigma_T \circ \Theta_T = -\sigma_T$, we obtain that
\beq{eq:fluxreverses}
J_{T}^{(\varepsilon)}\circ \Theta_T = - J_{T}^{(\varepsilon)}.	
\eeq
The following observation will be useful below: denoting by $S$ the involution of $\rr^2$ given by $(x_1,x_2)\mapsto(x_2,x_1)$, one easily check that $\PP^{(S\varepsilon)}=\PP^{(\varepsilon)}$, which further implies  $\sigma_T^{(S\varepsilon)}=\sigma_T^{(\varepsilon)}$ and 
\beq{JTsym}
J_T^{(S\varepsilon)}=SJ_T^{(\varepsilon)}.
\eeq
\begin{proposition}\label{prop:KS-def-fluxes}
For any $T\in\nn^\ast$, let $N_T\colon\Omega\to\nn$ be the random variable counting the number of `$S$' in $\omega_{\lbr1,T\rbr}$. Then, one has
\beq{eq:rexxprJt}
J_{T}=\lim_{\varepsilon\to0}J_{T}^{(\varepsilon)}=2(2N_T-T)\begin{bmatrix}1\\1\end{bmatrix}.
\eeq
\end{proposition}
\proof 
Let $\omega\in\Omega$ and $\xi\in\Xi$ be such that $\omega=F(\xi)$. Substituting  the relations~\eqref{eq:paramspijepsi} into~\eqref{eq:sigmathatnij} and expanding to first order in  $\varepsilon$, we find
\[
\sigma_T(\omega)=\widehat\sigma_T(\xi)
=-2(n_{++}+n_{--}-n_{+-}-n_{-+})(\varepsilon_1+\varepsilon_2) + O(|\varepsilon|^2),
\]
where $n_{ab}=n_{ab}(\xi_{\lbr1,T+1\rbr})$ (notice that the two hyperbolic cosines in~\eqref{eq:sigmathatnij} do not give any contribution at first order). The statement follows from the final observation that   
\[
n_{++}+n_{--}-n_{+-}-n_{-+}
\]
counts the number of `$K$' minus the number of `$S$' in $\omega_{\lbr1,T\rbr}$, and that the sum of these two numbers is $T$.\qed

The next proposition concerns the fluctuations of $J_T$ at equilibrium.
\begin{proposition}\label{prop:KS-CLT-fluxes}
For all $T\in\nn^\ast$ we have $\EE^{(0)}(J_{T})=0$ and 
\[
D_T=\EE^{(0)}\left(\frac{J_{T}J_T^\mathsf{T}}{T}\right)=4\begin{bmatrix}1&1\\1&1\end{bmatrix}.
\]
Moreover, the weak limit, as $T\to\infty$, of $\frac{J_{T}}{\sqrt{T}}$ with respect to $\PP^{(0)}$ exists and is a centered Gaussian random vector with covariance 
\[
D_\infty=4\begin{bmatrix}1&1\\1&1\end{bmatrix}.
\]
\end{proposition}
\proof By Remark~\ref{rem:KSber}, $\PP^{(0)}$ is the Bernoulli measure on $\Omega$ generated by $Q(K)=1/2$. Thus, the random variable $N_T$ in~\eqref{eq:rexxprJt} has a binomial law of parameters $\frac12$ and $T$, and the claims follow at once.\qed

We now compute the linear response. For that purpose, we define the averaged currents
\[
\overline  J^{(\varepsilon)}_T\defeq\frac 1 T \EE^{(\varepsilon)}(J_{T}^{(\varepsilon)}), \qquad \overline  J^{(\varepsilon)}\defeq\lim_{T\to\infty}\overline  J^{(\varepsilon)}_T,
\]
and the associated Onsager matrices
\[
L_{T}\defeq D_\varepsilon\overline  J^{(\varepsilon)}_{T}\Big|_{\varepsilon=0}, \qquad
L_{\infty}\defeq D_\varepsilon\overline  J^{(\varepsilon)}\Big|_{\varepsilon=0},
\] 
where $D_\varepsilon J\defeq[\partial_{\varepsilon_j}J_i]_{i,j\in\{1,2\}}$ denotes the Jacobian matrix of the vector field $J$.

\begin{proposition}\label{prop:KS-response-coeffs}
We have 
\[
L_T=\frac12 D_T, 
\]
for $T\in\nn^\ast$, but
\beq{eq:FDLinfty}
L_\infty=\begin{bmatrix}3&1\\1&3\end{bmatrix}\not=\begin{bmatrix}2&2\\2&2\end{bmatrix}=\frac12D_\infty.
\eeq
Thus, the finite-time fluctuation--dissipation relation holds, but the infinite-time one is violated.
\end{proposition}
\proof The assertion $L_T=\frac12 D_T$ is the usual finite-time fluctuation--dissipation theorem, which applies here thanks to the relations~\eqref{eq:fluxrelsigma} and~\eqref{eq:fluxreverses}  (see for example~\cite[Section 4]{JPRB-2011}).  For the reader's convenience, we outline the proof. For $\varepsilon$ small enough and $\alpha\in\rr^2$, set
\[
G_T(\varepsilon,\alpha)
\defeq\frac1T\log\EE^{(\varepsilon)}\left(\e^{-\alpha\cdot J_T^{(\varepsilon)}}\right).
\]
Since
\begin{align*}
\sum_{\omega\in\Omega_T}\PP^{(\varepsilon)}([\omega])
\e^{-\alpha\cdot J_T^{(\varepsilon)}(\omega)} 
&=\sum_{\omega\in\Omega_T}\wP^{(\varepsilon)}([\omega])
\e^{(\varepsilon-\alpha)\cdot J_T^{(\varepsilon)}(\omega)}\\
&=\sum_{\omega\in\Omega_T}\wP^{(\varepsilon)}([\Theta_T(\omega)])
\e^{(\varepsilon-\alpha)\cdot J_T^{(\varepsilon)}(\Theta_T(\omega))}\\
&=\sum_{\omega\in\Omega_T}\PP^{(\varepsilon)}([\omega])
\e^{-(\varepsilon-\alpha)\cdot J_T^{(\varepsilon)}(\omega)},
\end{align*}
we obtain the celebrated  Gallavotti  symmetry
\beq{eq:symmetryGT}
G_T(\varepsilon, \varepsilon-\alpha) = G_T(\varepsilon,\alpha).
\eeq
Starting with the identities
\[
D_T=[\partial_{\alpha_i}\partial_{\alpha_j}G_T(\varepsilon, \alpha)]|_{\alpha=\varepsilon=0},\qquad
L_T=-[\partial_{\varepsilon_j}\partial_{\alpha_i} G_T(\varepsilon,\alpha)]|_{\alpha =\varepsilon=0},
\]
the symmetry~\eqref{eq:symmetryGT}  implies that $L_T=\frac12D_T$, as claimed.

We now prove~\eqref{eq:FDLinfty}. By the definition~\eqref{eq:defEPsigma} of $\ep$, we find  
\[
\varepsilon\cdot\overline J^{(\varepsilon)}
=\lim_{T\to\infty}\frac1T\EE^{(\varepsilon)}(\sigma_T^{(\varepsilon)})=\ep^{(\varepsilon)}.
\]
Part~\ref{KS-iii} of Theorem~\ref{mainthm-KS}  yields 
\beq{eq:epjFD}
\varepsilon\cdot\overline J^{(\varepsilon)}
=\varepsilon_1(3\varepsilon_1+\varepsilon_2)+\varepsilon_2(3\varepsilon_2+\varepsilon_1)
+O(|\varepsilon|^3),
\eeq
which, taking~\eqref{JTsym} into account,\footnote{The symmetry \eqref{JTsym} allows to conveniently estimate $\overline J^{(\varepsilon)}$ in terms of $\ep^{(\varepsilon)}$, but its role is not fundamental. One can, in principle, also obtain \eqref{eq:FDLinfty} by writing $J_T^{(\varepsilon)}$ in terms of the vector $X_T$ introduced in Section~\ref{sec:FMEP} and then using the LDP that it obeys (with respect to $\PP^{(\varepsilon)}$).} implies~\eqref{eq:FDLinfty}.\qed

We finish with a brief  comment regarding  the failure of the infinite-time fluctuation--dissipation relation. Its usual derivation (see for example~\cite[Section 5]{JPRB-2011}) fails to apply here, because the limit 
\[
G(\varepsilon,\alpha) \defeq\lim_{T\to\infty} G_T(\varepsilon,\alpha)
\]
is not $C^2$ at $(0,0)$. In fact, the discrepancy between $\lim_{T\to\infty} L_T$ and $L_\infty$ comes from the hyperbolic cosines in~\eqref{eq:sigmathatnij}. At fixed $T$, their contribution to $\sigma_T^{(\varepsilon)}$ is only $O(|\varepsilon|^2)$. However, since $\log \cosh(x) \sim |x|$ when $x$ is large  (see~\eqref{eq:hatsigmaeta}), the contribution of the hyperbolic cosines in~\eqref{eq:sigmathatnij} to $T^{-1}\sigma_T^{(\varepsilon)}$ in the limit $T\to \infty$ becomes $O(\varepsilon)$. Thus, the limit $T\to\infty$ and the limit $\varepsilon\to0$ (or the derivative with respect to $\varepsilon$) cannot be interchanged.

\subsection{Proof of Theorem~\ref{KS-new}}
\label{sec-KS-new}

Theorem~\ref{KS-new} is proved in a very similar way to Theorem~\ref{mainthm-KS}, and we only outline the main differences here. We use the notation in Theorem~\ref{KS-new}, and we recall, in particular, that $\gamma$, $\chi$, $\eta$ and $\delta$ were defined in~\eqref{eq:xichigamma}-\eqref{eq:etadelta}. We also freely use the notation of the proof of Theorem~\ref{mainthm-KS}, and in particular, we assume throughout that $F(\xi)=\omega$, where $\xi\in\Xi_{T+1}$ and $\omega\in\Omega_T$, or $\xi\in\Xi$ and $\omega\in\Omega$.

By recalling the expression~\eqref{eq:PfnuFM} for $\PP(F(\xi))$, and noting that $\wP(F(\xi))$ is obtained in the same way with $q_i, r_i$ replaced by $\widehat q_i, \widehat r_i$, we obtain that 
\eqref{eq:sigmathatnij} is replaced by
\begin{align*}
\widehat\sigma_T&=\frac{n_{++}+n_{--}}2\log\frac{q_1q_2}{\widehat q_1\widehat q_2} 
+\frac {n_{+-}+n_{-+}}2\log\frac{r_1r_2}{\widehat r_1 \widehat r_2}
+\log\left(\frac{\cosh\left(\frac{n_{++}-n_{--}}{2}\log\frac{q_1}{q_2}\right)}
{\cosh\left(\frac{n_{++}-n_{--}}{2}\log\frac{\widehat q_1}{\widehat q_2}\right)}\right)+O(1)\\
&=\delta T+\eta A_T+\chi|B_T|+O(1),
\end{align*}
where we have used that $n_{+-}+n_{-+}=T-(n_{++}+n_{--})$, and where
\[
A_T\defeq n_{++}+n_{--},\qquad 
B_T\defeq n_{++}-n_{--},
\]
recalling that $n_{ab}=n_{ab}(\xi_{\lbr1, T+1\rbr})$. Using this, \eqref{eq:ealphautvtwt} is replaced by
\beq{eq:expressionealpha}
e(\alpha)=-\delta\alpha+ \lim_{T\to\infty}\frac1T\log\EE\left(\e^{-\alpha(\eta A_T+\chi|B_T|)}\right).
\eeq 
We introduce then
\[
q_T(\lambda)\defeq\frac1T\log\EE\left(\e^{\lambda_1A_T+\lambda_2B_T }\right)
=\frac1T\log\left({\bf p}P(\lambda)^T\bf 1\right), 
\]
where
\[
P(\lambda)\defeq\begin{bmatrix}
\e^{\lambda_1+\lambda_2}q_1&r_1\\
r_2&\e^{\lambda_1-\lambda_2}q_2
\end{bmatrix}.
\]
By computing the dominant eigenvalue $\kappa_+(\lambda)$ of $P(\lambda)$, we conclude that
\begin{align*}
Q(\lambda)&\defeq\lim_{T\to\infty}q_T(\lambda)=\log\kappa_+(\lambda)\\
&=\frac12\log(q_1q_2)+\lambda_1+\log\left(\cosh(\lambda_2+\gamma)
+\sqrt{\sinh^2(\lambda_2+\gamma)+\e^{-2(\lambda_1+\rho)}}\right),
\end{align*}
where 
\[
\rho\defeq\frac12\log\left(\frac{q_1q_2}{r_1r_2}\right).
\]
The gradient and Hessian matrix of $Q$ at $\lambda=0$ are given by
\begin{gather*}
\nabla Q(0)=\frac1{r_1+r_2}
\begin{bmatrix}r_1-2r_1r_2+r_2&r_2-r_1\end{bmatrix},\\[4pt]
D^2Q(0)=\frac{4r_1r_2}{(r_1+r_2)^3}\begin{bmatrix}
q_1r_2^2+q_2r_1^2 & r_2-r_1\\[2mm]
r_2-r_1 &  q_1+q_2\end{bmatrix}.
\end{gather*}
Now observing that $\lambda_2\mapsto Q(\lambda_1, \lambda_2)$ is even around $-\gamma$ and invoking Lemma~\ref{lem:pressurevalabs}, we deduce from~\eqref{eq:expressionealpha} that\footnote{Recall our convention $\sign(0)=1$.}
\beq{eq:formuleealphapairks}
e(\alpha)=-\delta\alpha+Q(-\eta\alpha,-\sign(\gamma)(|\gamma|\wedge\alpha\chi)).
\eeq
By computing $-e'(0)$ we obtain the formula for $\ep(\PP, \wP)$ in Part~\ref{point:iieppairks} of Theorem~\ref{KS-new}.

We now prove Part~\ref{point:part2ep0}. By the generalization of Theorem~\ref{thm-int}~\ref{it:ergoconvsigma} discussed at the end of Section~\ref{sec-setup}, since $(\Omega, \phi, \wP)$ is ergodic, 
we have $\ep(\PP, \wP)=0$ iff $\PP =\wP$. The equivalence with the remaining two conditions in Theorem~\ref{KS-new}~\ref{point:part2ep0} is then an easy exercise.

By~\eqref{eq:formuleealphapairks}, 
since $Q$ is real analytic, we obtain that $e$ is real analytic on $\rr$ if $\chi = 0$, and real analytic on $\rr \setminus \{|\gamma|/\chi\}$ if $\chi\neq 0$. In the latter case, explicit computations show that $e$ is differentiable at $\alpha=|\gamma|/\chi$ but not twice differentiable. Computing the jump in the second derivative gives
\[
(\partial^+e')(|\gamma|/\chi)-(\partial^-e')(|\gamma|/\chi) = -\chi|\chi|\e^{\rho-\eta|\gamma|/\chi}\neq 0.
\]
This proves Part~\ref{point:iiipartanalytic}. 

The random vectors $[A_T\ \ B_T]$ satisfy the LDP, the law of large numbers and the central limit theorem as in parts~\ref{prt:ldpXT} and~\ref{prt:cltXT} of Proposition~\ref{prop:conv_freeenergy_1D_Ising}. In particular,
\[
\frac{[A_T\ \ B_T]-T\nabla Q(0)}{\sqrt{T}}
\]
converges in law towards a normal, centered random vector $[X_1\ \ X_2]$ with covariance matrix $D^2Q(0)$.

We now turn to the CLT for $\sigma_T$. By the estimates above, $(\sigma_T-T\ep(\PP, \wP))/\sqrt{T}$ has the same limiting distribution, if any, as 
\[
\frac{{\delta T+\eta A_T+\chi|B_T|-T\ep(\PP, \wP)}}{{\sqrt{T}}}
=\frac{\eta(A_T-T(\partial_1Q)(0))+\chi(|B_T|-T|(\partial_2Q)(0)|)}{\sqrt{T}}.
\]
If $\chi = 0$, then obviously this converges in law to $Z=\eta X_1$, whose variance is $\eta^2(\partial_1^2Q)(0)$, which coincides with the formula given in Part~\ref{point:pairKSCLT1} of Theorem~\ref{KS-new}. If $\gamma\not=0$, then $(\partial_2Q)(0)\neq0$, and using the same arguments as in Section~\ref{sec:proofCLTsigmaKS}, we obtain that $(\sigma_T-T\ep(\PP, \wP))/\sqrt{T}$ has the same limiting distribution as 
\[
\frac{\eta(A_T-T(\partial_1Q)(0))+\sign(\gamma)\chi(B_T-T(\partial_2Q)(0))}{\sqrt{T}},
\]
which converges in distribution to $Z=\eta X_1+\sign(\gamma)\chi X_2$, whose variance is again given by the formula for ${\rm Var}(Z)$ in Part~\ref{point:pairKSCLT1} of Theorem~\ref{KS-new}.
 
We now turn to Part~\ref{point:pairKSCLT2}, and for this we assume that $\chi \not=0$ and $\gamma=0$ (note that this implies that $\chi < 0$). We then have $(\partial_2Q)(0)=0$, and  $(\sigma_T-T\ep(\PP, \wP))/\sqrt{T}$ has the same limiting distribution as
\[
\frac{\eta(A_T-T(\partial_1Q)(0))+\chi|B_T|}{\sqrt{T}},
\]
which converges in law to $Z_1- |Z_2|$ with $Z_1=\eta X_1$ and $Z_2=\chi X_2$. This implies the statements of Part~\ref{point:pairKSCLT2} and concludes the proof of Theorem~\ref{KS-new}.

\section{Spin instruments}
\label{spin-main}
In this section we provide proofs of our results on the various spin instruments described in Sections~\ref{spin-intro} and~\ref{sec-nevnnevxx}.

\subsection{XXZ-spin instruments}\label{xy-main}
We start with the proofs of Theorems~\ref{xy-ot-on}, \ref{thm-xxz-twotime}, \ref{thm-xxz-ran} and~\ref{thm-xxz-multi} pertaining to one- and two-time measurement protocols for XXZ-interaction. 


\subsubsection{One-time measurements}
\label{sec-xxz-one}
We first prove a slightly more general version of Theorem~\ref{xy-ot-on}, replacing the spin-$\frac12$ system $\cS$  with a
generic spin.

Let $\mathbf{S}\defeq(S_x,S_y,S_z)$  be a family of operators on the finite-dimensional Hilbert space $\cH$ satisfying the commutation relations
\beq{eq:comrelS123}
[S_z,S_\pm]=\pm S_\pm, \qquad [S_+,S_-]=2S_z,
\eeq
where $S_\pm\defeq S_x\pm \i S_y$.
Note that the family  $2{\bf S}$ provides a representation of the Lie algebra $\mathfrak{su}(2)$. We do not assume this representation to be irreducible and, as a consequence, the map $\Phi$ of the instrument constructed below will not be irreducible in general.

The only changes compared to the setting of Section~\ref{spin-intro} concern the definition of the system Hamiltonian $H_\cS$ in~\eqref{eq-xxz-ham} and that of the system-probe interaction $V$ in~\eqref{eq-xxz-v}. The system Hamiltonian
becomes $H_\cS\defeq\omega S_z$ and the system-probe interaction is given by
\[
V\defeq\lambda\left(S_x\otimes\sigma_x+S_y\otimes\sigma_y\right)+\mu S_z\otimes\sigma_z.
\]
In the following we  identify $\cH\otimes \cH_p$ with $\cH\oplus\cH$ so that the total Hamiltonian~\eqref{xy-ham} reads, in block-matrix form,
\beq{Hmatrix}
H = \begin{bmatrix}
(\omega+\mu) S_z+\tfrac\epsilon2&\lambda S_-\\
\lambda S_+& (\omega-\mu)S_z-\tfrac\epsilon2
\end{bmatrix}.
\eeq
The special case of Section~\ref{spin-intro} is recovered by setting ${\bf S}=\frac{1}{2}(\sigma_x, \sigma_y, \sigma_z)$.

\begin{lemma}\label{xy-u}
Let
\beq{eq:def_Omega_pm}
\Lambda_\pm\defeq\tfrac12(\epsilon-\omega)+\mu(S_z\pm\tfrac12), \qquad 
\Omega_\pm\defeq\sqrt{\Lambda_\pm^2+\lambda^2S_\mp S_\pm}.
\eeq
The propagator $U\defeq\e^{-\i t H}$ is given by
\beq{fri-ti00}
U=\e^{\frac{\i t\mu }{2} }\begin{bmatrix}
\e^{-\frac{\i t \omega }{2}}V_{++}&  -\i\e^{-\frac{\i t\omega}{2}}V_{+-}\\
-\i\e^{\frac{\i t \omega}{2}}V_{-+} & \e^{\frac{\i t \omega}{2}}V_{--}
\end{bmatrix},
\eeq
with
\beq{eq:Vpmpm}
V_{\pm\pm}\defeq\e^{-\i t\omega S_z}\left(\cos(t\Omega_\pm)\mp\i\Lambda_\pm\frac{\sin(t\Omega_\pm)}{\Omega_\pm}\right),
\qquad V_{\pm\mp}\defeq\lambda\e^{-\i t\omega S_z}\frac{\sin(t\Omega_\pm)}{\Omega_\pm}S_\mp.
\eeq
\end{lemma}

\proof We write $H= H_0 + \lambda W$, where
\[
H_0\defeq\begin{bmatrix}
(\omega+\mu) S_z+\tfrac\epsilon2&0\\
0& (\omega-\mu)S_z-\tfrac\epsilon2
\end{bmatrix}, \qquad W\defeq\begin{bmatrix}
0&S_-\\S_+ &0
\end{bmatrix}.
\]

The interaction-picture propagator $\Gamma^t\defeq\e^{\i tH_0}\e^{-\i t H}$ satisfies 
\beq{prop-C}
\i\partial_t \Gamma^t=\lambda \e^{\i t H_0}W\e^{-\i tH_0}\Gamma^t,\qquad \Gamma^0=\one.
\eeq
Invoking the commutation relations~\eqref{eq:comrelS123}, one shows that for continuous functions $f:\rr\to\rr$,
\[
f(S_z)S_\pm=S_\pm f( S_z\pm1), \quad
f(\Lambda_\mp)S_\pm=S_\pm f(\Lambda_\pm),\quad
f(\Omega_\mp)S_\pm=S_\pm f(\Omega_\pm),
\]
which leads to
\[
\e^{\i tH_0}W\e^{-\i tH_0}=\begin{bmatrix}
0 & \e^{\i t 2 \Lambda_+}S_-\\ \e^{-\i t 2\Lambda_-}S_+ & 0
\end{bmatrix}.
\]
One easily concludes that the solution of~\eqref{prop-C} is given by
\[
\Gamma^t=\begin{bmatrix}
\ds\e^{\i t \Lambda_+}\left(\cos(t\Omega_+)-\i\Lambda_+\frac{\sin(t\Omega_+)}{\Omega_+}\right) & \ds-\i\lambda\e^{\i t \Lambda_+}\frac{\sin(t\Omega_+)}{\Omega_+}S_-\\
\ds-\i \lambda\e^{-\i t\Lambda_-}\frac{\sin(t\Omega_-)}{\Omega_-}S_+ &\ds \e^{-\i t\Lambda_-}\left(\cos(t\Omega_-)+\i\Lambda_-\frac{\sin(t\Omega_-)}{\Omega_-}\right)
\end{bmatrix},
\]
and computing  $U=\e^{-\i tH_0}\Gamma^t$ yields the result.\qed

Lemma~\ref{xy-u} gives that the one-time instrument~\eqref{ancila} is given by
\beq{eq:PhipmX}
\Phi_{\pm}[X]=\left(\frac{1}{2}-\eta\right) V_{\pm +}^\ast XV_{\pm +}
+ \left(\frac{1}{2}+\eta\right) V_{\pm -}^\ast XV_{\pm-}.
\eeq
Recall that the probe state~\eqref{pr-st-xy} can be written as
\[
\rho_p=Z_p^{-1}\e^{-\beta_pH_p},
\]
with $Z_p\defeq\tr\,\e^{-\beta_pH_p}$ and $\beta_p\defeq\frac2\epsilon\atanh(2\eta)$.

\begin{lemma}\label{lem:invariant_state_su(2)_su(2)} 
The density matrix
\[
\rho\defeq Z_\cS^{-1}\e^{-\beta_\cS H_\cS},
\]
with $Z_\cS\defeq\tr\,\e^{-\beta_\cS H_\cS}$ and $\beta_\cS\defeq\frac\epsilon\omega\beta_p$, satisfies 
\beq{fev-1}
\Phi_+^\ast[\rho]=\left(\frac{1}{2}-\eta\right) \rho,\quad \Phi_-^\ast[\rho]=\left(\frac{1}{2}+\eta\right)\rho.
\eeq
\end{lemma}
\proof We will prove the first relation in~\eqref{fev-1}. A similar computation yields the second one. Invoking again the commutation relations~\eqref{eq:comrelS123}, we have 
\[
[S_z, S_\pm S_\mp] = 0,\qquad S_-\rho S_+=\e^{-\beta_p\epsilon}S_-S_+\rho,
\]
and it follows from~\eqref{eq:PhipmX} that
\begin{align*}
\Phi_+^\ast[\rho]&=\left(\frac{1}{2}-\eta\right)V_{++}\rho V_{++}^\ast+\left(\frac{1}{2}+\eta\right)V_{+-}\rho V_{+-}^\ast\\[1mm]
	&=\left(\frac{1}{2}-\eta\right) \left(\cos^2(t\Omega_+)+\Lambda_+^2\frac{\sin^2(t\Omega_+)}{\Omega_+^2}\right)\rho+\left(\frac{1}{2}-\eta\right) \lambda^2S_-S_+\frac{\sin^2(t\Omega_+)}{\Omega_+^2}\rho\\[1mm]
	&=\left(\frac{1}{2}-\eta\right)\left(\cos^2(t\Omega_+)+\frac{\Lambda_+^2+\lambda^2S_-S_+}{\Omega_+^2}\sin^2(t\Omega_+)\right)\rho\\[1mm]
	&=\left(\frac{1}{2}-\eta\right)(\cos^2(t\Omega_+)+\sin^2(t\Omega_+))\rho=\left(\frac{1}{2}-\eta\right)\rho.
\end{align*}
\qed

The relations~\eqref{fev-1} give that   $\Phi^\ast[\rho]=\rho$ and 
\begin{theorem}\label{thm:XXZot}
The unraveling $\PP$ of $((\Phi_-,\Phi_+), \rho)$ is the Bernoulli measure generated by the mass function  
$Q(\pm)=\frac{1}{2}\mp\eta$. 
\end{theorem}

\begin{remark}\label{rem:frops} If the representation $\bf S$ of the Lie algebra $\mathfrak{su}(2)$ is irreducible, then it is easy to show that $\cH$ has no non-trivial subspace left invariant by the family $(V_{ab})_{a,b\in\{-,+\}}$. It follows from~\cite[Theorem~2.1]{JPW-2014} that $\Phi$ is irreducible, and hence that $\rho$ is the only density matrix for which~\assref{(A)} holds. This applies, in particular, to the case 
${\bf S}=\frac12(\sigma_x,\sigma_y,\sigma_z)$ considered in Section~\ref{spin-intro}.
\end{remark} 
\subsubsection{Two-time measurements with a thermal probe}
\label{sec-xy-tttp}
In this section we consider the two-time measurement protocol for XXZ-spin interaction with thermal probes. We prove Theorem~\ref{thm-xxz-twotime} and further properties of the corresponding instrument.

In the case ${\bf S}=\frac{1}{2}(\sigma_x, \sigma_y, \sigma_z)$, the operators  in~\eqref{eq:def_Omega_pm} take the form 
\[ 
\Lambda_\pm=\frac12\nu_{\pm} P_\pm+ \frac12(\epsilon-\omega)P_{\mp},
\qquad\Omega_\pm^2=\frac14\nu_{\pm}^2P_\pm +\delta^2 P_\mp,
\]
where the projections $P_\pm$ are given by~\eqref{eq:Ppm}, and
\[
\nu_{\pm}\defeq(\epsilon-\omega)\pm2\mu, \qquad \delta\defeq\sqrt{\left(\frac{\epsilon-\omega}2\right)^2+\lambda^2}.
\]
We deduce that~\eqref{eq:Vpmpm} becomes
\beq{fri-ti0}
\begin{split}
V_{\pm\pm}&=\e^{\mp\i t(\omega+\nu_\pm)/2}P_{\pm}
+\e^{\pm\i t\omega/2}\left(\cos(t\delta)\mp\i\frac{\epsilon - \omega}2\frac{\sin(t\delta)}{\delta}\right)P_{\mp},\\[1mm]
V_{\mp\pm}&=\lambda\e^{\mp\i t\omega/2}\frac{\sin(t\delta)}{\delta}\sigma_\pm,
\end{split}
\eeq
where $\sigma_\pm\defeq(\sigma_x\pm\i\sigma_y)/2$.

\begin{theorem}\label{thm-xy-tp}
\hspace{0em}
\ben
\item\label{it:thone} The instrument modeling the two-time measurement protocol for the {\upshape XXZ}-spin interaction with thermal probe  is given by
\beq{fri-ti2}
\begin{split}
\Phi_{\pm\pm}[X] &\defeq\frac{\e^{\mp\beta_p\epsilon/2}}{2\cosh(\beta_p\epsilon/2)}  V_{\pm\pm}^\ast X V_{\pm\pm},\\
\Phi_{\mp\pm}[X] &\defeq\frac{\e^{\pm\beta_p\epsilon/2} }{2\cosh(\beta_p\epsilon/2)} V_{\pm\mp}^\ast XV_{\pm\mp}.
\end{split}
\eeq
\item\label{it:thtwo} $\Phi$ is irreducible and
\[ 
\rho\defeq\rho_p=\frac{1}{2\cosh(\beta_p \epsilon/2)}
\begin{bmatrix}\e^{-\beta_p\epsilon/2}&0\\0& \e^{\beta_p\epsilon/2}\end{bmatrix}
\]
is the unique density matrix satisfying $\Phi^\ast[\rho]=\rho$. In particular, all the conclusions of Theorem~\ref{thm-two-time} hold. 
\item\label{it:ththree} $\ep(\cJ,\rho)=0$ and $e\equiv 0$.
\een

\medskip\noindent
In the following, we set $\ds s\defeq\left(\frac\lambda\delta\sin(\delta t)\right)^2\in[0,1]$.

\medskip
\ben\setcounter{enumi}{3}
\item\label{it:thfour} 
The unraveling of $((\Phi_a)_{a\in\cA}, \rho)$ is the PMP measure $\PP$ generated by 
$((M_a)_{a\in\cA}, {\bf p})$, where 
\beq{bell-1}
\begin{aligned}
M_{++}&\defeq p\begin{bmatrix}1&0\\0&1-s\end{bmatrix},\quad &
M_{--}&\defeq(1-p)\begin{bmatrix}1-s&0\\0&1\end{bmatrix},\\
M_{+-}&\defeq p\begin{bmatrix}0&0\\s&0\end{bmatrix},\quad &
M_{-+}&\defeq(1-p)\begin{bmatrix}0&s\\0&0\end{bmatrix},
\end{aligned}
\eeq
and ${\bf p}\defeq[p\ \ 1-p]$, with
\[
p\defeq\frac{\e^{-\beta_p\epsilon/2}}{2\cosh(\beta_p\epsilon/2)}.
\]
\item\label{it:thfive} For  $s\in {]}0,1[$, the measure  $\PP$ is not weak Gibbs.
\item\label{it:thsix} If $s=0$, that is  if $\delta t\in\pi\nn^\ast$, then $\PP$  is a Bernoulli measure.
\item\label{it:thsev}  If $s=1$, that is if $\epsilon=\omega$ and $\lambda t\in\pi(\nn+1/2)$,  then $\PP$ is a Markov measure.
\een
\end{theorem}

\proof Parts~\ref{it:thone}--\ref{it:thtwo} follow from elementary calculations and Remark~\ref{rem:frops}. By Theorem~\ref{thm-two-time}~\ref{it:2tfour} one has
\[
\ep(\cJ,\rho)=\beta_p\epsilon\,\tr(\rho \Phi_{-+}[\one])-\beta_p\epsilon\,\tr(\rho \Phi_{+-}[\one])=0.
\]
Since $e$ is real analytic on $\rr$, convex, and $\e^\prime(0)=\e^\prime(1)=0$, we have $e\equiv 0$, which gives~\ref{it:ththree}. 
To prove~\ref{it:thfour}, observe that the two-dimensional space of diagonal $2\times2$ matrices is invariant under each map $\Phi_a$. Expressing the restriction of these maps in the basis $(P_+,P_-)$ yields the desired representation. Concerning~\ref{it:thfive}, consider, for each $T\in\nn^\ast$, the sequence $\omega\in\Omega$ where $\omega_k=(+, +)$ for $k\not=T+1$, $\omega_{T+1}= (-,+)$. Using~\ref{it:thfour} one derives 
\[
\frac{\PP_{2T+1}(\omega)}{\PP_{T+1}(\omega)\PP_T\circ\phi^{T+1}(\omega)}= \frac{(1-s)^T}{(1-p)(1-s)^T + p},
\]
from which we conclude that
\[
\lim_{T\rightarrow \infty}\frac 1 T \sup_{S\in [1,T-1]}\sup_{ \omega\in\supp\PP}\left|\log  \frac{\PP_T(\omega)}{\PP_{S}(\omega)
\PP_{T-S}(\phi^S(\omega))}\right|\ge\frac12\log\frac1{1-s}>0,
\]
and so, by Part~\ref{it:noetwo} of Theorem~\ref{noe-thm}, the measure $\PP$ is not weak Gibbs. 
Parts~\ref{it:thsix} and~\ref{it:thsev} are obvious.\qed
\subsubsection{Two-time measurements with random thermal probes}
\label{sec-xy-rtb}
We now turn to the proof of Theorem~\ref{thm-xxz-ran}, using the notation of the corresponding paragraph of Section~\ref{sec-two-time}. For each  $k\in\lbr1,K\rbr$ we set $\cH_k\defeq\cc^2$, 
\[  
\rho_k\defeq\frac{1}{2\cosh(\beta_k \epsilon/2)}
\begin{bmatrix}\e^{-\beta_k\epsilon/2}&0\\0& \e^{\beta_k\epsilon/2}\end{bmatrix},
\]
and the unitary $U_k$ is given by~(\ref{fri-ti00}, \ref{fri-ti0}). However, it will be convenient to use the following representation of the alphabet
\[
\cA\defeq\left\{kuv\mid k\in\lbr1,K\rbr\text{ and }u,v\in\{-,+\}\right\}.
\]

\begin{theorem}\label{thm-wbde}
\hspace{0em}
\ben
\item\label{it:rtone} The instrument modeling the two-time measurement for the {\upshape XXZ}-spin interaction with random thermal probe, is $\cJ\defeq(\Phi_a)_{a\in\cA}$,  where 
\[
\Phi_{kuv}\defeq w_k\Phi_{uv}^{(k)},
\]
$\Phi_{uv}^{(k)}$ being given by~\eqref{fri-ti2} with $\beta_p=\beta_k$.
\item\label{it:rttwo} $\Phi=\sum_{a\in \cA}\Phi_a$ is irreducible and the unique density matrix satisfying $\Phi^\ast[\rho]=\rho$ is given by
\[
\rho\defeq\begin{bmatrix} p&0\\0&1-p\end{bmatrix},
\quad \text{where} \quad p\defeq\sum_{k=1}^K w_k \frac{\e^{-\beta_k \epsilon/2}}{2\cosh (\beta_k \epsilon/2)}.\]
In particular, all the conclusions of Theorem~\ref{thm-two-time} hold. 
\item\label{it:rtthree} 
\[
\ep(\cJ,\rho)=\frac{s}{2}\sum_{k,l=1}^Kw_kw_l\frac{(\beta_k-\beta_l)\epsilon/2\,\sinh((\beta_k-\beta_l)\epsilon/2)}{\cosh((\beta_k+\beta_l)\epsilon/2)+\cosh((\beta_k-\beta_l)\epsilon/2)}.
\]
\item\label{it:rtfour} $\ep(\cJ,\rho)=0$ if and only if $s=0$ or   $\beta_1=\beta_2 = \dots=\beta_K$.
\item\label{it:rtfive}
\[
e(\alpha)=\log\left[1-\frac{s}{2}\left(1 -\sqrt{1+\Delta(\alpha)}\right)\right],
\]
where
\[
\Delta(\alpha)\defeq\sum_{k,l=1}^Kw_kw_l\frac{\cosh((1-2\alpha)(\beta_k-\beta_l)\epsilon/2)-\cosh((\beta_k-\beta_l)\epsilon/2)}{\cosh((\beta_k+\beta_l)\epsilon/2)+\cosh((\beta_k-\beta_l)\epsilon/2)}.
\]
\item\label{it:rtsix} The unraveling $\PP$ of the instrument $(\cJ,\rho)$ is the PMP measure generated by $((M_a)_{a\in\cA}, {\bf p})$, where ${\bf p}\defeq[p\ \ 1-p]$ and
\[
M_{kuv}\defeq w_kM_{uv}^{(k)}, 
\]
$M_{uv}^{(k)}$ being given by~\eqref{bell-1} with $p\defeq\e^{-\beta_k\epsilon/2}/2\cosh(\beta_k\epsilon/2)$.
\item\label{it:rtsev} For  $s\in {]}0,1[$ the measure  $\PP$ is not weak Gibbs.
\item\label{it:rteight} If $s=0$, that is  if $\delta t\in\pi\nn^\ast$, then $\PP$  is a Bernoulli measure.
\item\label{it:rtnin}  If $s=1$, that is if $\epsilon=\omega$ and $\lambda t\in \pi (\nn+1/2)$,  then $\PP$ is a Markov measure.
\een
\end{theorem}

\proof The proof of~\ref{it:rtsev} is the same as the proof of Part~\ref{it:thfive} of Theorem~\ref{thm-xy-tp}.  The remaining parts follow from the identifications~\eqref{rp-1} and~\eqref{rp-2}, and elementary computations that we omit.\qed
\subsubsection{Two-time measurements with multi-thermal probes}
\label{sec-tris}
In this section we study the general case of two-time measurements of XXZ-spin interaction with multi-thermal probes. In the end, we will specialize the discussion to the case $\omega=\epsilon$ and $\mu=0$, which will provide a proof of Theorem~\ref{thm-xxz-multi}.

By the well-known representation theory of $\mathfrak{su}(2)$, the Hilbert space $\cH_p$ and the associated tensor product of two spin-$\tfrac12$ representations split into the direct sum of a one-dimensional singlet sector $\cH_0$ carrying the trivial (spin-$0$) representation and a $3$-dimensional triplet sector $\cH_1$ carrying the spin-$1$ representation~\cite[Sections~XIII.26--27]{Messiah2}. Denoting the latter by ${\bf S}\defeq(S_x,S_y,S_z)$ and identifying $\cH\otimes\cH_p=\cH\otimes(\cH_0\oplus\cH_1)$ with $(\cH\otimes\cH_0)\oplus(\cH\otimes\cH_1)=\cH\oplus(\cH_1\oplus\cH_1)$, we can rewrite the Hamiltonian in block-matrix form
\[
H=\left[
\begin{array}{c|cc}
\frac{\omega}{2}\sigma_z&&\\
\hline
&(\epsilon+\mu)S_z+\frac{\omega}{2}&\sqrt{2}\lambda S_-\\
&\sqrt{2}\lambda S_+&(\epsilon-\mu)S_z-\frac{\omega}{2}
\end{array}
\right].
\]
Comparison with~\eqref{Hmatrix} allows us to apply Lemma~\ref{xy-u} to compute the unitary propagator
\[
U=\e^{-\i tH}=\left[
\begin{array}{c|cc}
\e^{-\i t\frac{\omega}{2}\sigma_z}&&\\
\hline
&U_{++}&U_{+-}\\
&U_{-+}&U_{--}
\end{array}
\right].
\]  
An elementary calculation gives
\begin{gather*}
\begin{split}U_{\pm\pm}\defeq\e^{\mp\i t(\frac\omega2+\epsilon\pm\mu)}\Pi_{\pm}
&+\e^{\i t\frac{\mu\mp\epsilon}2}  \left(\cos(t\delta_\pm)\mp\i\frac{\nu_\pm}{\delta_\pm}\sin(t\delta_\pm)\right)\Pi_0\\
&+\e^{\i t\frac{\mu\pm\epsilon}2} \left(\cos(t\delta_\mp)\mp\i\frac{\nu_\mp}{\delta_\mp}\sin(t\delta_\mp)\right)\Pi_\mp, 
\end{split}\\[2pt]
U_{\pm\mp}\defeq-\i\frac{\sqrt{2}\lambda}{\delta_\pm}\e^{\i t\frac{\mu\mp\epsilon}2} \sin(t\delta_\pm)\Pi_0S_\mp
          -\i\frac{\sqrt{2}\lambda}{\delta_\mp}\e^{\i t\frac{\mu\pm\epsilon}2}\sin(t\delta_\mp)\Pi_\mp S_\mp,
\end{gather*}
where $\Pi_{+/0/-}$ denote the spectral projections of $S_z$ and
\[
\nu_\pm\defeq\frac12\left(\omega-\epsilon\pm\mu\right),\qquad
\delta_\pm\defeq\sqrt{4\lambda^2+\nu_\pm^2}.
\]
To evaluate~\eqref{twotimeviolin} and derive the expression of the instrument $\cJ=(\Phi_a)_{a\in\cA}$, we need to express the projections $\one\otimes P_l$ as well as $X\otimes\one$ in the same basis, an elementary exercise which leads to
\begingroup  
\allowdisplaybreaks
\begin{gather*}
\one\otimes P_{++}=\left[\begin{array}{c|cc}0&&\\\hline&\Pi_+&\\&&\Pi_+\end{array}\right],\quad
\one\otimes P_{+-}=\frac12\left[\begin{array}{c|cc}\one&|+\rangle\langle0|&|-\rangle\langle0|\\\hline
|0\rangle\langle+|&\Pi_0&\\|0\rangle\langle-|&&\Pi_0\end{array}\right],\\
\one\otimes P_{-+}=\frac12\left[\begin{array}{c|cc}\one&-|+\rangle\langle0|&-|-\rangle\langle0|\\\hline
-|0\rangle\langle+|&\Pi_0&\\-|0\rangle\langle-|&&\Pi_0\end{array}\right],\quad
\one\otimes P_{--}=\left[\begin{array}{c|cc}0&&\\\hline&\Pi_-&\\&&\Pi_-\end{array}\right],\\
X\otimes\one=\left[\begin{array}{c|cc}X&&\\\hline&X_{++}\one&X_{+-}\one\\&X_{-+}\one&X_{--}\one\end{array}\right],
\end{gather*}
\endgroup 
where $|0\rangle$ is the eigenvector of $S_z$ corresponding to the eigenvalue $0$,  $|\pm\rangle$ the eigenvectors of $\sigma_z$, and
$X_{ij}$ the corresponding matrix elements of $X$.

Direct computations give that our two-time measurement  process with multi-thermal probe is described by the instrument 
\[
\Phi_a:X\mapsto w_a V_a^\ast X V_a,
\]
where
\begingroup  
\allowdisplaybreaks
\begin{gather*}
\begin{aligned}
V_{\pm\pm\pm\pm}&\defeq\e^{\pm\i t(\frac{\omega+\epsilon\pm3\mu}2)}P_\pm+\bar a_\pm P_\mp,&\qquad 
V_{\pm\pm\mp\mp}&\defeq0,\\
V_{++\pm\mp}&=V_{\pm\mp++}^\ast\defeq b_+\sigma_+,&
V_{--\pm\mp}&=V_{\pm\mp--}^\ast\defeq b_-\sigma_-,\\
\end{aligned}
\\
\begin{split}
V_{\pm\mp\pm\mp}&\defeq\frac12\left(\e^{-\i t\frac\omega2}+\e^{\i t\frac{\epsilon-\mu}2}a_+\right)P_+
+\frac12\left(\e^{\i t\frac\omega2}+\e^{\i t\frac{\epsilon+\mu}2}a_-\right)P_-,\\
V_{\pm\mp\mp\pm}&\defeq\frac12\left(\e^{-\i t\frac\omega2}-\e^{\i t\frac{\epsilon-\mu}2}a_+\right)P_+
+\frac12\left(\e^{\i t\frac\omega2}-\e^{\i t\frac{\epsilon+\mu}2}a_-\right)P_-,
\end{split}
\end{gather*}
\endgroup 
with\footnote{Note that $|a_\pm|^2+2b_\pm^2=1$.}
\[
a_\pm\defeq\cos(t\delta_\pm)\mp\i\frac{\nu_\pm}{\delta_\pm}\sin(t\delta_\pm),\qquad
b_\pm\defeq\frac{\sqrt{2}\lambda}{\delta_\pm}\sin(t\delta_\pm),
\]
and, for $a=(a_1,a_2,a_3,a_4)\in\cA$,
\[
w_a\defeq\frac{\e^{-(\beta_1a_1+\beta_2a_2)\epsilon/2}}{4\cosh(\beta_1\epsilon/2)\cosh(\beta_2\epsilon/2)}.
\]
Since all $\Phi_a$'s  preserve the two-dimensional subspace of diagonal matrices, we can achieve a PMP representation $(M_a)_{a\in\cA}$ of the instrument $\cJ$ in the same way as in the previous sections, with
\begingroup  
\allowdisplaybreaks
\begin{gather*}
\begin{aligned}
M_{\pm\pm\pm\pm}&\defeq w_{\pm\pm}(P_\pm+|a_\pm|^2P_\mp),&\qquad 
M_{\pm\pm\mp\mp}&\defeq0,\\
M_{++\pm\mp}&\defeq w_{++}b_+^2\sigma_-,&
M_{\pm\mp--}&\defeq w_{\pm\mp}b_-^2\sigma_-,\\
M_{--\pm\mp}&\defeq w_{--}b_-^2\sigma_+,&
M_{\pm\mp++}&\defeq w_{\pm\mp}b_+^2\sigma_+,\\
\end{aligned}
\\
\begin{split}
M_{\pm\mp\pm\mp}&\defeq\frac{w_{\pm\mp}}4
\left( |1+\e^{\i t\nu_+}a_+|^2P_++ |1+\e^{-\i t\nu_-}a_-|^2P_-\right),\\
M_{\pm\mp\mp\pm}&\defeq\frac{w_{\pm\mp}}4
\left( |1-\e^{\i t\nu_+}a_+|^2P_++ |1-\e^{-\i t\nu_-}a_-|^2P_-\right).\\
\end{split}
\end{gather*}
\endgroup
The matrix corresponding to the map $\Phi$ is
\[
M\defeq\sum_{a\in \cA}M_a=\begin{bmatrix}
1-\pi_- &\pi_-\\
\pi_+ &1-\pi_+
\end{bmatrix},
\qquad
\pi_\pm\defeq2w_{\pm\pm}b_\pm^2+(w_{+-}+w_{-+})b_\mp^2.
\]
In the trivial case\footnote{This case only happens when $b_-=b_+=0$.} $\pi_+=\pi_-=0$, one has $M=\one$, any state $\rho$ satisfies $\Phi^\ast[\rho]=\rho$, and the unraveling of $(\cJ,\rho)$ is a convex combination of two Bernoulli measures. In what follows, we shall assume that $\pi_\pm$ are not both zero, so that $M$ and hence $\Phi$ are irreducible.  The unique invariant state is
\[
\rho\defeq\frac{1}{\pi_-+\pi_+}\begin{bmatrix}\pi_+ &0\\0&\pi_-\end{bmatrix},
\]
and the unraveling $\PP$ of $(\cJ,\rho)$ is the PMP measure generated by $((M_a)_{a\in\cA},{\bf p})$ with probability vector
${\bf p}\defeq[\pi_+\ \ \pi_-]/(\pi_++\pi_-)$.

Suppose that $\pi_+\not=0$, and for any $T\in\nn^\ast$ let $\omega\in \Omega$ be such that $\omega_k={+}{+}{+}{+}$ for $k\not=T+1$ and $\omega_{T+1}={+}{-}{+}{+}$. It follows that
\[
\frac{\PP_{2T+1}(\omega)}{\PP_{T+1}(\omega)\PP_T\circ\phi^{T+1}(\omega)}
=\frac{1+\frac{\pi_-}{\pi_+}}{|a_+|^{-2T}+\frac{\pi_-}{\pi_+}},
\]
and hence
\[
\lim_{T\rightarrow \infty}\frac 1 T \sup_{S\in [1,T-1]}\sup_{ \omega\in\supp\PP}\left|\log  \frac{\PP_T(\omega)}{\PP_{S}(\omega)
\PP_{T-S}(\phi^S(\omega))}\right|\ge\log\frac1{|a_+|}>0.
\]
By  Part~\ref{it:noetwo} of Theorem~\ref{noe-thm}, the measure $\PP$ is not weak Gibbs. A completely similar argument holds when $\pi_-\not=0$.

Another elementary calculation, starting with Formula~\eqref{tbus-l}, gives
\begin{align*}
\ep(\cJ,\rho)&=\frac A2
\frac{\left((\beta_1-\beta_2)\epsilon/2\right)\sinh\left((\beta_1-\beta_2)\epsilon/2\right)}
{\cosh\left((\beta_1-\beta_2)\epsilon/2\right)+\cosh\left((\beta_1+\beta_2)\epsilon/2\right)},
\end{align*}
where
\[
A\defeq\frac{\e^{-(\beta_1+\beta_2)\epsilon/2}b_+^2c_+^2
+\cosh\left((\beta_1-\beta_2)\epsilon/2\right)\left(b_+^2c_-^2+b_-^2c_+^2\right)
+\e^{(\beta_1+\beta_2)\epsilon/2}b_-^2c_-^2}
{\e^{-(\beta_1+\beta_2)\epsilon/2}b_+^2
+\cosh\left((\beta_1-\beta_2)\epsilon/2\right)\left(b_+^2+b_-^2\right)
+\e^{(\beta_1+\beta_2)\epsilon/2}b_-^2},
\]
with $c_\pm^2\defeq2b_\pm^2+|1-\e^{\pm\i t\nu_\pm}a_\pm|^2$. In particular, $\ep(\cJ,\rho)=0$ if and only if $\beta_1=\beta_2$. Recalling~\eqref{tbus-l} again, one easily computes the matrix  $M(\alpha)$ and its largest eigenvalue which gives the entropic pressure
\[
e(\alpha)=\log\left( \frac12\tr M(\alpha)+ \sqrt{\left(\frac 12\tr M(\alpha)\right)^2-\det M(\alpha)}\right).
\]
We shall not write the general expression of this function, but restrict ourselves to the special case $\epsilon=\omega$ and $\mu=0$ considered in Theorem~\ref{thm-xxz-multi}. Using the relations
\[
a_+=a_-=\cos(2\lambda t), \qquad b_+=b_-=\frac1{\sqrt2}\sin(2\lambda t),\qquad c_+=c_-=2\sin(\lambda t),
\]
we observe that $\pi_++\pi_-=\sin^2(2\lambda t)$ so that $\pi_\pm$ are not both vanishing iff $\lambda t\not\in\frac\pi2\nn^\ast$. 
Moreover, an explicit evaluation of the previous formula gives
\[
e(\alpha)=2\log\left(\cos^2(\lambda t)+A(\alpha)\sin^2(\lambda t)\right),
\]
with
\[
A(\alpha)\defeq\left(
\frac{\cosh\left((2\alpha-1)(\beta_1-\beta_2)\epsilon/2\right)+\cosh\left((\beta_1+\beta_2)\epsilon/2\right)}
{\cosh\left((\beta_1-\beta_2)\epsilon/2\right)+\cosh\left((\beta_1+\beta_2)\epsilon/2\right)}
\right)^{1/2}.
\]
Differentiation at $\alpha=0$ further yields
\[
\ep(\cJ,\rho)=2\sin^2(\lambda t)
\frac{\left((\beta_1-\beta_2)\epsilon/2\right)\sinh\left((\beta_1-\beta_2)\epsilon/2\right)}
{\cosh\left((\beta_1-\beta_2)\epsilon/2\right)+\cosh\left((\beta_1+\beta_2)\epsilon/2\right)}.
\]
\subsection{X00-spin instruments}
\label{xx-main}
This  model has been described in  Section~\ref{sec-nevnnevxx}. We shall again identify $\cH\otimes\cH_p$ with 
$\cH\oplus \cH$, so that the Hamiltonian~\eqref{xx-ham} reads
\[
H=\frac{1}{2}\begin{bmatrix}\omega\sigma_z +\epsilon &\lambda\sigma_x\\ \lambda\sigma_x& \omega\sigma_z-\epsilon.
\end{bmatrix}.
\]
It is a simple exercise to check that the associated propagator is given by
\beq{u-nev} 
U\defeq\e^{-\i tH}=\begin{bmatrix}V_{++}&V_{+-}\\V_{-+} & V_{--}\end{bmatrix},
\eeq
where
\beq{v-xy-int}
\begin{split}
V_{\pm\pm}&\defeq\cos\left(\Omega_\pm t/2\right)
-\i(\omega\sigma_z\pm\epsilon)\Omega_\pm^{-1}\sin\left(\Omega_\pm t/2\right),\\[4pt]
V_{\pm\mp} &\defeq-\i\lambda\Omega_\pm^{-1}\sin\left(\Omega_\pm t/2\right)\sigma_x,
\end{split}
\eeq
and
\[
\Omega_\pm\defeq\left(\lambda^2+(\omega\sigma_z\pm\epsilon)^2\right)^{1/2}.
\]

\subsubsection{One-time measurements}
\label{xx-onetime-long}
We prove Theorem~\ref{xx-one-int}.
The instrument $\cJ=(\Phi_-,\Phi_+)$ describing one-time measurements is given by
\[
\Phi_\pm[X]\defeq\left(\frac12-\eta\right)V_{\pm+}^\ast XV_{\pm+}+\left(\frac12+\eta\right)V_{\pm-}^\ast XV_{\pm-}.
\]
Both $\Phi_+$ and $\Phi_-$ leave the two-dimensional space of diagonal matrices invariant. The matrices describing their action on this space are
\beq{Pu-eq}
M_\pm\defeq\begin{bmatrix}\left(\frac12\mp\eta\right)(1-s_\pm)&\left(\frac12\pm\eta\right)s_\mp\\
\left(\frac12\pm\eta\right)s_\pm&\left(\frac12\mp\eta\right)(1-s_\mp)\end{bmatrix},
\eeq
where $s_\pm$ is given by~\eqref{eq-upm}. Note that, as the parameters $\epsilon,\omega,\lambda$ and $t$ vary on ${]}0,\infty{[}$, the pair $(s_-,s_+)$ takes all values in the set $[0,1]\times[0,1{[}$. Thus, for any $\Phi^\ast$-invariant state $\rho$, the instrument $(\cJ,\rho)$ admits a PMP representation. 

If $s_+=s_-=0$, then $M_\pm=(\frac12\pm\eta)\one$, any state $\rho$ is invariant under $\Phi^\ast$, and the unraveling $\PP$ of $(\cJ,\rho)$ is the Bernoulli measure generated by the mass function $Q(\pm)=\frac12\pm\eta$. In the opposite cases,
an elementary calculation shows that the unique probability vector $\mathbf{p}$ satisfying $\mathbf{p}(M_-+M_+)=\mathbf{p}$ is given by $\mathbf{p}=[p\ \ 1-p]$ with $p$ given by~\eqref{str-3}. In particular, the density matrix~\eqref{str-2} is the unique state $\rho$ for which the instrument $(\cJ,\rho)$ satisfies Assumption~\assref{(A)}. 
This yields Parts~\ref{it:otone}--\ref{it:otthree}. To prove Part~\ref{it:otfour}, observe that, in the special case $(s_-,s_+)=(1,0)$, for any $T\in\nn^\ast$ one has
\[
\PP_T({-}{-}\cdots{-}{-})=\frac12+\eta,\qquad
\PP_T({+}{+}\cdots{+}{+})=\frac12-\eta.
\]
Parts~\ref{part:X00-KSp}--\ref{part:X00-KSm} follow by comparing~\eqref{Pu-eq} with~\eqref{ksdef}, while Part~\ref{it:otsev} follows from the fact that, for $\eta=0$, the probability vector $\mathbf{p}=[1/2\ \ 1/2]$ is invariant under both $M_\pm$. Part~\ref{it:oteight}
and the fact that Assumption~\assref{(C)} holds is a direct consequence of Lemma~\ref{PMP-props}~\ref{it:PMPfour} and Theorem~\ref{thm-int}~\ref{it:AssDdifferentiability_e_alpha}.

In view of Remark~\ref{rem:1} and Theorem~\ref{thm-int}~\ref{it:ergoconvsigma}, in order to prove Part~\ref{it:otnine} it suffices to show that, under the conditions $|\eta|\in{]}0,1/2{[}$ and $s_\pm\in{]}0,1{[}$,  the identity $\PP_T=\wP_T$ for all $T\in\nn^\ast$ implies $s_\pm=1/2$. From the Perron--Frobenius theorem for matrices with strictly positive entries, we infer that the eigenvalues $r_\pm$ and $u_\pm$ of $M_\pm$ satisfy $0\le|u_\pm|<r_\pm$, the spectral projection associated to the dominant eigenvalue $r_\pm$ having strictly positive entries. Writing the identity ${\bf p}M_+^T {\bf 1}={\bf p}M_-^T{\bf 1}$, which  holds for all $T\in\nn$, in terms of the spectral representation of $M_\pm$ yields
\beq{thenose}
a_+r_+^T+(1-a_+)u_+^T=a_-r_-^T+(1-a_-)u_-^T
\eeq
for some $a_\pm>0$. Taking the logarithm on both sides of this identity, dividing by $T$ and letting $T\to\infty$, we deduce $r_+=r_-=r>0$. Dividing both sides of~\eqref{thenose} by $r^T$ and letting again $T\to\infty$ gives $a_+=a_-=a>0$. If $a\not=1$ then, considering again~\eqref{thenose}, $u_+=u_-$ and
hence
\[
0=\tr(M_--M_+)=2\eta(2-(s_++s_-)),
\]
which contradicts our hypotheses. Thus, $a=1$ and Relation~\eqref{thenose} with $T=1$ yields
\[
r=\frac12\pm\eta\left(1-\frac{4s_+s_-}{s_++s_-}\right),
\]
\ie $r=1/2$ and $4s_+s_-=s_++s_-$. Inserting the last relation into $\PP_3({+}{-}{+})=\wP_3({+}{-}{+})$ further yields
\[
0=\mathbf{p}(M_+M_-M_+-M_-M_+M_-)\mathbf{1}
=-2\eta^3\left((s_++s_-)^2-3(s_++s_-)+2\right),
\]
which implies $s_++s_-=1=4s_+s_-$ and hence $s_\pm=1/2$.

Finally, to prove Part~\ref{it:otten}, set
\[
W\defeq\begin{bmatrix}
(1+4\eta^2)(s_++s_-)-4\eta(s_+-s_-)&-(1-4\eta^2)(s_++s_-)\\
-(1-4\eta^2)(s_++s_-)&(1+4\eta^2)(s_++s_-)+4\eta(s_+-s_-)
\end{bmatrix},
\]
and, observing that 
\[
M_\pm^\mathsf{T}=W^{-1}M_\pm W,\qquad
{\bf p}W=\frac{16\eta^2 s_+s_-}{s_++s_-}{\bf 1}^\mathsf{T},\qquad
\frac{16\eta^2 s_+s_-}{s_++s_-}W^{-1}{\bf 1}={\bf p}^\mathsf{T},
\]
we conclude that for any $T\in\nn^\ast$ and $\omega\in\Omega$,
\begin{align*}
\PP_T(\omega)=
{\bf p}M_{\omega_1}\cdots M_{\omega_T}{\bf 1}
&={\bf p}WW^{-1}M_{\omega_1} W\cdots W^{-1}M_{\omega_T} WW^{-1}{\bf 1}\\
&={\bf 1}^\mathsf{T}M_{\omega_1}^\mathsf{T}\cdots M_{\omega_T}^\mathsf{T}{\bf p}^\mathsf{T}
={\bf p}M_{\omega_T}\cdots M_{\omega_1}{\bf 1}
=\wP_T(\omega).
\end{align*}
The proof of Theorem~\ref{xx-one-int} is complete.

To conclude the discussion of the one-time measurements of the X00-spin system, we note that by Parts~\ref{part:X00-KSp}--\ref{part:X00-KSm} the  failure of the fluctuation--dissipation relations for the Keep--Switch instrument discussed in Section~\ref{sec:FDR} translates to the failure of these relations for the X00-spin instrument with pure probe state, \ie  $|\eta|=1/2$.

\subsubsection{Two-time measurements with a thermal probe}
\label{sec-xx-ttm}
We prove Theorem~\ref{thm:X00tp}. One easily determines the two-time X00-spin instrument with thermal probe $\cJ=(\Phi_a)_{a\in\cA}$ to be 
\beq{fri-ma2}
\begin{split}
\Phi_{\pm\pm}[X] &\defeq\frac{\e^{\mp\beta\epsilon/2} }{2\cosh(\beta \epsilon/2)} 
V_{\pm\pm}^\ast X V_{\pm\pm},\\
\Phi_{\mp\pm}[X] &\defeq\frac{\e^{\pm\beta\epsilon/2}}{2\cosh(\beta \epsilon/2)} 
V_{\pm\mp}^\ast X V_{\pm\mp},
\end{split}
\eeq
where $V_{\pm\pm}$ and $V_{\pm\mp}$ are given by~\eqref{v-xy-int}. Recall that $s_\pm$ are given by~\eqref{eq-upm}. All the maps $\Phi_a$ leave the two-dimensional space of diagonal matrices invariant, and their action on this space is given by the matrices $(M_a)_{a\in\cA}$, where
\beq{equ:X002M}
M_{\pm\pm}\defeq\frac{\e^{\mp\beta \epsilon/2}}{2\cosh (\beta \epsilon/2)}\begin{bmatrix}
1-s_{\pm}&0\\0&1-s_{\mp}
\end{bmatrix},\qquad
M_{\mp\pm}\defeq\frac{\e^{\pm\beta \epsilon/2}}{2\cosh (\beta \epsilon/2)}\begin{bmatrix}
0&s_{\mp} \\s_{\pm}&0
\end{bmatrix}.
\eeq

\begin{remark}\label{rem:rtp}
Recalling the construction of the random thermal probes in Section~\ref{sec-two-time} and observing that 
\[
\cK_\pm=\left(\begin{bmatrix} 1-s_\pm&0\\0&1-s_\mp\end{bmatrix},\begin{bmatrix} 0 &s_\pm\\s_\mp&0\end{bmatrix}\right),
\]
defines two pairs of Keep--Switch matrices, we conclude that the  two-time X00-spin instrument can be viewed as a randomization of the Keep--Switch instruments defined by $\cK_+$ and $\cK_-$ with respective weights $w_\pm\defeq\e^{\mp\beta \epsilon/2}/2\cosh (\beta \epsilon/2)$.
\end{remark}

Consider first the special case $s_+=s_-=0$. Since $M_{\mp\pm}=0$, $M_{\pm\pm}$ are both multiples of the identity, and $M\defeq\sum_{a\in\cA}M_a=\one$, any probability vector ${\bf p}$ is left-invariant with respect to $M$ and the PMP measure generated by $((M_a)_{a\in\cA},{\bf p})$ is the Bernoulli measure concentrated on $\{{-}{-},{+}{+}\}^{\nn^\ast}\subset\Omega$ and associated to the mass function $Q({\pm\pm})=\e^{\mp\beta\epsilon/2}/2\cosh(\beta \epsilon/2)$. This yields Part~\ref{it:ttone}.

Excluding the preceding case, one easily deduces that the matrix $M$ and hence the map $\Phi$ are irreducible. Setting $\eta\defeq\frac12\tanh(\beta\epsilon/2)$, a simple calculation shows that ${\bf p}\defeq[p\ \ 1-p]$, with $p$ given by~\eqref{str-3}, is the unique invariant probability vector for $M$. This settles Parts~\ref{it:tttwo}--\ref{it:ttthree}. Parts~\ref{it:ttfour} and~\ref{it:ttfive} are easily established by applying Formulas~\eqref{tbus-l}.

If $s_-=s_+=1/2$, direct calculation shows that $\PP$ is the Bernoulli measure on $\Omega$ generated by the mass function 
\[
Q({+}{+})=Q({+}{-})=\frac12\frac{\e^{-\beta\epsilon/2}}{2\cosh(\beta\epsilon/2)},\qquad
Q({-}{+})=Q({-}{-})=\frac12\frac{\e^{\beta\epsilon/2}}{2\cosh(\beta\epsilon/2)}.
\]
 In the opposite cases, given the connection with the Keep--Switch instrument mentioned in Remark~\ref{rem:rtp}, the proof of Part~\ref{it:ttsix} is the same as in the Keep--Switch case given in Section~\ref{sec-KS-i}. This concludes the proof of Theorem~\ref{thm:X00tp}.

\subsubsection{Two-time measurements with random thermal probes}
\label{sec-xx-rttm}
We prove Theorem~\ref{thm:X00random}.
The setting and notation are  the same as in Section~\ref{sec-xy-rtb}, except that now each $U_k$ is given by~\eqref{u-nev}. It follows that the relevant instrument $\cJ\defeq(\Phi_a)_{a\in\cA}$ is given by
\[
\Phi_{kuv}\defeq w_k\Phi_{uv}^{(k)},
\]
where the map $\Phi_{uv}^{(k)}$ is given by~\eqref{fri-ma2} with $\beta=\beta_k$. Its action on diagonal $2\times2$ matrices is described by $M_{kuv}\defeq w_kM_{uv}^{(k)}$,  $M_{uv}^{(k)}$ being given by~\eqref{equ:X002M} with $\beta=\beta_k$. 

If $s_+=s_-=0$, then the argument of the previous section carries over, and we conclude that for any diagonal density matrix $\rho$, the unraveling of $(\cJ,\rho)$ is the Bernoulli measure on $(\lbr1,K\rbr\times\{{-}{-},{+}{+}\})^{\nn^\ast}$ generated by the mass function $Q(k{\pm\pm})=w_k\frac{\e^{\mp\beta_k\epsilon/2}}{2\cosh(\beta_k\epsilon/2)}$. This yields Part~\ref{it:rmone}.

Assuming now that $s_++s_->0$, we again observe that $M\defeq\sum_{a\in \cA}M_a$ and hence
the map $\Phi$ are irreducible. The unique left-invariant probability vector ${\bf p}\defeq[p\ \ 1-p]$ of $M$ is easily seen to be given by~\eqref{str-3} with $\eta$ as in~\eqref{equ:X00eta2}. The remaining parts of Theorem~\ref{thm:X00random} are proved in a similar way to their counterparts of Theorem~\ref{thm:X00tp}.


\section{Rotational instruments}
\label{sec-ex-ro}

In this section we prove Theorem~\ref{rot-theorem}. We assume throughout that  $\Delta\in\cI\defeq[0,2{[}\setminus\qq$.

Concerning our assumptions, we note that since  $\rho=\frac{1}{2}\one$ and $\Phi[\one]=\sum_{a\in\cA}\Phi_a[\one]=\one$, Assumption~\assref{(A)} holds. Moreover, for any $\omega,\nu\in\Omega_{\rm fin}$, one has
\beq{eq:proofasscrot}
\PP([\omega 3\nu])
=\frac12\tr(\Phi_{\omega}\circ\Phi_3\circ\Phi_{\nu}[\one])
=\frac1{12}\tr(\Phi_{\omega}[\one])\tr(\Phi_{\nu}[\one])=\frac13\PP([\omega])\PP([\nu]),
\eeq
and similarly for $\wP$ (recall that $\theta(3) = 3$).  Thus, Assumption~\assref{(C)} holds with $\tau=1$. Since $\Phi_1^2=0$, we have $\PP([\omega])=0$ whenever $\omega\in\Omega_{\rm fin}$ contains the string $11$. In particular,
\beq{eq:P110}
\PP([11]) = 0
\eeq 
shows that $\tau=1$ is the smallest integer one can take in Assumption~\assref{(C)}.

We will now prove that, for $T \in \nn^*$,
\beq{eq:OmetaTplus}
\Omega_T^+\defeq\supp\PP_T=\{\omega\in\Omega_T\mid\PP([\omega])>0\}
\eeq
consists of all the words in $\Omega_T$ that do not contain the string $11$. In particular, in view of the choice of the involution $\theta$, Assumption~\assref{(B)} follows from \eqref{eq:OmetaTplus}.

Before we prove \eqref{eq:OmetaTplus}, we make the following additional observations.

First, since $\Phi_2\geq\frac 12\Phi_3$, it follows from~\eqref{eq:proofasscrot} that for $\omega,\nu\in\Omega_{\rm fin}$,
\beq{eq:presdec2}
 \PP([\omega2\nu])\geq\frac12\PP([\omega3\nu])=\frac16\PP([\omega])\PP([\nu]).
\eeq

Moreover, for all $T,T'\in \nn$ and $\omega\in\Omega_{\rm fin}$,
\beq{eq:rotpadded0}
\PP([0^T\omega0^{T'}])=3^{-T-T'}\PP([\omega]).
\eeq
 Next, denoting by $(e_1,e_2)$ the canonical basis of $\cc^2$ and noting that $V=|e_2\rangle\langle e_1|$, we find that for $\omega,\nu\in\Omega_{\rm fin}$,
\beq{eq:w1v}
\begin{split}
\PP([\omega1\nu])&=\frac1{24}\tr(\Phi_\omega\big[|e_2\rangle\langle e_1|\Phi_\nu[\one]e_1\rangle\langle e_2|\big])=\frac1{24}\langle e_1|\Phi_\nu[\one]e_1\rangle\tr(\Phi_\omega[|e_2\rangle\langle e_2|])\\
&=6\tr(\Phi_{1\nu}[\one] )\tr(\Phi_{\omega1}[\one])=24\PP([\omega1])\PP([1\nu]).
\end{split}	
\eeq
Finally, the central identity in this section is that, for all $T\in \nn^\ast$,
\beq{eq:P10t1}
\begin{split}
\PP([10^T1])&=2^{-1}\tr(\Phi_1\circ\Phi_0^T\circ\Phi_1[\one])\\
&=(288)^{-1}3^{-T}\tr(|e_2\rangle\langle e_1|R_{T\Delta}e_2\rangle\langle e_2| R_{T\Delta}^\mathsf{T}e_1\rangle\langle e_2|)\\
&=(288)^{-1}3^{-T}\sin^2(T\pi\Delta).
\end{split}	
\eeq
The assumption $\Delta\in\cI$ guarantees that $\sin^2(T\pi\Delta)>0$ for all $T\in\nn^\ast$.

We now return to the proof of \eqref{eq:OmetaTplus}. We have already seen that if $\omega \in \Omega_{\rm fin}$ contains the string 11, then $\PP([\omega])=0$. It remains to prove the converse. To this end, suppose, by contradiction, that there is $\omega_0\in\Omega_{\rm fin}$, not containing $11$, such that $\PP([\omega_0])=0$. From~\eqref{eq:proofasscrot} and~\eqref{eq:presdec2}, we deduce that $\omega_0$ contains a subword $\xi\in\{0,1\}^m$ for some $m\geq 1$ such that $\PP([\xi])=0$. Using then \eqref{eq:rotpadded0},
we further see that $\xi$ can be taken of the form 
\[
\xi=10^{n_1}10^{n_2}1\cdots10^{n_r}1,
\]
for some $r\in \nn^\ast$ and $n_1,\ldots,n_r\in\nn^\ast$. Using now \eqref{eq:w1v}, we conclude that there must be $i\in \lbr 1 , r\rbr$ such that $\PP([10^{n_i}1])=0$, which contradicts \eqref{eq:P10t1}. We have thus established \eqref{eq:OmetaTplus}, and the proof of Part~\ref{prt:rot-ass} of Theorem~\ref{rot-theorem} is complete.
 
It should be clear from the previous discussion that the strings $101, 1001, 10001, \dots$ play an important role. In order to enumerate them, we introduce the following notation.
\begin{definition}\label{def:defNr}
Given $\omega\in\Omega_T$, let $r\in \nn$ be maximal such that there exist $\ell_1<\ell_2<\dots<\ell_r$ and $n_1,n_2,\dots,n_r\in\nn^\ast$ such that $\omega_{\lbr\ell_i,\ell_i+n_i+1\rbr}=10^{n_i}1$, $i\in\lbr1,r\rbr$. We write then $N_{\omega}=(n_1,n_2,\dots,n_r)$. If $\omega$ contains no subword of the kind $10^n1$ with $n\geq 1$, then $r=0$ and $N_\omega=()$.
\end{definition}

 For example, if $\omega = 1031\underline{00}1\underline{000}10321\underline{0}1\underline{000}10$, we have $N_\omega = (2,3,1,3)$ (corresponding to the length of the strings of zeroes underlined). 

Let $\ell(x)\defeq\min_{p\in\zz} |x-p|$. It is then immediate that
\beq{eq:compsin2}
\ell(T\Delta) \leq |\sin(T\pi\Delta)| = |\sin(\pi \ell(T\Delta))| \leq \pi\ell(T\Delta).
\eeq

\begin{lemma}\label{lem:Pnlb} There is a constant $C>0$, depending on $\Delta$ only, such that for any $T\in\nn^\ast$ and $\omega \in \Omega_T^+$,
\[
\PP([\omega])\geq\e^{-CT}\prod_{i=1}^r\sin^2(n_i\pi\Delta)
\geq\e^{-CT}\prod_{i=1}^r(\ell(n_i\Delta))^2,
\]
with $(n_1, \dots, n_r)=N_\omega$ (when $r=0$, the products above are taken to be 1).
\end{lemma}
\proof
The second inequality follows from~\eqref{eq:compsin2}. We now prove the first one.
An easy induction argument relying on~\eqref{eq:proofasscrot} and~\eqref{eq:presdec2} shows that it suffices to prove the result in the case where $\omega$ contains only 0's and 1's. And in that case, the result immediately follows from~\eqref{eq:rotpadded0}, \eqref{eq:w1v} and~\eqref{eq:P10t1}.
\qed

Part~\ref{prt:rot-reg} of Theorem~\ref{rot-theorem} is the contents of
\begin{proposition}\label{prop:finiteeverywhereas} 
For Lebesgue-almost all $\Delta\in \cI$, we have $e(\alpha )<\infty$ for all $\alpha\in\rr$.
\end{proposition}
\proof
First assume that $\Delta$ has the following property: there exists a constant $C>0$ such that for all $T\in\nn^\ast$,
\beq{eq:proplnzeta}
\ell(T\Delta) \geq C\e^{-T}.	
\eeq
Then, by Lemma~\ref{lem:Pnlb}, since $\sum_{i=1}^r n_i < T$, there exists $C'>0$ such that $|\sigma_T(\omega)| \leq C' T$ for all $T\in\nn^\ast$ and $\omega\in \Omega_T^+$, so that $e(\alpha) \leq C' |\alpha|$. {\tiny }

We now give a direct proof that~\eqref{eq:proplnzeta} is satisfied for almost all $\Delta$ (this is of course well known, since in particular~\eqref{eq:proplnzeta} holds for all Diophantine numbers $\Delta$). Denote by $\lambda$ the normalized Lebesgue measure on $[0,2]$ and consider the sets $A_T\defeq\{\Delta\in[0,2]\mid\ell(T\Delta)<\e^{-T}\}$ with $T\in\nn^\ast$. Then $\lambda(A_T)\le2\e^{-T}$ so that $\sum_{T\in\nn^\ast}\lambda(A_T) <\infty$. By the Borel--Cantelli lemma there is a set $E\subset[0,2]$ with $\lambda(E)=1$ such that any $\Delta\in E$ belongs at most to a finite number of $A_T$'s. Since clearly $E\subset \cI$, any $\Delta \in E$ satisfies $\ell(T\Delta) >0$ for all $T\in \nn^\ast$, and thus also \eqref{eq:proplnzeta}  for some $C>0$. The proof is complete.
\qed

We now prove Parts~\ref{prt:rot-nost} and~\ref{prt:rot-st} of Theorem~\ref{rot-theorem}. Part~\ref{prt:rot-nost} follows from Propositions~\ref{prop:infiniteoutside01} and~\ref{prop:rotationalderfinite} below, and Part~\ref{prt:rot-st} follows from  Propositions~\ref{prop:infiniteoutside01} and~\ref{prop:rotationalderINfinite}.

Let $\Gamma\colon\nn^\ast\to[1,\infty{[}$ be an increasing function such that
\[
\lim_{T\to\infty}\Gamma(T)=+\infty,\qquad\sup_{T\in\nn^\ast} T\e^{-\Gamma(T)}<\infty
\] 
(we shall consider the cases $\Gamma(T) = T^2$ and $\Gamma(T) = \e^{T^2}$ below). We prove in Lemma~\ref{lem:continuedfrac} (with $\psi=\e^{-\Gamma}$) that there exists a dense set $I_\Gamma\subset[0,2{[}$ such that for all $\Delta\in I_\Gamma$,
\beq{eq:liminfqre}
0 < \liminf_{T\to\infty} \ell(T\Delta)\e^{\Gamma(T)}  <\infty.
\eeq
Note in particular that~\eqref{eq:liminfqre} implies that $\Delta$ is irrational, so that $I_\Gamma\subset\cI$.

We make the convention that $c>0$ is a constant depending on $\Delta$ and $\Gamma$ only, which can be different each time it appears. 

For further reference, we note that if $\Delta \in I_\Gamma$, then for all $T\in\nn^\ast$,
\beq{eq:logellgtrmgammaT}
\log\ell(T\Delta)\geq-\Gamma(T)-c,
\eeq
and there exists a sequence $T_i\to\infty$ such that
\beq{eq:qizetai}
\log\ell(T_i\Delta)\leq-\Gamma(T_i)+c.
\eeq

\begin{proposition}\label{prop:infiniteoutside01} Assume that $\Gamma$ is such that
$\lim_{T\to\infty}T^{-1}\Gamma(T)=\infty$ and that $\Delta\in I_\Gamma$.
Then $e(\alpha ) = +\infty$ for all $\alpha\notin [0,1]$.
\end{proposition}
\proof
By the symmetry~\eqref{es-s}, it is enough to prove the result for $\alpha > 1$. By~\eqref{eq:P10t1}, \eqref{eq:compsin2} and~\eqref{eq:presdec2}, we have
\beq{eq:P10t122}
\PP([10^{T}1]) \leq \e^{-cT}(\ell (T\Delta))^2, \qquad \wP([10^{T}1])=\PP([12^{T}1]) \geq  \e^{-cT}. 
\eeq
As a consequence, we obtain
\[
\begin{split}
\frac1{T+2}\log\sum_{\omega\in\Omega^+_{T+2}}\e^{(1-\alpha)\log\PP([\omega])
+\alpha\log\wP([\omega])}
&\geq\frac1{T+2}\log\e^{(1-\alpha)\log\PP([10^{T}1])+\alpha\log\wP([10^{T}1])}\\
&\geq\frac{(1-\alpha)(-cT+2\log\ell(T\Delta))-cT\alpha}{T+2}.
\end{split}	
\]
By our assumption on $\Gamma$, the right-hand side diverges along the sequence $T_i$ of~\eqref{eq:qizetai}, and hence the proof is complete.\qed

By~\eqref{eq:leftrightder}, we have
\[
(\partial^-e)(1)=-(\partial^+e)(0)=\ep(\cJ,\rho)=-h_\phi(\PP)
-\lim_{T\to\infty}\frac1T\sum_{\omega\in\Omega_T^+}\PP([\omega])\log\wP([\omega]).
\]
(Recall that the Kolmogorov--Sinai entropy satisfies $h_\phi(\PP)\in[0,\log4]$, since $|\cA| = 4$.)
For later convenience, we note that the above can also be expressed as
\beq{eq:partialem1}
(\partial^-e)(1)=-(\partial^+e)(0)=-h_\phi(\PP)
-\lim_{T\to\infty}\frac1T\sum_{\omega\in\Omega_T^+}\wP([\omega])\log\PP([\omega]).
\eeq

\begin{proposition}\label{prop:rotationalderINfinite} Let $\Gamma(T)=\e^{T^2}$ for all $T\in \nn^\ast$ and let $\Delta \in I_\Gamma$.
Then $e(\alpha)=+\infty$ for all $\alpha\notin[0,1]$, and $(\partial^-e)(1)=-(\partial^+e)(0)=+\infty$.
\end{proposition}
\proof
By Proposition~\ref{prop:infiniteoutside01}, $e(\alpha) = +\infty$ when $\alpha \notin [0,1]$. Moreover, by~\eqref{eq:partialem1} and~\eqref{eq:P10t122}, we find
\begin{align*}
(\partial^-e)(1) &\geq-\log4-\liminf_{T\to\infty}\frac1{T+2}\wP([10^T1])\log \PP([10^T1])\\
&\geq-\log4-\liminf_{T\to\infty}\frac {\e^{-cT}}{T+2}\log(\ell(T\Delta)).
\end{align*}
In view of~\eqref{eq:qizetai}, and by our choice of $\Gamma$, we conclude that $(\partial^-e)(1) = +\infty$.\qed

\begin{proposition}\label{prop:rotationalderfinite} Let $\Gamma(T) = T^2$ for all $T\in \nn^\ast$ and let $\Delta \in I_\Gamma$. Then $e(\alpha) = + \infty$ for all $\alpha \notin [0,1]$, and $(\partial^-e)(1)=-(\partial^+e)(0) < \infty$.
\end{proposition}
\proof
We have by Proposition~\ref{prop:infiniteoutside01} that $e(\alpha) = + \infty$ for all $\alpha \notin [0,1]$.  We prove here that $(\partial^-e)(1)<\infty$. For $\omega\in\Omega_T$, let $N_\omega=(n_1,\ldots, n_r)$, and note that if $\omega \in \Omega_T^+$, then
\[
-\log\PP([\omega])\leq cT+2\sum_{i=1}^r\Gamma(n_i),
\]
as a consequence of Lemma~\ref{lem:Pnlb} and~\eqref{eq:logellgtrmgammaT}. From this and~\eqref{eq:partialem1}, we obtain
\beq{eq:rprime0}
(\partial^-e)(1)\leq c+2\limsup_{T\to\infty}\frac  {u_T}T,
\eeq
where, for $ T\in\nn^\ast$, we have set
\beq{eq:defuts}
u_T\defeq\sum_{\omega\in\Omega_T}f_T(\omega)\wP([\omega]),
\qquad f_T(\omega)\defeq\sum_{i=1}^r\Gamma(n_i)
\eeq
(recall that Assumption~\assref{(B)} holds, so that $\wP([\omega]) = 0$ when $\omega \in \Omega_T \setminus \Omega_T^+$).

We thus need to show that $u_T$ increases at most linearly.  For $T\geq 3$ and $n\in\lbr1,T-2\rbr$, consider the sets
\[
A_{T,n}\defeq\{\omega\in\Omega_T\mid\omega_{\lbr1,n+2\rbr}=10^n1\},
\]
and let
\[
A_{T,0}\defeq\Omega_T\setminus\bigcup_{n=1}^{T-2}A_{T,n}.
\]
We observe that for all $T\geq 3$, $n\in\lbr0,T-2\rbr$ and $\omega \in A_{T,n}$,
\beq{eq:ftomegagamman}
f_T(\omega)=\Gamma(n)+f_{T-1}(\omega_{\lbr2,T\rbr}),
\eeq
with the convention $\Gamma(0) = 0$. Since for $T\geq 3$ the family $(A_{T,n})_{n\in\lbr0,T-2\rbr}$ is a partition of $\Omega_T$, we find, using the invariance of $\wP$,
\begin{align*}
u_T&=\sum_{n=0}^{T-2}\sum_{\omega\in A_{T,n}}f_T(\omega)\wP([\omega]),\\
u_{T-1}&=\sum_{\omega\in\Omega_T}f_{T-1}(\omega_{\lbr2,T\rbr})\wP([\omega])
=\sum_{n=0}^{T-2}\sum_{\omega\in A_{T,n}}f_{T-1}(\omega_{\lbr2,T\rbr})\wP([\omega]).
\end{align*}
Thus, using~\eqref{eq:ftomegagamman} and recalling that $\Gamma(0) = 0$, we find, for $T\geq 3$,
\begin{align*}
u_T-u_{T-1}&=\sum_{n=1}^{T-2}\sum_{\omega\in A_{T,n}}\Gamma(n)\wP([\omega])
=\sum_{n=1}^{T-2}\Gamma(n)\wP([10^n1])\leq\sum_{n=1}^{T-2}\Gamma(n)3^{-n-2}\leq c,
\end{align*}
where the next-to-last inequality relies on
\[
\wP([10^n1])=\PP([12^n1])\le\PP([13^n1])\le\PP([3^{n+2}])
=\frac12 3^{-n-2}.
\]
The right-hand side of~\eqref{eq:rprime0} is thus finite, which completes the proof.\qed


\appendix 



\section{Continued fractions}
\label{sec-cont-fra}

In this appendix we prove Lemma~\ref{lem:continuedfrac}, which was used in the last section. To this end, we start with a brief summary of some properties of continued fractions (see for example~\cite{khinchin_continued_1964,burger_exploring_2000} for more details).

Let $(a_n)_{n\in\nn}\subset\zz$ be such that $a_n\in \nn^\ast$ for all  $n\in \nn^\ast$.\footnote{We recall the convention chosen in Section~\ref{sec-setup} about $\nn$ (which includes $0$) and $\nn^\ast$ (which does not).} Define
\[
[a_0; a_1]\defeq a_0 + \frac 1 {a_1}	, \quad [a_0; a_1,a_2] = a_0 + \frac 1 {a_1 + \frac 1 {a_2}},
\]
and, more generally, for $i\in \nn^\ast$,
\[
[a_0; a_1, \dots, a_i]\defeq a_0 + \dfrac{1}{a_1 + \dfrac{1}{a_2 + \dfrac{1}{\substack{\displaystyle a_3+\\\big.} ~ \ddots ~   \dfrac{1}{a_{i-1} + \dfrac{1}{a_i}}}}}.
\]

It is well known that the limit
\[
[a_0;a_1, a_2, \dots]\defeq \lim_{i\to\infty} [a_0; a_1, \dots, a_i]
\]
exists and is irrational. There is a bijection between the sequences $(a_n)_{n\in\nn}$ such that $a_n\in \nn^\ast$ for all  $n\in \nn^\ast$ and the irrational numbers. Moreover, for each $\zeta\in\rr\setminus\qq$, we have
\beq{eq:continuedfracexp}
\zeta=[a_0;a_1, a_2, \dots],
\eeq
where 
\begin{align*}
\zeta_0&\defeq\zeta, &\qquad  a_0&\defeq\lfloor \zeta_0 \rfloor,\\
\zeta_{i+1}&\defeq\frac1{\zeta_i-a_i},&\qquad a_{i+1} &\defeq \lfloor \zeta_{i+1} \rfloor, \qquad i\in \nn.
\end{align*}
The right-hand side of~\eqref{eq:continuedfracexp} is called the {\em continued fraction expansion} of $\zeta$, and this expansion is unique.

For $i\in \nn^\ast$, let $p_i \in \zz$ and $q_i \in \nn^\ast$ be such that  the fraction
\beq{eq:piqidef}
\frac {p_i}{q_i} =  [a_0; a_1, \dots, a_i]
\eeq
is irreducible, and let $p_{-1} \defeq 1$, $q_{-1} \defeq 0$, $p_0 \defeq a_0$ and $q_0 \defeq 1$.  We then have for $i\in \nn^\ast$,
\begin{align}\label{eq:reqrelm}
	\begin{bmatrix}
	p_i & p_{i-1}\\
q_i & q_{i-1}
	\end{bmatrix}& = 	\begin{bmatrix}
	p_{i-1} & p_{i-2}\\
q_{i-1} & q_{i-2}
	\end{bmatrix} \begin{bmatrix}
		a_i & 1 \\ 1& 0
	\end{bmatrix},
\end{align}
and, in particular, $q_{i+1} > q_i$. 
It is also well known that if $\zeta$ is given by~\eqref{eq:continuedfracexp} (so that $\zeta = \lim_{i\to\infty} \frac {p_i}{q_i}$), then
\beq{eq:sandwichzeta}
\frac {p_{0}}{q_{0}}< \frac {p_{2}}{q_{2}}	< \frac {p_{4}}{q_{4}}< \dots < \zeta < \dots < \frac {p_{5}}{q_{5}} <\frac {p_{3}}{q_{3}}< \frac {p_{1}}{q_{1}},
\eeq
and 
\beq{eq:twoboundsxiqi}
\frac 1 {2  q_{i+1}}\leq \left|{q_i}\zeta  -   {p_i}\right|  \leq \frac 1 {q_{i+1}}, \qquad i\in \nn.
\eeq

We recall here the {\em best approximation property} (see for example \cite[Theorem 5.9]{burger_exploring_2000}).
\begin{lemma}\label{lem:bestapprox}
Fix $i\in \nn$, and let $(p,q)\in (\zz \times \lbr 1, q_{i+1}\rbr) \setminus \{(p_{i}, q_{i}), (p_{i+1}, q_{i+1})\}$.
Then
\beq{zetaqmp}
|\zeta q - p|  > |\zeta q_{i}-p_{i}|	.
\eeq
\end{lemma}
\proof
Let $x,y \in \zz$ be such that
\[
\begin{bmatrix}
	p_{i+1} & p_{i}\\
q_{i+1}& q_{i}
\end{bmatrix} \begin{bmatrix}
	x\\y
\end{bmatrix} = \begin{bmatrix}
 p\\q
 \end{bmatrix}.
\]
Such $x,y\in \zz$ exist, as the determinant of the matrix here is $\pm 1$ by~\eqref{eq:reqrelm}.
We consider two cases. 

First, if $x=0$ then $(p,q) = (yp_i, yq_i)$. Clearly $y>0$, since $q, q_i > 0$, and in fact the condition $(p,q) \neq (p_i, q_i)$ implies that $y \geq 2$. The result is then obvious, since then $|\zeta q - p|   \geq  2|\zeta q_{i}-p_{i}|$.

We now assume that $x\neq 0$. The condition
\beq{eq:conditionqxy}
q=xq_{i+1}+yq_{i}\in  \lbr 1, q_{i+1}\rbr
\eeq 
implies that $xy<0$. Indeed, clearly \eqref{eq:conditionqxy} implies that $xy\leq 0$, and if we had $y=0$, then by \eqref{eq:conditionqxy} we would find $x=1$, which contradicts the condition $(p,q)\neq (p_{i+1}, q_{i+1})$. This shows that $xy<0$.

Since also $(\zeta q_{i+1}-p_{i+1})(\zeta q_{i}-p_{i}) < 0$ by~\eqref{eq:sandwichzeta}, we conclude that $x(\zeta q_{i+1}-p_{i+1})$ and $y(\zeta q_{i}-p_{i})$ have the same sign. But then,
\begin{align*}
|\zeta q - p| &= |x(\zeta q_{i+1}-p_{i+1})+y(\zeta q_{i}-p_{i})|	\\
&= |x||\zeta q_{i+1}-p_{i+1}|+|y||\zeta q_{i}-p_{i}|	 > |\zeta q_{i}-p_{i}|,
\end{align*}
which completes the proof.\qed

Let $\ell(x)\defeq\min_{p\in \zz} |x-p|$ as in Section~\ref{sec-ex-ro}.
Let $i\in \nn^\ast$. Since $q_{i+1}>q_i$, Lemma~\ref{lem:bestapprox} applies to the pairs $(p, q_i)$ for all $p \neq p_i$,
from which we conclude that\footnote{Note that \eqref{eq:ellzetaqi} does not hold for $i=0$ in general, because we may have $q_1 = q_0=1$ (for example if $\zeta = \pi/4$) and so Lemma~\ref{lem:bestapprox} may not apply to all pairs $(p, q_i)$, $p \neq p_i$.}
\beq{eq:ellzetaqi}
\ell(\zeta q_i) =|\zeta q_i -  {p_{i}}|.
\eeq
In addition, for all $i\in \nn$, applying Lemma~\ref{lem:bestapprox} to all pairs $(p,q)\in (\zz \times  \lbr 1, q_{i+1}-1\rbr)\setminus \{(p_i, q_i)\}$ shows that 
\beq{eq:twoboundsxiqi2}
\ell(\zeta q)\geq \left|\zeta q_i -  {p_{i}}\right|,  \qquad q\in \lbr 1, q_{i+1}-1\rbr .
\eeq

\begin{lemma}\label{lem:continuedfrac}
Let $\psi\colon\nn^\ast\to{]}0,1]$  be a decreasing function such that $\sup_{q\in\nn^\ast}q\psi(q)<\infty$. Then, there exists a dense subset $U\subset\rr$ such that for all $\zeta\in U$,
\beq{eq:liminfq}
0<\liminf_{q\to\infty}\frac{\ell(q\zeta)}{\psi(q)}<\infty.	
\eeq
\end{lemma}
\proof
Fix any open interval $I \subset \rr$ and fix $x \in I\setminus \qq$. Let $(\widehat a_i)_{i\in \nn}$ be such that $x=[\widehat a_0; \widehat a_1, \widehat  a_2, \dots]$, and let $\widehat p_i, \widehat q_i$ be as in~\eqref{eq:piqidef} with $(a_i)$ replaced by $(\widehat a_i)$. Then, since $x = \lim_{i\to\infty}\frac {\widehat p_i}{\widehat q_i}$, one can choose $N$ large enough so that $\frac {\widehat  p_{N-1}}{\widehat  q_{N-1}} \in I$ and $\frac {\widehat p_N}{\widehat q_N} \in I$. We then consider $\zeta = [a_0; a_1, a_2, \dots ]$, where $a_i = \widehat a_i$ for $i\leq N$, and $a_{i+1}=\min\{n\in\nn\mid nq_i\psi(q_i)\ge1\}$ for $i\geq N$. Then, since $p_i =\widehat  p_i$ and $q_i =\widehat  q_i$  for $i\leq N$, and by~\eqref{eq:sandwichzeta}, we obtain that $\zeta \in I$. It remains to show that $\zeta$ satisfies~\eqref{eq:liminfq}. For all $i\geq N$, we have by~\eqref{eq:ellzetaqi}, \eqref{eq:twoboundsxiqi} and~\eqref{eq:reqrelm} that
\[
 \ell(\zeta q_i) = \left|\zeta q_i -  {p_{i}}\right|  \leq \frac 1 {q_{i+1}} \leq \frac 1 {a_{i+1}q_i} \leq \psi(q_i),
\]
so that the second inequality in~\eqref{eq:liminfq} holds. Moreover, by~\eqref{eq:twoboundsxiqi2}, \eqref{eq:twoboundsxiqi} and~\eqref{eq:reqrelm}, we have for all $i\geq N$ and all $q\in \lbr q_i,  q_{i+1}-1\rbr$ that 
\begin{align*}
\ell(\zeta q) &\geq  \left|\zeta q_i -  {p_{i}}\right|\geq \frac 1 {2 q_{i+1}} \geq \frac 1 {2(a_{i+1}+1)q_i} \geq \frac 1 {2(\frac 1 {q_i\psi(q_i)}+2)q_i}\\
&   = \frac {\psi(q_i)} {2(1+2 q_i \psi(q_i))}   \geq C^{-1} \psi(q_i) \geq  C^{-1} \psi(q),
\end{align*}
where $C = 2(1+2 \sup_{q\in \nn}q\psi(q))$. This establishes the first inequality in~\eqref{eq:liminfq}, hence the proof is complete.\qed

\addcontentsline{toc}{section}{References}
\newcommand{\etalchar}[1]{$^{#1}$}
\providecommand{\bysame}{\leavevmode \hbox to3em{\hrulefill}\thinspace}
\providecommand{\og}{``}
\providecommand{\fg}{''}
\providecommand{\smfandname}{and}
\providecommand{\smfedsname}{eds.}
\providecommand{\smfedname}{ed.}
\providecommand{\smfmastersthesisname}{Master Thesis}
\providecommand{\smfphdthesisname}{Thesis}

\end{document}